\documentclass[12pt]{article}

\usepackage[toc,page,header]{appendix}

\usepackage{minitoc}
\usepackage{framed}

\usepackage{comment}

\usepackage{multibib}
\newcites{apx}{Appendix References}
\usepackage{amsthm}
\usepackage[framemethod=TikZ]{mdframed} 
\usepackage{enumitem}
\setitemize{noitemsep,topsep=0pt,parsep=0pt,partopsep=0pt}
\usepackage{caption}
\usepackage{xcolor}
\usepackage{tikz}
\usetikzlibrary{tikzmark}
\usepackage{array}
\usepackage{multirow}

\usepackage{dcolumn}
\usepackage{placeins}
\usepackage{cancel}
\usepackage[authoryear]{natbib}
\usepackage[colorlinks = true,
            linkcolor = black,
            urlcolor  = blue,
            citecolor = black,
            anchorcolor = black]{hyperref}
\usepackage{subcaption}
\usepackage{setspace}
\usepackage{amsmath,amsfonts,bm,mathtools,physics,graphicx,longtable,booktabs}
\usepackage{lscape}

\DeclareMathOperator*{\argmin}{arg\,min}

\newcommand{\cN}{\mathcal{N}}
\newcommand{\bA}{\bm{A}}
\newcommand{\ba}{\bm{a}}
\newcommand{\bD}{\bm{D}}
\newcommand{\bI}{\bm{I}}
\newcommand{\bM}{\bm{M}}
\newcommand{\bQ}{\bm{Q}}
\newcommand{\bR}{\bm{R}}
\newcommand{\bU}{\bm{U}}
\newcommand{\bV}{\bm{V}}
\newcommand{\bX}{\bm{X}} 
\newcommand{\bW}{\bm{W}}
\newcommand{\bS}{\bm{S}}
\newcommand{\bs}{\bm{s}}
\newcommand{\bY}{\bm{Y}}
\newcommand{\bz}{\bm{z}}
\newcommand{\bZ}{\bm{Z}}
\newcommand{\inty}{\infty}

\newcommand{\bbeta}{\bm\beta}
\newcommand{\bgamma}{\bm\gamma}
\newcommand{\bdelta}{\bm\delta}
\newcommand{\bepsilon}{\bm\epsilon}
\newcommand{\bLambda}{\bm\Lambda}
\newcommand{\btheta}{\bm\theta}
\newcommand{\bnu}{\bm\nu}

\newcommand{\bxi}{\bm\xi}
\newcommand{\bzeta}{\bm\zeta}
\newcommand{\bOmega}{\bm\Omega}
\newcommand{\oracle}{\mathrm{oracle}}
\newcommand{\ridge}{\mathrm{penalty}}
\newcommand{\sdf}{\mathrm{subset}}
\newcommand{\wndf}{\mathrm{wn}}
\newcommand{\pndf}{\mathrm{pm}}
\newcommand{\full}{\mathrm{full}}
\newcommand{\deconf}{\mathrm{deconf}}
\newcommand{\naive}{\mathrm{na\ddot{\i}ve}}
\newcommand{\nonlinear}{\mathrm{nonlinear}}
\newcommand{\E}{\mathbb{E}}
\newcommand{\Var}{\mathrm{Var}}
\newcommand{\Cov}{\mathrm{Cov}}
\DeclareMathOperator*{\plim}{\mathit{p}-lim}

\newtheorem{lem}{Lemma}
\newtheorem{prop}{Proposition}
\newtheorem{thm}{Theorem}
\newtheorem{defn}{Definition}

\newcommand{\indep}{\mbox{$\perp\!\!\!\perp$}}
\DeclarePairedDelimiterX{\inp}[2]{\langle}{\rangle}{#1, #2}

\usepackage[margin = 1in]{geometry}

\newcommand{\blind}{0}

\newcommand{\titl}{
Na\"ive regression requires weaker assumptions than factor models to adjust for multiple cause confounding
}

\if1\blind {
  \title{\titl} 
  \date{\today}
  }\fi

\if0\blind {
  \title{\titl\thanks{For helpful discussion and feedback we thank David Blei, Matias Cattaneo, Guilherme Duarte, Chris Felton, Sandy Handan-Nader, Gary King, Apoorva Lal, Chris Lucas, Ian Lundberg, Jason Mian Luo, Jonathan Mummolo, Clayton Nall, Betsy Ogburn, Marc Ratkovic, Fredrik S\"avje, Eric Tchetgen Tchetgen, Roc\'io Titiunik, Matt Tyler, Daniel Thompson, Yixin Wang, Sean Westwood, Yiqing Xu. Research reported in this publication was supported by The Eunice Kennedy Shriver National Institute of Child Health \& Human Development of the National Institutes of Health under Award Number P2CHD047879.}
 }

\author{Justin Grimmer\thanks{Professor, Department of Political Science, Stanford University and Senior Fellow at the Hoover Institution. \href{mailto:jgrimmer@stanford.edu}{jgrimmer@stanford.edu}.} \and Dean Knox\thanks{Assistant Professor, Operations, Information and Decisions Department, Wharton School, University of Pennsylvania. \href{mailto:dcknox@upenn.edu}{dcknox@upenn.edu}.} \and Brandon M. Stewart\thanks{Assistant Professor and Arthur H. Scribner Bicentennial Preceptor, Department of Sociology and Office of Population Research, Princeton University. \href{mailto:bms4@princeton.edu}{bms4@princeton.edu}.}}
 \date{Draft Version:
 \today
  }
    }\fi

\doparttoc %

\begin{document}

\def\spacingset#1{\renewcommand{\baselinestretch}%
{#1}\small\normalsize} \spacingset{1}

  \maketitle\thispagestyle{empty}
  \begin{abstract} %
\noindent The empirical practice of using factor models to adjust for shared, unobserved confounders, $\bZ$, in observational settings with multiple treatments, $\bA$, is widespread in fields including genetics, networks, medicine, and politics. Wang and Blei (2019, WB) formalizes these procedures and develops the ``deconfounder,'' a causal inference method using factor models of $\bA$ to estimate ``substitute confounders,'' $\hat\bZ$, then estimating treatment effects---regressing the outcome, $\bY$, on part of $\bA$ while adjusting for $\hat\bZ$. WB claim the deconfounder is unbiased when there are no single-cause confounders and $\hat{\bZ}$ is ``pinpointed.'' We clarify pinpointing requires each confounder to affect \textit{infinitely} many treatments. We prove under these assumptions, a na\"ive semiparametric regression of $\bY$ on $\bA$ is asymptotically unbiased. Deconfounder variants nesting this regression are therefore also asymptotically unbiased, but variants using $\hat{\bZ}$ and subsets of causes require further untestable assumptions. We replicate every deconfounder analysis with available data and find it fails to consistently outperform na\"ive regression. In practice, the deconfounder produces implausible estimates in WB's case study to movie earnings: estimates suggest comic author Stan Lee's cameo appearances causally contributed \$15.5 billion, most of Marvel movie revenue. We conclude neither approach is a viable substitute for careful research design in real-world applications.
  \end{abstract}

\noindent%
\vfill

\newpage

\tableofcontents

\clearpage

\spacingset{1.45} %

\section{Introduction}
Machine learning methods are increasingly applied across statistics, social science, and industry.  Recently, they have been touted as providing new and flexible methods for causal inference research.  For example, \cite{WanBle19} (WB) introduces the \textit{deconfounder}, an approach for causal inference on multiple treatments, $\bA$, that affect an outcome, $\bY$, in observational settings where shared, unobserved confounders, $\bZ$, affect both $\bA$ and $\bY$. The deconfounder fits a factor model to the treatments to estimate a \textit{substitute confounder},  $\hat\bZ=f(\bA)$, a function of the observed treatments; it then estimates treatment effects with a regression of $\bY$ on $\hat\bZ$ and some part of $\bA$. This procedure generalizes a popular genetics estimator \citep{Price06}; it closely parallels well-established empirical practices in research on social networks, medicine, legislative politics, and beyond.

WB offers a theoretical justification, claiming that adjusting for the substitute confounder in this way will provide unbiased estimates of causal effects under certain conditions: ``the theory finds confounders that are \textit{effectively observed}, even if not explicitly so, and embedded in the multiplicity of the causes'' (p. 7). This is the basic intuition underlying the use of factor models for causal inference: \textit{if the following assumptions are satisfied}, multi-cause confounding has observable implications for the joint distribution of the treatments, which the analyst can capture with dimension reduction of $\bA$ and then adjust for. These are (A.1) there are no confounders that affect only one treatment, (A.2) there is a multi-cause confounder $\bZ$ satisfying weak unconfoundedness, (A.3) the substitute confounder $\hat{\bZ}=f(\bA)$ \textit{pinpoints} the confounder $\bZ$ as defined in Section~\ref{s:deconfounder_definition}, and (A.4) $\bA$ follow a probabilistic factor model and are conditionally independent given $\bZ$.\footnote{\citet{WanBle20} condenses into two assumptions: (1) there is a $\bZ$ which is a smallest $\sigma$-algebra such that $P(\bA|\bZ) = \prod_{j=1}^J P(\bA_j|\bZ)$ and $P(\bA_j|\bZ)$ is not a point mass for any value of $j$.  Conditioning on $\bZ$ must yield $\bA \indep \bY(\mathbf{a}) | \bZ$. (2) $\bZ$ is pinpointed by $\bA$ s.t. P($\bZ|\bA)=\delta_{f(\bA)}$. (These are stronger than ours due to the minimal $\sigma$-algebra condition. \citet{ogburn2019comment} gives alternative assumptions.}

We demonstrate that in this setting, multiple treatments can in fact address unobserved confounding, but only under strong conditions. We clarify that pinpointing (A.3) requires \textit{strong infinite confounding}---every confounder must asymptotically affect an infinite number of treatments, and each confounder's effects on those treatments must not go to zero too quickly \citep{d2019multi}. Thus, beyond the previously stated ``no single-cause confounding'' assumption (A.1), consistent causal inference further requires \textit{no finite-cause confounding}.  In Theorem~\ref{thm:deconfounder_inconsistency_nonlinear}, we show that when only a finite treatments are available, the deconfounder is inconsistent across a broad range of data-generating processes.

Under the above assumptions, we further prove Theorem~\ref{thm:deconfounder_naive_equality}, our main theoretical result: semiparametric regressions of $\bY$ on $\bA$, ignoring confounding---a baseline that WB calls \textit{na\"ive} regression---in fact, surprisingly, asymptotically approach consistency as the number of treatments, $m$, grows large. Versions of the \textit{full deconfounder}---using the na\"ive regression along with $\hat{\bZ}$---also achieve consistency under these conditions, but only because they use the same information as a na\"ive regression. In other words, everything but the treatments is unnecessary: the added factor-model machinery is extraneous. The \textit{subset deconfounder}---a variant popular in genetics \citep{Price06}, regressing $\bY$ on only $\hat{\bZ}$ and a subset of $\bA$---requires strong and unverifiable additional assumptions about treatment effects to yield asymptotically unbiased estimates, and so is asymptotically weakly dominated by na\"ive regression. Together, these results clarify the limits of existing empirical research using factor-model adjustment.

To investigate finite-sample performance, we replicate all six simulation designs used in the deconfounder papers. We reach substantively different conclusions, demonstrating neither the deconfounder nor the na\"ive regression reliably outperforms the other. We show that these conflicting results are largely due to our stabilization of estimation and varying of key simulation parameters. Our analysis further reveals that the factor-model machinery introduces new complications in estimation which can be avoided with na\"ive regression.

To assess the viability of these methods in real-world problems, we revisit WB's main case study, the effects of actor appearances on movie revenue.  We show both the deconfounder and na\"ive regression produce results that are implausible. For example, both WB's model and na\"ive regression estimate that Stan Lee---the legendary comic writer who appears for 200 seconds in Marvel superhero films---caused over an eight-fold increase in movie revenue with his cameos, more than any of the 900 other actors analyzed, and contributed \$15.5 billion of revenue. We show simply adjusting for film budget yields far more reasonable results, even though budget is a quintessential example of the multi-cause confounding that factor modeling is intended to address.

While machine learning methods do allow for more flexible specifications, they do not alter the basic assumptions needed for unbiased causal inference. In contrast, our results show thoughtful research designs \textit{do} weaken the assumptions needed---such as when researchers use proxy variables associated with confounders, but conditionally independent of treatments and outcome \citep[e.g.][]{miao2018identifying}.  While multiple treatments can reveal the imprint of an underlying confounder, we show this requires both strong confounding and a large number of associated, but not causally ordered, treatments.  Given the implausibility of all conditions jointly holding in applied settings, deconfounder-type algorithms are no substitute for explicitly measuring confounders and then adjusting.  Machine learning methods are useful, but the foundational challenges of causal inference remain.

We provide background for the deconfounder in Section~\ref{s:deconfounder}.  Section~\ref{s:asymptotic} characterizes the asymptotic behavior of the deconfounder and the na\"ive regression, followed by a discussion of finite-sample implications and performance in Section~\ref{s:finite}. We discuss implications for real-world studies in Section~\ref{s:applications}, then conclude. Supplements include a notation guide (Supplement~\ref{a:notation}), proofs (Supplement~\ref{a:derivation}), and additional empirical details including replications of every deconfounder simulation and case study with available data (Supplements~\ref{app:sim}--\ref{a:actors}). 
All replication code and data is available on Code Ocean.\footnote{Each simulation and application has its own replicable capsule at the following links:  \href{https://codeocean.com/capsule/6317432/tree}{Medical Deconfounder Simulation 1}, \href{https://codeocean.com/capsule/1786302/tree}{Medical Deconfounder Simulation 2}, \href{https://codeocean.com/capsule/7721162/tree}{Smoking Simulation}, \href{https://codeocean.com/capsule/4729471/tree}{Smoking Simulation Best ELBO},  \href{https://codeocean.com/capsule/9121886/tree}{GWAS}, \href{https://codeocean.com/capsule/6053879/tree}{Logistic Tutorial}, \href{https://codeocean.com/capsule/9195430/tree}{Quadratic Tutorial}, \href{https://codeocean.com/capsule/9344175/tree}{Posterior Predictive Check Simulation}, \href{https://codeocean.com/capsule/7026471/tree}{Subset Deconfounder Simulation}, \href{https://codeocean.com/capsule/0482695/tree}{Actor Case Study}, and \href{https://codeocean.com/capsule/8916869/tree}{Breast Cancer Case Study}.
}

\section{The Deconfounder and Multiple Causal Inference}
\label{s:deconfounder}
WB formalizes and provides statistical theory for a procedure that has been used extensively in genetics, social science, network science, medicine, and industry \citep[e.g.][]{Price06}---estimating a factor model of treatments in an attempt to adjust for shared confounding among those treatments. In this section, we offer background and preview our results. Throughout, we use $\bA$ to denote the set of $m$ treatments for which we have $n$ observations,  $\bZ$ is the true shared unobserved confounder, and $\hat{\bZ}$ is the estimated substitute confounder.

\subsection{The Deconfounder}
\label{s:deconfounder_definition}
WB seeks to ``develop the deconfounder algorithm, prove that it is unbiased, and show that it requires weaker assumptions than traditional causal inference'' (p. 1574). The recommended algorithm is to fit a factor model to the treatments, check its fit, and then run a regression of the outcome on all treatments and low-dimensional representations of each unit extracted from the factor model. WB offer three settings (their Theorems 6--8) which leverage parametric assumptions (Theorem 6) and limitations on what can be estimated (Theorems 7--8) to achieve identification.\footnote{Theorem 6 provides identification of average treatment effects by making parametric assumptions, including that the substitute confounder is piecewise constant and there can be no confounder/cause interactions.  Theorem 7 identifies the average treatment effects of a subset of the treatments.  Finally, Theorem 8 restricts estimation to only those treatments which map to the same value of the substitute confounder.   In short, Theorem 6 leverages functional form assumptions for identification, while Theorems 7 and 8 narrow the set of causal questions the method can answer.}  
All rely on the assumption of ``no single cause confounders'' and what is called ``consistency of the substitute confounder'' in WB, but called ``pinpointing'' in subsequent papers \citep{WanBle20}.  This later assumption requires that the substitute confounder, $\hat{\bZ}$---which is a deterministic function of the observed treatments---is a bijective transformation of $\bZ$. %
While the pinpointing assumption is stated as an exact equality, any method to consistently estimate $\bZ$ requires asymptotics in $n$ and $m$. In Definition~\ref{def:consistency_substitute_confounder} of Supplement~\ref{a:consistency_deconfounder}, we offer a redefinition of pinpointing as an asymptotic property. We analyze the deconfounder in the strongest possible asymptotic regime, where $m \rightarrow \infty$ and for each treatment, $n \rightarrow \infty$.   

\subsection{Related Work}
Even before publication, WB generated considerable conversation focused almost exclusively on its theoretical claims.  In response to a working paper version of WB, \citet{d2019multi} showed that general non-parametric identification is impossible.%
These issues arise because the factor model and the no unobserved single cause confounders assumptions only partially constrain the observed data distribution.%

Commentaries, published alongside WB, clarified implications of several key theoretical assumptions. \citet{imai2019discussion} notes that $\hat{\bZ}$ converges to a function of the observed treatments rather than the true $\bZ$, a random variable \citep[1607]{imai2019discussion}. This is problematic because the adjustment criteria implicitly assumes the support of $p(\hat{\bZ}_i | \bA_i =\mathbf{a})$ is the same as that of $p(\hat{\bZ}_i)$---which cannot be true because pinpointing implies that $p(\hat{\bZ}_i | \bA_i =\mathbf{a})$ is degenerate. Both \citet{ogburn2019comment} and \citet{d2019multi} emphasize that pinpointing generally requires $m$ going to infinity. We build on this point in Theorem~\ref{thm:deconfounder_inconsistency_nonlinear}, which proves the deconfounder is inconsistent for broad classes of data-generating processes with finite numbers of treatments.

\subsection{Takeaways and Contributions}
Because the substitute confounder is a function of the observed treatments, $\hat\bZ = f(\bA)$, the deconfounder estimates $\E[\bY | \bA, \hat\bZ] = \E[\bY | \bA, f(\bA)]$, which reduces to $\E[\bY | \bA]$. In other words, the deconfounder is only a method to learn a transformation of the treatments. There are several important restrictions implicit in the deconfounder assumptions including, notably, that the treatments, $\bA$, cannot causally depend on each other.  We maintain these assumptions and return to them in Section~\ref{s:assumptions}.   

Our contribution is twofold. First, we present Propositions~\ref{prop:bias_naive}--\ref{prop:bias_ridge_nonlinear} and Theorem~\ref{thm:deconfounder_naive_equality}, showing that under the deconfounder assumptions, na\"ive regression is asymptotically unbiased and that every variant deconfounder estimator only also achieves asymptotic unbiasedness if it uses the same information as a na\"ive regression. When the deconfounder uses only a subset of information, additional untestable and strong assumptions must be made about the treatment effects.  Second, we show the theoretical concerns raised here and in prior papers make the deconfounder and na\"ive regression unsuitable for current real-world applications.  %

\section{Asymptotic Theory Justifies the Na\"ive Regression Whenever The Deconfounder Can Be Used}
\label{s:asymptotic}
The most basic data-generating process for multi-cause confounding is a linear factor model and a linear outcome model---a case we call the \textit{linear-linear} model (WB defines this process in Equations 8, 9 and 20). 
We first develop asymptotic theory for this simple setting before generalizing to nonlinear factor and outcome models. %

\subsection{The Linear-Linear Model and Strong Infinite Confounding}
Consider $n$ observations drawn i.i.d. from the following data-generating process:
\begin{align}
   k \text{ unobserved confounders:}& &\bZ_i &\sim \cN(\bm{0}, \bI); \label{e:confounder}\\
   m \ge k \text{ observed treatments:}& &\bA_i &\sim \cN(\bZ_i^\top \btheta, \sigma^2 \bI); \label{e:treatment} \\
   \text{scalar outcome:}& & Y_i &\sim \cN(\bA_i^\top \bbeta + \bZ_i^\top \bgamma, \omega^2); \label{e:outcome}
\end{align}
We assume elements of $\btheta$ and $\bgamma$ are finite and $\sigma^2$ is nonzero. Our goal is to estimate $\bbeta = [\beta_1, \ldots, \beta_m]$, the causal effects of increasing the corresponding $[A_{i,1}, \ldots, A_{i,m}]$ by one unit; following WB, effects are assumed constant. Results are collected in $\bZ$, $\bA$, and $\bY$.\footnote{We occasionally denote simultaneous sampling of all $n$ observations with $\bZ \sim \cN(\bm{0}, \bI)$ or similar.}

The variable $\bZ$ is unobserved and therefore confounds our inferences about the causal effect of $\bA$ when both $\bgamma$ and $\btheta$ are nonzero. However, if the analyst could observe $\bZ$ and adjust for it, they would have the \textit{oracle} estimator,
\begin{align}
\left[ \ 
\hat\bbeta^{\oracle \top} ,\ \hat\bgamma^{\oracle \top}
\ \right]^\top
&\equiv
\left(
\left[ \bA, \bZ \right]^\top \left[ \bA, \bZ \right]
\right)^{-1}
\left[ \bA, \bZ \right]^\top \bY. \label{e:oracle}
\end{align}
It follows directly from the properties of ordinary least squares that the oracle is an unbiased and consistent estimator of treatment effects for any $m$.

No other estimator that we will consider is consistent for finite $m$. However, we define an asymptotic regime in $m$, called strong infinite confounding, under which the na\"ive regression will approach consistency.\footnote{A recent preprint, \citet{guo2020doubly}, defines a related ``dense confounding'' condition.}  
\begin{mdframed}
  \begin{defn}{(Strong infinite confounding under the linear-linear model.)}
    \label{def:infinite_confounding}
    A sequence of linear-linear data-generating processes with a fixed number of
    confounders, $k$, and growing number of causes, $m$, is said to
    be \textbf{strongly infinitely confounded} if as $m \to \infty$, all
    diagonal elements of $\btheta \btheta^\top$ tend to infinity.
\end{defn}
\end{mdframed}
The $j$-th diagonal element of $\btheta \btheta^\top$ contains the sum of the squared coefficients relating confounder $j$ to each of the $m$ treatments.\footnote{Lemma~\ref{l:pinpointing} of  Supplement~\ref{a:derivation} proves strong infinite confounding is necessary for pinpointing, and Lemma~\ref{l:infinite_confounding} connects this to the conditions for unbiased estimation of a na\"ive regression.} Intuitively, strong infinite confounding says that as $m$ grows large, the finite $k$ confounders continue to strongly affect a growing number of treatments. We discuss the practical implication of this condition for finite samples in Section~\ref{s:finite}. In Supplement~\ref{a:weakinfconf}, we build on an example from \cite{d2019comment} to show an example of ``weak'' infinite confounding, where the number of treatments grows but the diagonal elements of $\btheta \btheta^\top$ do not tend towards infinity. %

\subsection{The Na\"ive Regression in the Linear-Linear Setting}
As a baseline for the deconfounder, WB present the \textit{na\"ive} estimator,
which simply ignores $\bZ$. In Proposition~\ref{prop:bias_naive}, we characterize the asymptotic properties of na\"ive regression with finite $m$ and, perhaps surprisingly, establish it is asymptotically unbiased for the linear-linear model as both $n$ and $m$ go to infinity under strong infinite confounding.
\begin{mdframed}
\begin{prop}
\label{prop:bias_naive}
(Asymptotic Bias of the Na\"ive Regression in the Linear-Linear Model.)

Under the linear-linear model, the asymptotic bias of the na\"ive estimator,\\
$\hat{\bbeta}^\naive \ \equiv \ 
\left(
\bA^\top \bA
\right)^{-1}
\bA^\top \bY$,
follows
$\plim_{n \to \infty} \hat\bbeta^\naive - \bbeta 
\ = \ (\btheta^\top \btheta + \sigma^2 \bI)^{-1}
  \btheta^\top \bgamma$,
 indicating that the na\"ive estimator is inconsistent. However, when applied to a sequence of data-generating processes with growing $m$ which satisfy strong infinite confounding,\\
$\lim_{m \to \infty} \plim_{n \to \infty} \hat\bbeta^\naive - \bbeta
\ = \ \bm{0}$.

\end{prop}
\end{mdframed}

Intuitively, the na\"ive regression is unbiased as the number of treatments grow, because linear regression adjusts along the eigenvectors of the covariance matrix of the treatments, and in this setting the most consequential eigenvectors are the confounders. This connects to the core intuition of the deconfounder and a prior literature in genetics---under deconfounder assumptions, shared confounding leaves an imprint on the observed data distribution \citep{Price06,WanBle19,WanBle20}---and, as it turns out, this imprint is useful for the na\"ive regression.

\subsection{Deconfounder Under the Linear-Linear Model}
\label{s:declinlin}
In place of the na\"ive estimator, WB recommend the \emph{full deconfounder}, which under the linear-linear model proceeds in three steps: (1) take the singular value decomposition $\bA = \bU \bD \bV^\top$, (2) extract the first $k$ components, $\hat\bZ \equiv \sqrt{n} \bU_{1:k}$, and (3) adjust using
\begin{align}
\left[ \ 
\hat\bbeta^{\full \top} ,\ \hat\bgamma^{\full \top}
\ \right]^\top
&\equiv
\left(
\left[ \bA, \hat\bZ \right]^\top \left[ \bA, \hat\bZ \right]
\right)^{-1}
\left[ \bA, \hat\bZ \right]^\top \bY. \label{e:deconfounder}
\end{align}
But there is a problem: adding these new terms to the na\"ive regression renders Equation \eqref{e:deconfounder} inestimable. The substitute confounder is merely a linear transformation of the original treatments, $\hat\bZ = \sqrt{n} \bA \bV_{1:k} \bD_{1:k}^{-1}$, meaning that $\left[ \bA, \hat\bZ \right]^\top \left[ \bA, \hat\bZ \right]$ is always rank-deficient. This perfect collinearity is a consequence of that fact that the inclusion of $\hat\bZ$ brings no new information beyond that contained in the original treatments, $\bA$.  

The deconfounder papers and tutorials deploy variants of the linear-linear model throughout their simulations and empirical examples.  To render the model estimable, these examples use two modifications to the full deconfounder to break the perfect collinearity between the treatments and the substitute confounder: (1) the \textit{penalized full deconfounder}, which uses penalized outcome models to estimate treatment effects and adjust for the substitute confounder, and (2) the \textit{posterior full deconfounder}, which adds random noise to the substitute confounder $\hat\bZ$ by sampling it from an approximate posterior. We analyze both strategies and demonstrate that while both render the full deconfounder technically estimable, adding the substitute confounder does not help recover the quantity of interest.%

\subsubsection{Asymptotic Theory for Full Deconfounder Variants}
We analyze the \emph{penalized full confounder}, an estimator used in \citet{WanBle19} and \citet{Zha19} through their use of normal priors on regression coefficients. We analyze the frequentist version of this estimator, which uses a ridge penalty. In Supplement~\ref{a:bias_ridge}, we prove Proposition~\ref{prop:bias_ridge}, which gives the asymptotic bias of the penalized full confounder.
\begin{mdframed}
\begin{prop}
  \label{prop:bias_ridge} (Asymptotic Bias of the Penalized Full Deconfounder.)

  The penalized deconfounder
  estimator, as implemented in WB,
  is
  \begin{align*} 
  \left[ \ 
  \hat\bbeta^{\ridge \top} ,\ \hat\bgamma^{\ridge \top} 
  \ \right]^\top
  &\equiv \left( \left[ \bA, \hat\bZ \right]^\top \left[ \bA, \hat\bZ \right]
  + \lambda(n) \bI \right)^{-1} \left[ \bA, \hat\bZ \right]^\top \bY,
  \end{align*}  
  where $\hat\bZ$ is obtained by taking the singular value decomposition $\bA
  = \bU \bD \bV^\top$ and extracting the first $k$ components,
  $\hat\bZ \equiv \sqrt{n} \bU_{1:k}$, and $\lambda(n)$ is a ridge penalty assumed to be sublinear in $n$. Under the linear-linear model, the asymptotic bias of this estimator is
  given by
\begin{align*}
\plim_{n \to \infty} \hat\bbeta^\ridge - \bbeta 
  &=
  \overbrace{
  -
  \bQ_{1:k}
  \ \mathrm{diag}_j \left( \frac{
  1
  }{
  \sigma^2 + \Lambda_j + 1
  } \right)
  \bQ_{1:k}^\top \bbeta
  }^{\text{Regularization}} \\
  &\qquad
  \underbrace{+ \bQ_{1:k}
  \ \mathrm{diag}_j \left( \frac{
    \Lambda_j
  }{
    \sigma^2 + \Lambda_j + 1
  } \right)
  \bQ_{1:k}^\top \btheta^\top (\btheta \btheta^\top)^{-1} \bgamma}_{\text{Omitted Variable Bias}},
  \intertext{where $\bQ$ and $\bLambda = [\Lambda_1, \ldots, \Lambda_k, 0, \ldots]$ are respectively eigenvectors and eigenvalues obtained from decomposition of $\btheta^\top \btheta$. Under strong infinite confounding,}
\lim_{m \to \infty} \plim_{n \to \infty} \hat\bbeta^\ridge - \bbeta &= \bm{0}.
\end{align*}
\end{prop}
\end{mdframed}
Proposition~\ref{prop:bias_ridge} shows that the finite-$m$ bias in $\hat\bbeta^\ridge$ comes from two sources: regularization of coefficients and omitted-variable bias from excluding the true confounders, $\bZ$.  Under strong infinite confounding, as $m$ and $n$ grow large, both regularization bias and omitted-variable bias go to zero; the latter is true because $(\btheta\btheta^\top)^{-1}$ goes to $\bm{0}$. Therefore, like a na\"ive regression, the $\hat\bbeta^\ridge$ estimator is asymptotically consistent in $m$, but only because it effectively nests the na\"ive regression.

Briefly, the proof proceeds by examining the singular value decomposition of the augmented data matrix, $\left[\bA, \hat\bZ \right] = \bU^\ast \bD^\ast \bV^{\ast\top}$, and using the facts that (1) the first $m$ components of $\bU^\ast$ remain unchanged from $\bU$, since $\hat\bZ$ are merely rescaled versions of the original left-singular vectors; (2) the first $k$ diagonal elements of $\bD^{\ast 2}$ are equal to $\bD^2 + n\bI$ due to the additional variance of $\hat\bZ$; and (3) the last $k$ diagonal elements of $\bD^\ast$ are zero.

The second strategy employed by the deconfounder papers to address perfect collinearity in the linear-linear case is to integrate over an approximate posterior.  This renders the deconfounder estimable, since the resulting samples are no longer perfectly collinear with $\bA$. Proposition \ref{prop:bias_noise_posterior} gives the asymptotic bias (proof in Supplement~\ref{a:bias_noise_posterior}) and shows that sampling the substitute confounder from a posterior is inconsistent with a finite $m$ but converges to a na\"ive regression as $m$ grows large.\footnote{In Supplement~\ref{a:bias_noise_white} we also offer a proof of Proposition \ref{prop:bias_noise_white}, that estimators adding a fixed amount of white noise remain asymptotically inconsistent as $m$ grows, even under strong infinite confounding.} 

\begin{mdframed}
\begin{prop}
  \label{prop:bias_noise_posterior}
    (Asymptotic Bias of the Posterior-Mean Deconfounder.) 
    
    The posterior-mean deconfounder estimator is
  \begin{align*}
    \left[ \ 
    \hat\bbeta^{\pndf \top} ,\  \hat\bgamma^{\pndf \top}
    \ \right]^\top
    &\equiv
    \int
    \left(
    \left[ \bA, \bz \right]^\top \left[ \bA, \bz \right]
    \right)^{-1}
    \left[ \bA, \bz \right] \bY \ 
    f(\bz | \bA) \dd{\bz}
    ,
  \end{align*}
  where $f(\bz | \bA)$ is a posterior obtained from Bayesian principal component
  analysis.\footnote{While the regression cannot be estimated when $\bz = \E[\bZ
      | \bA]$, it is almost surely estimable for samples $\bz^\ast \sim f(\bz |
    \bA)$ due to posterior uncertainty, which eliminates perfect collinearity
    with $\bA$. The posterior-mean implementation of WB evaluates the integral by Monte Carlo methods and thus is able to compute the regression coefficients for each sample.} Under the linear-linear model, the asymptotic bias of this
  estimator is given by
  \begin{align*}
    \plim_{n \to \infty} \hat\bbeta^\pndf - \bbeta &=
 (\btheta^\top \btheta + \sigma^2 \bI)^{-1} \btheta^\top \bgamma,
 \intertext{and under strong infinite confounding,}
     \lim_{m \to \infty} \plim_{n \to \infty} \hat\bbeta^\pndf - \bbeta &=
    \bm{0}
  \end{align*}
\end{prop}
\end{mdframed}

\subsubsection{Asymptotic Theory for the Subset Deconfounder}
Theorem 7 of WB suggests an alternate version of the deconfounder---the subset deconfounder---which estimates the effect of some treatments, ignores others, and adjusts for the substitute confounder. After extracting a substitute confounder, $\hat\bZ$, this estimator designates a finite number, $m_F$, of the $m$ treatments as ``focal'' (we denote this column subset as $\bA_F$) and sets aside the remaining $m_N$ ``non-focal'' treatments ($\bA_N$). It then regresses the outcome, $\bY$, on only $\bA_F$ and $\hat\bZ$. The subset confounder avoids the collinearity issue if $m_{F} + k < m$.
This generalizes a popular estimator in genetics \citep{Price06} which estimates the effect of one treatment at a time.

In Proposition~\ref{prop:bias_subset}, we show that the asymptotic bias of the subset deconfounder remains non-zero \textit{even under strong infinite confounding}.  To approach consistency, the subset deconfounder requires additional strong conditions on the effects of the non-focal treatments.  

\begin{mdframed}
\begin{prop}
  \label{prop:bias_subset} 
 (Asymptotic Bias of the Subset Deconfounder.)  
 
  The subset deconfounder
  estimator, based on Theorem 7 from WB, is
  \begin{align}
      \left[ \ 
      \hat\bbeta_F^{\sdf \top} ,\ \hat\bgamma^{\sdf \top}
      \ \right]^\top
    &\equiv
    \left(
    \left[ \bA_F, \hat\bZ \right]^\top \left[ \bA_F, \hat\bZ \right]
    \right)^{-1}
    \left[ \bA_F, \hat\bZ \right]^\top \bY. \label{e:subset}
  \end{align}
  where the column subsets $\bA_F$ and $\bA_N$ respectively partition $\bA$ into a finite number of focal causes of interest and non-focal causes.
  The substitute confounder, $\hat\bZ$, is obtained by taking the singular value decomposition $\bA =
  \bU \bD \bV^\top$ and extracting the first $k$ components, $\hat\bZ \equiv
  \sqrt{n} \bU_{1:k}$. Under the linear-linear model, the asymptotic bias of this estimator is given by
  \begin{align*}
    \plim_{n \to \infty} \hat\bbeta_F^\sdf - \bbeta_F 
    &=
    \left(
    \bI - \btheta_F^\top (\btheta \btheta^\top)^{-1} \btheta_F
    \right)^{-1}
    \btheta_F^\top (\btheta \btheta^\top)^{-1} \btheta_N \bbeta_N, \\
    \intertext{
  with $\btheta_F$ and $\btheta_N$ indicating the column subsets of $\btheta$
  corresponding to $\bA_F$ and $\bA_N$, respectively. The subset deconfounder is unbiased for $\bbeta_F$ (i) if $\btheta_F = \bm{0}$, (ii) if $\lim_{m \rightarrow \infty} \btheta_N \bbeta_N = \bm{0}$ and $\lim_{m \rightarrow \infty} \left[
  \bI - \btheta_F^\top (\btheta \btheta^\top)^{-1} \btheta_F
  \right]^{-1}$ is convergent, or (iii) if both strong infinite confounding holds and $(\btheta \btheta^\top)^{-1} \btheta_{N} \bbeta_{N}$ goes to $\bm{0}$ as $m \rightarrow \infty$.  If one of these additional conditions hold, 
}
  \lim_{m \to \infty} \plim_{n \to \infty} \hat\bbeta_F^\sdf - \bbeta_F 
    &= \bm{0}
  \end{align*}
\end{prop}
\end{mdframed}
A proof is given in Supplement~\ref{a:bias_subset}; we interpret conditions (i--iii) below. Intuitively, the subset deconfounder is a biased estimator of the effect of $\bA_F$ because of mismodeling of the dependence structure among the causes. Though $\bA_F$ and $\bA_N$ would be conditionally independent if the true $\bZ$ could be observed and adjusted for, Lemma~\ref{l:residual_dependence} shows that they are not conditionally independent given $\hat\bZ$. This mismodeled dependence leads to omitted variable bias when excluding $\bA_N$.  But, unlike na\"ive regression, strong infinite confounding does not resolve this omitted variable bias for the subset deconfounder. In Supplement~\ref{a:subset_add} we provide further intuition for this result, leveraging properties of PCA regression to show the subset confounder is effectively a regularized regression that only adjusts along the first $k$ eigenvectors \citep{hastie2013elements}.  

The subset deconfounder and similar estimators in genetics \citep[e.g.][]{Price06} rely on more than the intuition that there is shared confounding.\footnote{In Supplement~\ref{a:subset_add} we provide an example where strong infinite confounding holds, but the subset deconfounder has bias that cannot be eliminated by adding more treatments.}
Proposition~\ref{prop:bias_subset} provides three stronger conditions, of which at least one is required for the subset deconfounder to approach unbiasedness.  
Condition (i), $\btheta_F = \bm{0}$, states that the focal treatment is unconfounded, and therefore no adjustment is needed. Condition (ii) is $\lim_{m \rightarrow \infty} \btheta_{N} \bbeta_{N} = \bm{0}$ and $\lim_{m \rightarrow \infty} \left[
  \bI - \btheta_F^\top (\btheta \btheta^\top)^{-1} \btheta_F
  \right]^{-1}$ is convergent. This will hold if, for example, each element of $\bbeta_{N}$, the treatment-outcome effects, and each column of $\btheta_N$, the confounder-treatment relationships, are drawn from zero-expectation distributions with finite variance and zero covariance between $\bbeta_N$ and $\btheta_N$. 
  Condition (iii) says that the subset deconfounder will hold if strong infinite confounding holds and  $\lim_{m \rightarrow \infty} \left(\btheta \btheta^{'}\right)^{-1} \btheta_{N} \bbeta_{N}  = 0$.  This could be satisfied, if, for example, the infinite sum $\sum_{j=1}^m \theta_{N,k',j} \beta_{N,j}$ is convergent for all latent confounders $k' \in \{1, \ldots, k\}$.  A necessary (but not sufficient) condition for this to hold is if $\left( \theta_{N, k', j} \ \beta_{N, j} \right)_{j = 1}^{m} $ converges to zero for each $k'$ as $m \rightarrow \infty$. For example, analysts might assume that treatment effects in the non-focal set, $\left(\beta_{N,j} \right)_{j = 1}^{m}$, go to zero fast enough to ensure the infinite series converge.  All of these conditions are extremely strong---even more than strong infinite confounding---because they either require assuming characteristics about the treatment effects (when the very reason for estimation is that they are unknown) or assuming there is no confounding at all. %

\subsubsection{Takeaways}
In the linear-linear setting, the default full deconfounder is rank deficient because the substitute confounder is a linear projection of the treatments.   There are three ways of addressing this issue: penalizing the outcome regression, sampling the substitute confounder from its posterior distribution and analyzing a subset of the causes.  The penalized and posterior full deconfounders converge asymptotically in $n$ and $m$ under strong infinite confounding to the correct solution only by virtue of the fact that they contain the na\"ive regression. By contrast, the subset deconfounder requires strong and unverifiable additional assumptions about the values of the treatment effects or the nonexistence of confounding.  

\subsection{Extensions to Separable Nonlinear Settings}
\label{s:nonlinear_analysis}

In this section, we extend our linear-linear results to settings with nonlinear factor and outcome regression models under separable confounding and strong infinite confounding. Under constant treatment effects, we connect the deconfounder with a partially linear regression \citep{robinson1988} to demonstrate that a semi-parametric na\"ive regression approaches asymptotic unbiasedness for the same reason that the deconfounder does.  We then relax the assumption of constant treatment effects to show in our most general setting that infinite $m$ is required for consistency of the deconfounder.

\subsubsection{Convergence of Deconfounder and Na\"ive Regressions}

Following \citet{WanBle19,WanBle20,Zha19}, we first study constant-effects outcome models of the form $\E[Y_i | \bA_i, \bZ_i] = \bA_i^\top \bbeta + g_Y(\bZ_i)$. Theorem~\ref{thm:deconfounder_naive_equality} shows any consistent deconfounder converges to a flexible na\"ive regression, which is also consistent.

\begin{mdframed}
\begin{thm}
\label{thm:deconfounder_naive_equality}
(Deconfounder-Na\"ive Convergence under Strong Infinite Confounding.)

Consider all data-generating processes in which (i) treatments are drawn from a
factor model with continuous density that is a function $\bZ$, (ii) $\bZ$ is pinpointed, and (iii) the
outcome model contains constant treatment effects and treatment assignment is ignorable nonparametrically given $\bZ$. Any consistent
deconfounder converges to a na\"ive estimator for any finite subset of
treatment effects.
\end{thm}
\end{mdframed}
A proof is given in Supplement~\ref{a:deconfounder_naive_equality}. The proof proceeds by noting that pinpointedness implies that the function $g_{\bA}(\bz) \equiv \E[ \bA_i | \bZ_i=\bz ]$ is asymptotically recoverable and invertible. (Here, pinpointedness hinges on a generalization of strong infinite confounding to nonlinear settings: the conditional entropy of $\bZ$ given $\bA$ must approach zero as $m$ grows large.) This makes the deconfounder equivalent to the partially linear regression:
\begin{align}
  \left(\hat\bbeta^\deconf, \hat{g}_Y^\deconf \right)
  &= \argmin_{\bbeta^\ast, g_Y^\ast} \ \sum_{i=1}^n \left(
  Y_i - \bA_{i}^\top \bbeta^\ast - g_Y^\ast(\hat{g}_{\bA}^{-1}(\bA_i))
  \right)^2.
\end{align}
The above is consistent for $\bbeta$ \citep{robinson1988}. Intuitively, this shows that what the deconfounder buys is part of the function mapping the treatments to the outcome, $g_Y^\ast(\hat{g}_{\bA}^{-1}(\cdot))$.

We can instead directly fit the following partially linear regression.  As in Section~\ref{a:bias_naive}, we partition the $m$ causes, $\bA_i$, into finite
$m_F$ focal causes of interest, $\bA_{i,F}$, and $m_N$ nonfocal causes,
$\bA_{i,N}$. Then the conditional expectation function of the outcome can be
rewritten $\E[ Y_i | \bA, \bZ ] = \bA_{i,F}^\top \bbeta_F + \bA_{i,N}^\top
\bbeta_N + g_Y(\bZ_i)$. The associated conditional expectations of the treatments are $g_{\bA_F}(\bz) \equiv \E[ \bA_{i,F} | \bZ_i=\bz
]$ and $g_{\bA_N}(\bz) \equiv \E[ \bA_{i,N} | \bZ_i=\bz ]$. The semiparametric na\"ive regression is
\begin{align}
  \hat{h}_{\bA_F}^\naive
  &= \argmin_{h_{\bA_F}^\ast} \sum_{i=1}^n \left\| \bA_{i,F} - h_{\bA_F}^\ast(\bA_{i,N}) \right\|_F^2
  \label{e:naive_nonlinear_af} \\
  \hat{h}_Y^\naive
  &= \argmin_{h_Y^\ast} \sum_{i=1}^n \left( Y_i - h_Y^\ast(\bA_{i,N}) \right)^2 
  \label{e:naive_nonlinear_y} \\
  \hat\bbeta_F^\naive
  &=
  \left( \tilde\bA_F^{\naive \top} \tilde\bA_F^\naive \right)^{-1}
  \tilde\bA_F^{\naive \top} \tilde\bY^\naive
  \label{e:naive_nonlinear}
\end{align}
where $\tilde\bA_F^\naive$ collects
$\bA_{i,F} - \hat{h}_{\bA_F}^\naive(\bA_{i,N})$
and $\tilde\bY^\naive$ collects
$Y_i - \hat{h}_Y^\naive(\bA_{i,N})$.  

These two approaches asymptotically converge to one another. In short, the key problem is estimating the composite function $g_Y(\hat{g}_{\bA}^{-1}(\cdot))$. This can be done directly by a flexible na\"ive regression without the added complication or factor-model functional form assumptions of the deconfounder.  While a flexible na\"ive regression guarantees that we recover the correct control function without parametric information about $f(\bA | \bZ)$, we note that in applied settings, considerable data is required to make use of this result.

\subsubsection{Inconsistency of the Nonlinear Deconfounder in Finite $m$}
Next, we generalize to the class of continuous factor and outcome models with additively separable confounding, $\E[Y_i | \bA_i, \bZ_i] = f(\bA_i) + g_Y(\bZ_i)$. This class nests all models considered in this paper, covering nonlinear factor models, nonlinear and interactive treatment effects, and arbitrary nonlinear confounding. Theorem~\ref{thm:deconfounder_inconsistency_nonlinear} states that the deconfounder is inconsistent for all such data-generating processes with finite $m$.

\begin{mdframed}
\begin{thm}
\label{thm:deconfounder_inconsistency_nonlinear}
(Inconsistency of the Deconfounder in Nonlinear Settings.)

Consider all data-generating processes in which a finite number of conditionally ignorable treatments are drawn from
a factor model with continuous density that is a function of the confounding variable $\bZ$. Then the deconfounder is inconsistent for
any outcome model with additively separable confounding.
\end{thm}
\end{mdframed}
A proof is given in Supplement~\ref{a:deconfounder_inconsistency_nonlinear}. This result follows from a simple premise: because $\bA_i$ is a stochastic function of the confounder, $\bZ_i$, analysts can never recover the exact value of $\bZ_i$ using a \textit{finite} number of treatments \citep{imai2019discussion}, even if given the function $g_{\bA}^{-1}$ mapping $\E[ \bA_i | \bZ_i ]$ to $\bZ_i$---because $\E[ \bA_i | \bZ_i ]$ is unknown. Next, the error in $\hat\bZ_i$ depends on $\bA_i$, and the outcome $Y_i$ is in part dependent on this component (i.e., $Y_i$ depends on $\bZ_i$, which is only partially accounted for by $\hat\bZ_i$). Therefore, outcome analyses that neglect the unobserved mismeasurement, $\hat\bZ_i - \bZ_i$, will necessarily suffer from omitted variable bias.

\subsection{Takeaways}
Every estimator considered in this paper, save the oracle, is inconsistent for finite $m$.  The deconfounder's pinpointing assumption requires \textit{strong infinite confounding}---an asymptotic regime for $m$ that is a very stringent assumption. For the subset deconfounder, except in knife-edge cases, \textit{strong infinite confounding} is insufficient and requires further strong assumptions.  When the deconfounder does work (is estimable and satisfies conditions for asymptotic unbiasedness), we prove that a suitably flexible na\"ive regression converges to the deconfounder asymptotically in $m$.  Thus the deconfounder works in limited settings.  When the deconfounder works, the factor-model machinery of the deconfounder is unnecessary, because a na\"ive regression asymptotically produces the same result.

\section{Deconfounder Does Not Consistently Outperform Na\"ive Regression in Finite Samples}
\label{s:finite}

Having established that the deconfounder offers no gains over na\"ive regression in asymptotic bias, we now reconsider the simulation evidence for finite sample performance. Results demonstrate the deconfounder cannot in general improve over na\"ive regression. We note that these findings conflict with the positive simulation evidence presented in the deconfounder papers. The divergence in our findings largely stems from (i) improvements that we make in estimation, including substantial gains in stability, and (ii) our extension of simulations to more thoroughly probe changes in key parameters. Supplement~\ref{a:common_deviations} provides a thorough discussion of these and other deviations.  This section concludes with a brief overview of additional simulation evidence in the supplement.

\subsection{Linear-Linear Deconfounders Only Help When Biases Cancel}
In the linear-linear setting, the substitute confounder is a linear function of the treatments---information already captured by including the treatments in the linear outcome model. We show that in the linear-linear setting, deconfounder can sometimes outperform the na\"ive regression in subsets of the parameter space where differing na\"ive and deconfounder biases align in the right way.  However, these situations always rely on parameters that would be unknown to the analyst. We also show that given estimation instability in near-collinear estimators, the full deconfounder with a linear factor model is never appropriate.

\subsubsection{Medical Deconfounder} 
\cite{Zha19} presents an application of the deconfounder to the analysis of electronic health records. The first simulation study presented in the paper considers a situation where there are two treatments, of which only one has a true non-zero coefficient.%
The true data generating process draws $n=1,000$ patients from a linear-linear model.\footnote{This process is $Z_i \sim \cN(0, 1)$, $A_{i,1} \sim \cN(0.3 Z_i, 1)$, $A_{i,2} \sim \cN(.4Z_i, 1)$, $Y_i \sim \cN(0.5 Z_i + 0.3 A_{i,2}, 1)$.}
They estimate a one-dimensional substitute confounder using probabilistic principal component analysis.\footnote{\citet{Zha19} uses black box variational inference \citep{ranganath2014black}, then estimates the outcome model with automatic differentiation variational inference \citep{kucukelbir2017automatic}. We use \texttt{Stan} code for probabilistic PCA and the outcome model. Further details are in Supplement~\ref{a:med1}.}
We introduce a faster and more accurate variant deconfounder which performs PCA, extracts the top component, and then runs ridge regression with a penalty chosen by cross-validation. \citet{Zha19} report a single sample. We repeat this process 1,000 times, assessing bias, variance and root mean squared error (RMSE) in Table~\ref{tab:med1}.

\begin{table}[ht]
\centering
\begin{tabular}{|cr|cc|cc|cc|}
 \hline\hline
 && \multicolumn{2}{c|}{Bias} &  \multicolumn{2}{c|}{Std. Dev.} & \multicolumn{2}{c|}{RMSE} \\ 
 & Model & $\beta_1$ & $\beta_2$ & $\beta_1$ & $\beta_2$ & $\beta_1$ & $\beta_2$ \\
  \hline
 &  Na\"ive & 0.120 & 0.160 & 0.033 & 0.033 & 0.125 & 0.164 \\ 
 \textbf{Orig. simulation}
 &  Oracle & 0.000 & 0.000 & 0.032 & 0.033 & 0.032 & 0.033 \\ 
 ($\beta_1=0$, $\beta_2 = 0.3$)
 &  Deconfounder & 0.145 & 0.189 & 0.675 & 0.877 & 0.690 & 0.897 \\ 
 &  PCA+CV-Ridge & 0.028 & -0.146 & 0.014 & 0.029 & 0.031 & 0.149 \\ 
   \hline
 \textbf{Our simulation}
 &  Deconfounder & 0.150 & 0.197 & 0.685 & 0.893 & 0.701 & 0.914 \\ 
 ($\beta_1=-0.3$, $\beta_2 = 0.3$) 
 &  PCA+CV-Ridge & 0.188 & -0.099 & 0.028 & 0.037 & 0.191 & 0.106 \\ 
   \hline
   \hline
\end{tabular}
\caption{\textbf{Simulation Study 1 of the Medical Deconfounder}.  The main ``Deconfounder'' estimation procedure is from \citet{Zha19} and uses probabilistic PCA and Bayesian linear regression. ``PCA + CV-Ridge'' is an improved deconfounder estimator we developed. ``Na\"ive''and ``Oracle'' estimators are as described in the main text.}
\label{tab:med1} 
\end{table}

The original deconfounder performs poorly, with higher bias and variance than the na\"ive estimator due to near collinearity driving estimation instability. Our PCA+CV-Ridge deconfounder appears to perform better than na\"ive, but this does not hold across the parameter space. Under the original data generating process, the effect of $A_1$ is zero and therefore, the ridge penalty drives the coefficient of $A_1$ towards the truth. This results in apparent good performance for the simplified deconfounder variant. In the last two rows of Table~\ref{tab:med1}, we repeat the same simulation switching the true effect of $A_1$ to $-0.3$: the simplified deconfounder now performs slightly worse than the na\"ive regression. 

\subsubsection{Subset Deconfounder} Proposition 4 shows strong infinite confounding is insufficient for the subset deconfounder to provide unbiased estimates of treatment effects, even when na\"ive regression can achieve oracle-like performance. Only under strong assumptions about treatment effects will the subset deconfounder be unbiased. We design a simulation to demonstrate this fact for different sequences of treatment effects, $\beta$, even when strong infinite confounding is satisfied ($\theta_{j} = 10 \ \forall \ j$). The full simulation is included in Supplement~\ref{a:subsetsim}. %

Table \ref{t:sim_best} provides the average RMSE for treatment effect estimates.  When treatment effects are constant ($\beta_{j} = 10$ or $\beta_{j} = 100$) the subset deconfounder's performance fails to improve as more treatments are added.  This is true even though na\"ive regression's average RMSE converges on the oracle's performance.  Similarly, we see that when $\beta_{j} \sim \cN(1,2)$, the subset deconfounder converges on an average bias of $1$---the mean of $\beta_j$ (see Case ii from Proposition 4).
In Table \ref{t:sim_best} the subset deconfounder is unbiased only when the sequence of treatment effects converge to 0 as more treatments are added (e.g., if $\beta_{j} = \frac{1}{j}$).  
\begin{table}[h!!]
  \centering
  \small

\centering
\begin{tabular}{l|l|lllll}
  \hline
 & Method & $m$=3 & $m$=10 & $m$=50 & $m$=100 & $m$=200 \\ 
  \hline
\multirow{3}{1in}{$\beta_{j} = 10$}
& Oracle & 0.010 & 0.010 & 0.010 & 0.010 & 0.010 \\ 
& Na\"ive & 0.333 & 0.100 & 0.022 & 0.014 & 0.011 \\ 
& Deconfounder & 10.000 & 10.000 & 10.000 & 10.000 & 10.000 \\ 
\hline 
\multirow{3}{1in}{$\beta_{j} = 100$}& Oracle & 0.010 & 0.010 & 0.010 & 0.010 & 0.010 \\ 
& Na\"ive & 0.333 & 0.100 & 0.022 & 0.014 & 0.011 \\ 
& Deconfounder & 100.000 & 100.000 & 100.000 & 100.000 & 100.000 \\ 
\hline 
\multirow{3}{1in}{$\beta_{j} \sim \cN(1, 2) $}  & Oracle & 0.010 & 0.010 & 0.010 & 0.010 & 0.010 \\ 
& Na\"ive & 0.333 & 0.101 & 0.022 & 0.014 & 0.011 \\ 
 & Deconfounder & 1.465 & 1.283 & 1.362 & 1.195 & 1.026 \\ 
\hline
\multirow{3}{1in}{$\beta_{j} = \frac{1}{m} $} & Oracle & 0.011 & 0.010 & 0.010 & 0.010 & 0.010 \\ 
& Na\"ive & 0.333 & 0.101 & 0.022 & 0.014 & 0.011 \\ 
& Deconfounder & 0.611 & 0.293 & 0.091 & 0.054 & 0.033 \\ 
\hline
   \hline
\end{tabular}
    \normalsize
\caption{\textbf{The Subset Deconfounder is Unbiased Only under Strong Assumptions about Treatment Effects}  As the number of treatments grow (columns moving from left to right), both the na\"ive regression converges on the oracle's average RMSE (entries in table average over RMSE of each treatment effect), while the subset deconfounder's performance depends on the treatment's effects. The subset deconfounder is unbiased only when the sequence of treatments converges to zero.  Even when the treatments are random draws from a normal distribution, the bias of the subset deconfounder converges on the average treatment effect. }\label{t:sim_best}
\end{table}

\subsubsection{Takeaways} In finite-sample linear-linear settings, the deconfounder cannot improve performance over the na\"ive regression. Due to estimation instability, variants of the full deconfounder with a linear factor model are never preferable to na\"ive regressions.  Subset deconfounders only outperform na\"ive regressions if strong assumptions about treatment effects are satisfied.

\subsection{Nonlinear Deconfounder and Na\"ive Approaches Can Sometimes Exploit Parametric Information}

The best-case scenario for the deconfounder is it may efficiently exploit known parametric information about a nonlinear data generating process. We examine one such case, drawing on a simulation posted on the \texttt{blei-lab} GitHub page \citep{wang2019github}. We find that even in this ideal setting, the deconfounder is weakly dominated by a correctly specified nonlinear na\"ive regression.

We simulate $n=10,000$ draws from the following data-generating process,
\begin{align}
\left[\begin{array}{l} A_{i,1} \\ A_{i,2} \\ Z_i \end{array} \right]
&\sim \cN\left(
\left[\begin{array}{l} 0 \\ 0 \\ 0 \end{array} \right]
,
\left[\begin{array}{lll}
 1 & \rho & \rho \\
 \rho & 1 & \rho \\
 \rho & \rho & 1
\end{array} \right]
\right)
\nonumber \\[1ex]
Y_i &\sim
\cN(0.4 + 0.2 A_{i,1}^2 + 1 A_{i,2}^2 + 0.9 Z_i^2,\ 1) \label{e:quadratic_outcome}
\end{align}          
for $\rho=0.4$. %
As before, these are collected in $\bZ$, $\bA$, and $\bY$. A substitute confounder is obtained by taking the singular value decomposition $\bA = \bU \bD \bV^\top$ and extracting the first component, $\hat\bZ \equiv \bA \bV_{1}$, so for $\hat Z_{i} = \frac{\sqrt{n}}{D_{1}}(V_{11}A_{i1}  + V_{21} A_{i2})$.  Following \citet{wang2019github}, we then estimate a linear regression of $\bY$ on three predictors: $\bA_1^2$, $\bA_2^2$, and $\hat\bZ^2$.%

 \begin{figure}[h!!!]
\centering
\includegraphics[width=.48\textwidth]{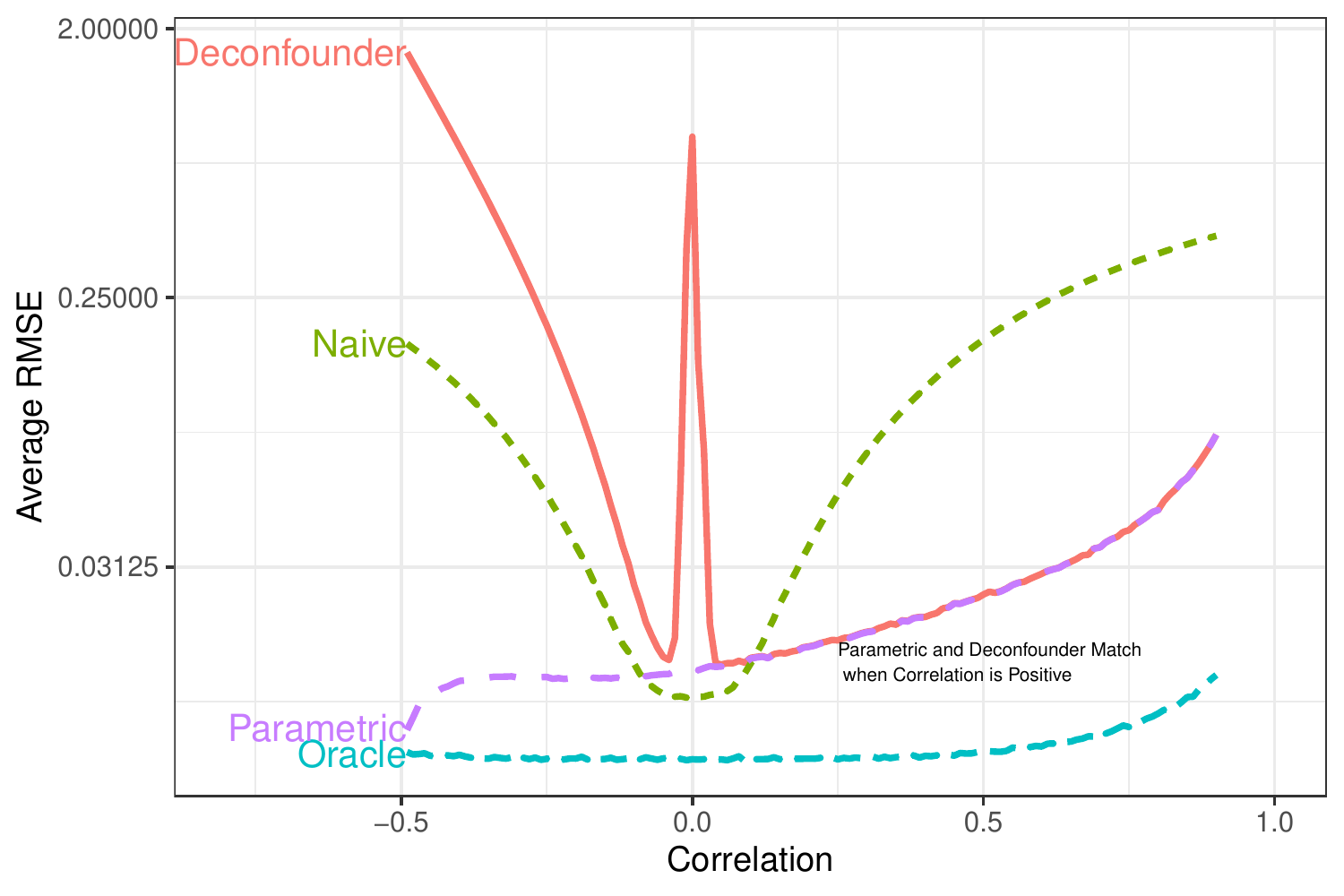}
\hfill
\includegraphics[width=.48\textwidth]{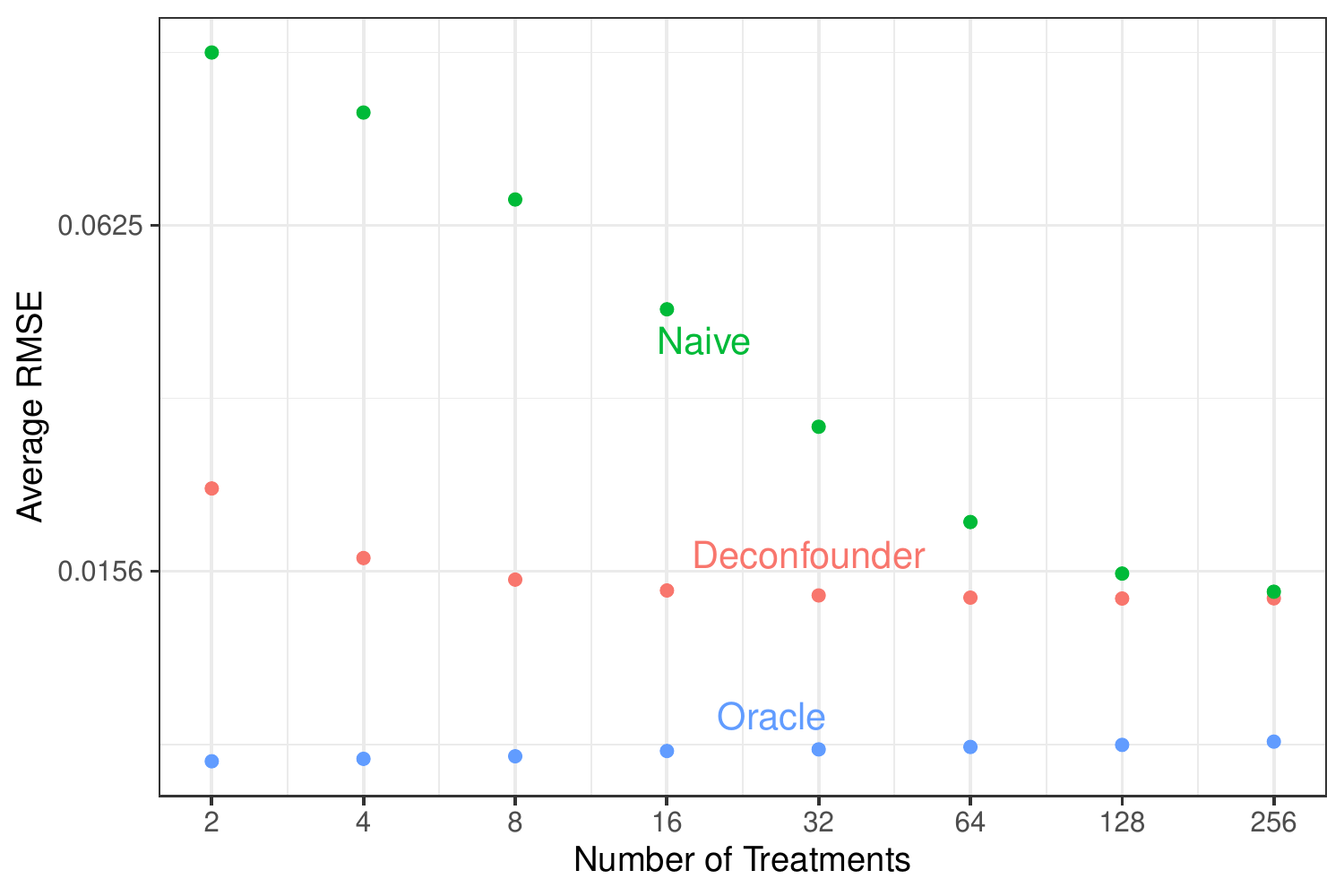}
\caption{\textbf{Under Strong Infinite Confounding, a Parametric Na\"ive Model Weakly Dominates the Deconfounder (left); Na\"ive Regression Converges to the Same Performance as the Number of Treatments Increases (right).} The left plot shows average RMSE for varying $\rho$, the off-diagonal covariances in Equation~\eqref{e:quadratic_outcome}. The right plot shows as $m$ increases, deconfounder and na\"ive RMSEs converge.}
\label{f:quadratic}
\end{figure}

The deconfounder captures the interaction of the treatments, as can be seen by expanding the polynomial $\hat{Z}_{i}^2 = \left(\frac{\sqrt{n}}{D_{1}}(V_{11}A_{i1}  + V_{21} A_{i2})\right)^2 \ = \ \frac{n}{D_1^2} \left( V_{11}^2 A_{i1}^2 + V_{21}^2 A_{i2}^2  + V_{11} V_{21} A_{i1} A_{i2} \right)$.
The expansion is not a linear combination of $A_{i,1}^2$ and $A_{i,2}^2$ due to the inclusion of the interaction $A_{i,1} A_{i,2}$. However, the deconfounder only incorporates partial information about the true functional form of the outcome model, Equation~\eqref{e:quadratic_outcome}. By using the same information more carefully, a better parametric na\"ive estimator can be derived. By properties of the multivariate normal distribution, $\frac{1}{\rho^2} \E[Z_i|\bA_i]^{2} = A_{i,1}^2 + A_{i,2}^2 + 2 A_{i,1} A_{i,2}$. Therefore, the causal effects can also be estimated by a regression of $\bY$ on $\bA_1^2$, $\bA_2^2$, and $(\bA_1^2 + \bA_2^2 + 2\bA_1 \circ \bA_2)$, where $\circ$ denotes the elementwise product. We refer to this approach as the ``parametric" alternative; it is also a na\"ive regression, as it does not seek to estimate $\bZ$. As Figure~\ref{f:quadratic} (left panel) shows, this substantially improves over the deconfounder for negative $\rho$, in terms of average root mean squared error, while capturing the same information for positive $\rho$. Moreover, when the latent confounder satisfies strong infinite confounding, the na\"ive regression will approach the deconfounder's performance as the number of treatments grows.  The right panel demonstrates this point for the original setting of $\rho=0.4$.

While the quadratic example shows it may be theoretically possible to design a simulation where nonlinear information helps the deconfounder improve over an incorrectly specified na\"ive regression, it is difficult.  In this tutorial simulation, the factor model performs poorly for negative correlations while the parametric model does well.  In more complex simulations involving factor model nonlinearity, reported gains are often modest.  \citet{WanBle19} presents a simulation based on Genome Wide Association Study (GWAS) data which show a maximum RMSE reduction of merely 3\% for the deconfounder over the na\"ive.  (In our replication in Supplement~\ref{a:gwas}, we show that in fact, the deconfounder actually performs worse than na\"ive regression for the non-zero coefficients---an unfortunate pattern for applied genetics, where the primary interest is detecting nonzero coefficients and assessing their size.)  Similarly, a second simulation study in \citet{Zha19} report deconfounder RMSE that is only slightly better than na\"ive RMSE (our simulations point to the opposite conclusion, that the deconfounder does slightly \textit{worse}; see Supplement~\ref{a:med2}).  

\subsubsection{Takeaways} 
It is sometimes possible for the deconfounder to improve over incorrectly specified na\"ive regression in finite samples---if the deconfounder learns the correct nonlinearity. However, in practice, making use of this fact is impossible without extensive knowledge of the data generating process. Moreover, even the deconfounder papers show, at best, marginal improvements.  When such extensive knowledge is available, a well-specified na\"ive regression making use of that knowledge will also perform well. We conjecture that if the high-dimensional treatments lie on a low-dimensional manifold \textit{and} the correct factor model specification is known, it might be more efficient to model the relationship between $\hat{\bZ}$ and $\bY$ semiparametrically (as in the deconfounder) rather than directly modeling high-dimensional $\bA$ and $\bY$ semiparametrically (as in  na\"ive regressions).  However, this has yet to be demonstrated in any simulation.  

\subsection{Takeaways and Additional Empirical Results}
We have replicated every simulation across the deconfounder papers for which data is available, and we find no evidence that the deconfounder consistently outperforms the na\"ive regression.  Several simulations---like the medical deconfounder and quadratic examples highlighted above---perform better at some parameter values but worse at others. Full details of all six replicated simulations are in the supplement.

Separately, \citet{WanBle19} use posterior predictive checks (PPCs) of the factor model, arguing these will assess when the deconfounder can improve estimates.  If this claim were true, it would allow highly flexible density estimation to be used, even when the true parametric form of the factor model was unknown---as is always the case in practice. However, it is not. Theoretically, Proposition~\ref{prop:bias_subset} proves that for subset deconfounders, this is impossible because the performance depends on untestable assumptions about the treatment effects, not the factor model. And for the full deconfounder, PPCs are ill-suited to evaluating conditional independence of $\bA$ given $\hat{\bZ}$, perhaps the most relevant observable property of the factor model \citep{imai2019discussion}. Empirically, we further present a new simulation in Supplement~\ref{a:ppc} demonstrating that the PPC does not reliably indicate whether a deconfounder will perform well, either in absolute terms or relative to na\"ive regression.

Our asymptotic results suggested that the deconfounder would not outperform the na\"ive regression, and simulations have shown this to hold in finite samples.  In nonlinear settings it is possible to exploit parametric information in the factor model, but it is both difficult to do in practice and can also be used to comparably perform the na\"ive regression.  It remains possible that a simulation could establish a particular data generating process where the deconfounder performs better than na\"ive regression, but this has yet to be demonstrated.

\section{Neither Na\"ive Regression nor Deconfounder is Currently Suitable For Real World Applications}
\label{s:applications}
After deriving the deconfounder's properties, WB recommend it for empirical research in in social science, neuroscience, and medicine.  \citet{Zha19} propose the deconfounder as a solution for assessing drug treatment effects using observational health records, 
writing that the deconfounder is ``guaranteed to capture all multi-medication confounders, both observed and unobserved'' \citep[][p. 2]{Zha19}.

As we have shown, however, this is not a property of the deconfounder.  In practice, both the deconfounder and na\"ive regression will fail to capture confounders unless they affect an infinite number of treatments.  Therefore, we cannot recommend either current approach for use in real-world applications. This is true even though we have shown that (i) under the deconfounder's assumptions, the na\"ive estimator is asymptotically unbiased across many settings given strong infinite confounding; and (ii) the full deconfounder inherits this property because it includes the same information as na\"ive regression. We emphasize that the required assumptions are \textit{exceedingly strong}. To demonstrate the consequences of their violation, we now investigate the real-world case study in WB---of actors' effects on box office revenue---and show that both produce implausible estimates. We then highlight some of the explicit and implicit assumptions of the deconfounder (and na\"ive regression) which lead us to be skeptical that credible applications can be found.

\subsection{Actor Case Study Reveals Limitations of the Deconfounder}
\label{s:actor}
WB's case study investigates how the cast of a movie causally affects movie revenue. The deconfounder is applied to the TMDB 5,000 data set, estimating how much each of the $m=901$ actors affected the revenue of $n=2,828$ movies. WB presents results from a full deconfounder in which substitute confounders are estimated using the leading $k=50$ dimensions of a poisson matrix factorization (PMF) of the binary matrix of movie-actor appearance indicators. A linear regression of log revenue on actor appearance and substitute confounders is used to estimate what is described as the causal effect of the cast.

\begin{table}[!ph]
\caption{\textbf{The Deconfounder Estimates Implausible Effects for Actors.}
Estimated causal effect of each actor's casting on movie revenue. Following WB, estimates are computed by linear regression of log revenue on actor indicators and additional covariates. Each row reports a different specification (for example, ``deconfounder'' rows each adjust for a 50-dimensional substitute confounder, and the ``controls'' row adjusts for budget, a multi-cause confounder). The top panel contains estimators that analyze all actors simultaneously, including the full deconfounder; the bottom panel contains estimators that analyze each actor $j$ in isolation, including the subset deconfounder and the univariate na\"ive estimator $\hat\beta_j = \Cov(\bA_j, \bY) / \Var(\bA_j)$. Two versions of each deconfounder estimator are used, one relying on a cached poisson matrix factorization (PMF) provided by WB and another using a re-estimated PMF. For each estimator, the top five actors and associated estimates are presented in the form ``Actor ($\times e^{\hat\beta_j}$),'' indicating an estimate that Actor causally modifies revenue by a multiplicative factor of $e^{\hat\beta_j}$.}
\label{t:actors}
\begin{tabular}{p{1in} | p{.8in} | p{4in}}
\hline
\multicolumn{3}{c}{\bf Estimating all actor effects simultaneously (full deconfounder)} \\
\hline
Na\"ive & Standard & Stan Lee ($\times$9.31), John Ratzenberger ($\times$9.26), Sacha Baron Cohen ($\times$7.09), Leonardo DiCaprio ($\times$5.50), Josh Hutcherson ($\times$5.19)
 \\
\hline
Deconfounder & Cached PMF & Stan Lee ($\times$9.29), John Ratzenberger ($\times$8.29), Sacha Baron Cohen ($\times$8.12), Josh Hutcherson ($\times$5.02), Corey Burton ($\times$4.91)
 \\
\hline
Deconfounder & Rerun PMF & Courteney Cox ($\times$15.32), Tom Cruise ($\times$14.74), John Ratzenberger ($\times$11.13), Vera Farmiga ($\times$9.86), Sacha Baron Cohen ($\times$9.83)
 \\
\hline
Controls & Adjusting for budget & Sacha Baron Cohen ($\times$6.93), Brian Doyle Murray ($\times$4.08), Conrad Vernon ($\times$4.04), Julie Andrews ($\times$3.84), Tomas Arana ($\times$3.83)
 \\
\hline
\multicolumn{3}{c}{\bf Estimating effects one actor at a time (subset deconfounder)} \\
\hline
Na\"ive & Univariate & Jess Harnell ($\times$12.28), Ava Acres ($\times$10.16), Warwick Davis ($\times$10.09), Stan Lee ($\times$9.85), Orlando Bloom ($\times$9.50)
 \\
\hline
Deconfounder & Cached PMF & Jess Harnell ($\times$13.49), Ava Acres ($\times$10.49), Chris Miller ($\times$9.19), Orlando Bloom ($\times$9.02), Stan Lee ($\times$8.77)
 \\
\hline
Deconfounder & Rerun PMF & Lasco Atkins ($\times$5.64), Sacha Baron Cohen ($\times$4.24), John Ratzenberger ($\times$4.01), Desmond Llewelyn ($\times$3.91), Will Smith ($\times$3.64)
 \\
\hline
\end{tabular} 
\end{table}

We replicate this analysis in Table~\ref{t:actors}, using cached PMF output from WB to ensure that our conclusions are unaffected by random seed. The five largest estimates from this model as well as several alternatives are reported in Table~\ref{t:actors} (expanded results, including appearance-weighted log-scale coefficients,\footnote{WB rank actors by multiplying each actor's log-scale coefficients by their number of movie appearances. This transformation is difficult to interpret substantively but produces similar results.} are in Supplement~\ref{a:actors}).

According to WB's results, the single most valuable actor is Stan Lee---whose appearances are cameos in movies based primarily on his Marvel comic books\footnote{And a 40 second appearance in ``Mallrats."} 
totaling 200 seconds of screen time. These unreported estimates suggest that with his casting, Marvel Cinematic Universe (MCU) producers causally increased their movies' revenue by 831\%---more than nine times the box-office haul of their counterfactual Stan-less versions, a total of \$15.5 billion in additional earnings. The subset deconfounder's estimates are similarly implausible, suggesting that Jess Harnell causally increases a movie's revenue by 1,128\%. This is driven by his appearance as a voice actor in the high-budget ``Transformers'' series, as his credits are otherwise in peripheral roles not included in this data set, such as a supporting role in the animated series \textit{Doc McStuffins}. %
WB's subset deconfounder suggests that his appearances collectively increased revenue by \$2.5 billion.

The deconfounder produces implausible estimates because it fails to capture important multi-cause confounders. This is clearest when we explicitly adjust for a movie's budget---the quintessential multi-cause confounder, enabling the casting of big-name stars and also reflecting the studio's underlying belief in the viability of the film.\footnote{In \citet{WanBle19} and replication code generously shared with us, actor analyses did not condition on any observed covariates. After we shared our draft with Wang and Blei in July 2020, a reference implementation conditioning on budget and runtime was posted.} 
This simple adjustment produces dramatically different assessments of actor value that are far more reasonable in scale, though likely still overstated.  The deconfounder claims to capture all multi-cause confounding---not only from budget but also genre, series, directors, writers, language, and release season.  WB explicitly argue that including observed covariates with the deconfounder is not necessary---yet this example shows that it is.%

\subsection{Strong Assumptions Rule Out Other Applications}
\label{s:assumptions}

The deconfounder's exceedingly strong assumptions often make it unsuitable for many uses. Although there are other embedded assumptions---see \citet{ogburn2019comment} for more---we focus on four that rule out important applications, including the case study above.

First, the deconfounder requires that treatments arise from a factor model with a low-dimensional confounder.  Practically speaking, analysts must know enough about the functional form of this factor model to feasibly estimate it.

Second, the deconfounder requires treatments to be independently drawn given $\bZ$, ruling out settings where treatments cause other treatments. This implies that casting one actor cannot influence whether another actor is cast later. This alone excludes many realistic settings---except perhaps genetics, where many of these ideas originated.

Third, pinpointing confounders with a factor model requires strong infinite confounding.  In practice, this means analysts must record a very large number of treatments, which are contaminated by comparatively few confounders. In the actor setting, this would be violated by producers who regularly work with the same sets of actors. Furthermore, the mere fact that all movies have finite casts implies limits to the information learned about confounding and means that, by Theorem~\ref{thm:deconfounder_inconsistency_nonlinear}, the deconfounder will always be inconsistent.

Fourth,  even when a pinpointable factor model of the proper class exists, parametric assumptions such as separable confounding or constant treatment effects are used in many proofs, both here and in WB.  Often these conditions help to address failures of positivity by leveraging functional form assumptions.  Yet, particularly in social and medical problems, causal heterogeneity is the rule, not the exception. 

It is unknown how sensitive the na\"ive and deconfounder families of methods are to slight violations of these assumptions.  Until there is a way to relax these assumptions or otherwise evaluate the severity of the consequences of violating them, we cannot recommend either the na\"ive regression or the deconfounder for real-world applications.

\subsection{Takeaways}

Assumptions used to prove deconfounder properties, like pinpointing, are extremely strong and unlikely to hold in real applications.  While analysts cannot know whether the deconfounder estimates are accurate, results from the actor case study are highly implausible.  Given the high-stakes nature of many proposed applications, we think a great deal more evidence is warranted before these methods are put into practice.

\section{Discussion}
WB investigated causal inference for multiple causes under shared confounding.  We have re-examined the theory for every variant deconfounder estimator, as well as every empirical application and simulation for which data are available.  We prove new results showing that for any finite $m$, the deconfounder is inconsistent. As $m \rightarrow \infty$, under strong infinite confounding, the na\"ive regression and full deconfounder both approach asymptotic unbiasedness, but the subset deconfounder requires strong additional assumptions.  We also examined finite-sample properties through simulation, finding no evidence that the deconfounder systematically outperforms the na\"ive regression---or that analysts could possibly identify when it might. Finally, we show that the deconfounder's estimates in existing real-world case studies are not credible and highlight the strong assumptions embedded in the deconfounder framework. In every simulation and empirical study in \citet{WanBle19}, \citet{Zha19} and \citet{wang2019github} for which data was available, our replications show---as predicted by our theory---that no deconfounder consistently improves over the na\"ive regression across the parameter space.  

We note that all theory in this paper is in an asymptotic regime where $n\rightarrow \infty$ for each treatment.  This is helpful for clarifying the strong assumptions necessary for the deconfounder and na\"ive regression to hold, but because $n$ does not grow as a function of $m$, empirical practice in high-dimensional settings likely requires even stronger assumptions.

Collectively, our findings suggest that if assumptions hold, there is no reason to prefer the deconfounder to the na\"ive regression.  However, we ultimately think that the strength of the assumptions is such that neither method should be used in practice.

\bibliography{bless}

\clearpage

\part{Supplemental Information}
\appendix
\parttoc

\clearpage
\FloatBarrier
\section{Notation}
\label{a:notation}
\begin{table}[hbt!]
  \caption{\textbf{Notation reference.} Table of notation for indices, parameters, random variables, and estimators used throughout the manuscript (continued on following page).}
  \label{t:notation}
  \begin{center}
    \begin{tabular}{rp{5in}}
      \hline
      \multicolumn{2}{l}{\bf Indices:} \\
      $n$
      & number of observations
      \\
      $k$
      & dimensionality of latent confounders
      \\
      $m$
      &
      dimensionality of treatments
      \\
      $m_F$, $m_N$
      &
      dimensionality of focal and non-focal treatments, respectively
      \\
      \hline
      \multicolumn{2}{l}{\bf Random variables:} \\
      $\underset{n \times k}{\bZ}$
      &
      unobserved confounders
      \\
      $\underset{n \times m}{\bnu}$
      & random component of treatments
      \\
      $\underset{n \times m}{\bA}$
      &
      observed treatment
      \\
      $\underset{n \times m_F}{\bA_F}, \underset{n \times m_N}{\bA_N}$
      &
      observed subsets of focal and non-focal treatments
      \\
      $\underset{n \times 1}{\bepsilon}$
      & random component of outcome
      \\
      $\underset{n \times 1}{\bY}$
      &
      observed outcome
      \\
      \hline
      \multicolumn{2}{l}{\bf Parameters:} \\
      $\btheta$
      & coefficients linearly mapping confounders to treatments
      \\
      $\bgamma$
      & coefficients linearly mapping confounders to outcome
      \\
      $\bbeta$
      & coefficients linearly mapping treatments to outcome
      \\
      $\bbeta_F, \bbeta_N$
      & subset of $\bbeta$ corresponding to focal and non-focal treatments
      \\
      $\sigma^2$
      & conditional variance of treatments
      \\
      $\omega^2$
      & conditional variance of outcome
      \\
      \hline
    \end{tabular}
  \end{center}
\end{table}

\begin{table}
  \caption{\textbf{Notation reference (continued).} Table of notation for indices, parameters, random variables, and estimators used throughout the manuscript (continued from previous page).}
  \label{t:notation2}
  \begin{center}
    \begin{tabular}{rp{5in}}
      \hline
      \multicolumn{2}{l}{\bf Estimators:} \\
      $\hat\bZ$
      & substitute confounder estimated by factor modeling of treatments
      \\
      $\hat\btheta$
      & implicitly learned mapping from substitute confounder to outcome
      \\
      $\hat\bbeta^\oracle$
      & infeasible oracle regression estimator using unobserved latent confounder
      \\
      $\hat\bbeta^\full$
      & infeasible full deconfounder estimator using collinear substitute confounder
      \\ 
      $\hat\bbeta^\sdf$
      & subset deconfounder estimator using substitute confounder and focal treatments only
      \\
      $\hat\bbeta^\naive$
      & na\"ive regression of outcome on treatments, ignoring confounding
      \\
      $\hat\bbeta^\ridge$
      & ridge deconfounder estimator using collinear substitute confounder and penalization
      \\
      $\hat\bbeta^\nonlinear$
      & nonlinear ridge deconfounder estimator using orthogonal polynomials of substitute confounder and penalization
      \\
      $\hat\bbeta^\wndf$
      & white-noise deconfounder estimator using collinear substitute confounder with generated noise
      \\
      $\hat\bbeta^\pndf$
      & posterior-mean deconfounder estimator using probabilistic principal components
      \\
      $\bS$
      & generated noise (posterior noise) used in white-noise (posterior-mean) deconfounder
      \\
      $\bM^\ast$
      & annihilator matrix corresponding to any regression estimator $\ast$
      \\
      \hline
      \multicolumn{2}{l}{\bf Derived variables:} \\
      $\bU \bD \bV^\top$
      & output of singular-value decomposition of $\bA$
      \\
      $\bQ, \bLambda$
      & output of eigendecomposition $\btheta^\top \btheta = \bQ \bLambda \bQ^\top$
      \\
      \hline
    \end{tabular}
  \end{center}
\end{table}

\FloatBarrier
\section{Derivations}
\label{a:derivation}
\subsection{Minor Asymptotic Results}
\label{a:minor_asymptotics}

In this section, we collect a number of minor results that will be useful in
bias derivations. We begin by reviewing properties of probabilistic principal
component analysis, then present lemmas relating to the asymptotic behavior of
the deconfounder and na\"ive estimators.

\subsubsection{Properties of Probabilistic Principal Component Analysis}
\label{a:ppca}

For convenience, we review properties of probabilistic principal component
analysis used in the remainder of this section. The generative model is
\begin{align*}
  \bZ &\sim \cN(\bm{0}, \bI) \\
  \bA &\sim \cN(\bZ \btheta, \sigma^2 \bI).
\end{align*}

For compactness, we use $\bZ \sim \cN(\bm{0}, \bI)$ to denote independently sampling of $\bZ_i$ from a normal distribution centered on the $i$-th row of the mean matrix and the given covariance matrix, then collecting samples in $\bZ$. \citet{TipBis99} show that this data-generating process implies 
\begin{align}
(\bZ | \bA) \sim \cN\left( \bA \btheta^\top \left( \btheta \btheta^\top + \sigma^2 \bI \right)^{-1},\ \sigma^2 \left( \btheta \btheta^\top + \sigma^2 \bI \right)^{-1} \right). \label{eq:post_z_given_a}
\end{align}

We now examine asymptotic relationships between the singular value decomposition
$\bA = \bU \bD \bV^\top$ and the eigendecomposition $\btheta^\top \btheta = \bQ
\bLambda \bQ^\top = \bQ_{1:k} \bLambda_{1:k} \bQ_{1:k}^\top$; note that the
trailing $m-k$ eigenvalues are zero. (Subscripts of the form $\bX_{i:j}$
generally indicate column subsets of matrix $\bX$ from column $i$ to column $j$,
except when indexing diagonal matrices where they indicate the corresponding
diagonal element.)
\begin{align*}
  \plim_{n \to \infty} \frac{1}{\sqrt{n}} \bD_{1:k}
  &= \left(\bLambda_{1:k} + \sigma^2 \bI \right)^{\frac{1}{2}} \\
  \plim_{n \to \infty} \frac{1}{\sqrt{n}} \bD_{(k+1):m}
  &= \sigma \bI \\
  \plim_{n \to \infty} \bV_{1:k}
  &= \bQ_{1:k} \\
  \plim_{n \to \infty} \bV_{(k+1):m} \bV_{(k+1):m}^\top
  &= \bQ_{(k+1):m} \bQ_{(k+1):m}^\top
\end{align*}
where the last equality follows from $\plim_{n \to \infty} \bV \bV^\top = \bQ
\bQ^\top = \bI$.

\subsubsection{Consistency and Inconsistency Results for the Deconfounder}
\label{a:consistency_deconfounder}

In this section, we present minor results relating to the consistency of the deconfounder. Consider $n$ observations drawn from a data-generating process with $k$
unobserved confounders, $\bZ \sim \cN(\bm{0}, \bI)$ and $m \ge k$ observed
treatments, $\bA \sim \cN(\bZ \btheta, \sigma^2 \bI)$. 

\begin{mdframed}
  \begin{defn}{(Pinpointedness of the substitute confounder.)}
    \label{def:consistency_substitute_confounder}
    A substitute confounder, $\bZ$ is said to be pinpointed if its posterior distribution, $f(\hat{\bz} | \bA)$, collapses to a Dirac delta, $\delta(g(\bZ))$, where $g(\bZ)$ is a bijective transformation of $\bZ$.
\end{defn}
\end{mdframed}
Specifically, pinpointedness \citep[previously referred to as ``consistency of the substitute confounder'']{WanBle19} does not require convergence of $\hat\bZ$ to $\bZ$, as consistency; for example, convergence to a rotation or rescaling will suffice. Below, we show pinpointing requires an infinite number of stochastic causes.

\begin{mdframed}
\begin{lem}
  \label{l:pinpointing}
  In the linear-linear setting, strong infinite confounding is necessary for $\hat\bZ$ to asymptotically pinpoint $\bZ$ as the number of causes goes to infinity.
\end{lem}
\end{mdframed}

\textit{Proof of Lemma~\ref{l:pinpointing}.}

\citet{TipBis99} show that under the probabilistic principal components analysis
model (see Supplement~\ref{a:ppca}), the posterior of the confounder, $f(\bz |
\bA)$, follows \eqref{eq:post_z_given_a}. The substitute confounder is a summary
statistic, such as the mode, of this posterior. We examine the best-case
scenario in which $\btheta$ and $\sigma^2$ are known. In this setting, the posterior
variance of the $k'$-th confounder is $\Var\left( Z_{i,k'} |
\bA_i, \btheta, \sigma^2 \right) = \sigma^2 \left( \btheta_{k'}
\btheta_{k'}^\top + \sigma^2 \right)^{-1}$. Because we assume $\sigma^2>0$, if the variance goes to zero, which pinpointing implies, then each $\btheta_{k'} \btheta_{k'}$, the $k'$-th diagonal element of $\btheta \btheta^\top$, must go to infinity. (We rule out the case of $\sigma^2=0$, by the assumption that the causes are a nondeterministic function of the latent confounders.) Thus, pinpointing of $\bZ_i$ implies strong infinite confounding. \qed

It is easy to see that this argument generalizes to all factor models with
continuous density, i.e. models with continuous $f(\bz_i)$ and $f(\ba_i | \bz_i)
= \prod_{j=1}^m f(\ba_{i,j}| \bz_i)$. Because $f(\bz|\ba) \propto f(\ba|\bz)
f(\bz)$ maintains nonzero variance for all finite $m$ when $f(\ba_i | \bz_i)$ is
nondegenerate, pinpointing requires infinite causes. We return to this point in
Theorem~\ref{thm:deconfounder_inconsistency_nonlinear}.

Next, we present results relating to the inconsistency of various components of the deconfounder in the linear-linear setting. To review, the deconfounder proceeds as follows. It takes
the singular value decomposition $\bA = \bU \bD \bV^\top$, then extracts the
first $k$ components to form $\hat\bZ \equiv \sqrt{n} \bU_{1:k}$. It then
computes $\hat\E[\bA | \hat\bZ] = \hat\bZ \hat\btheta$, where $\hat\btheta
\equiv \frac{1}{\sqrt{n}} \bD_{1:k} \bV_{1:k}^\top$.

\begin{mdframed}
\begin{lem}{(Inconsistency of $\hat\btheta$.)}
  \label{l:theta_hat}
  The asymptotic behavior of $\hat\btheta$ is governed by
  \begin{align*}
    \plim_{n \to \infty} \hat\btheta
    &= \left( \bLambda_{1:k} + \sigma^2 \bI \right)^{\frac{1}{2}} \bLambda_{1:k}^{-\frac{1}{2}}
    \bR^\top \btheta,
  \end{align*}
  where $\bR$ and $\bLambda_{1:k}$ are given by the eigendecomposition $\btheta
  \btheta^\top = \bR \bLambda_{1:k} \bR^\top $.
\end{lem}
\end{mdframed}

\begin{proof}
  We begin with the eigendecomposition $\btheta^\top \btheta = \bQ \bLambda
  \bQ^\top = \bQ_{1:k} \bLambda_{1:k} \bQ_{1:k}^\top$, where the last step
  follows from the fact that the trailing $m-k$ diagonal entries of $\bLambda$ are
  zero.

  We now turn to $\hat\btheta = \frac{1}{\sqrt{n}} \bD_{1:k} \bV_{1:k}^\top$. By
  the properties of probabilistic PCA, $\plim_{n \to \infty} \frac{1}{\sqrt{n}} \bD_{1:k} = \left(
  \bLambda_{1:k} + \sigma^2 \bI \right)^{\frac{1}{2}}$ and $\plim_{n \to
    \infty} \bV_{1:k} = \bQ_{1:k}$ \citep{TipBis99}. The lemma then
  follows from the singular value decomposition $\btheta = \bR
  \bLambda^{\frac{1}{2}} \bQ^\top = \bR \bLambda_{1:k}^{\frac{1}{2}}
  \bQ_{1:k}^\top$ by solving for $\bQ_{1:k}$ and substituting.
\end{proof}

It will be the case that the white-noised and subset deconfounder implicitly rely on $\hat\btheta^\top \hat\btheta$ to adjust for dependence between causes. In Lemma~\ref{l:residual_dependence}, we show that a consequence of Lemma~\ref{l:theta_hat} (inconsistency of $\hat\btheta$) is that $\hat\Cov(\bA) \equiv \hat\btheta^\top \hat\btheta$ is a poor estimator of the covariance of $\bA$; the dependence will be incorrectly modeled even as $n$ goes to infinity.

\begin{mdframed}
\begin{lem}{(Mismodeled dependence structure in $\bA$.)}
  \label{l:residual_dependence}
  When $\widehat\Cov(\bA) = \hat\btheta^\top \hat\btheta$ is used as an
  estimator for $\Cov(\bA)$, the unmodeled residual dependence among causes is
  asymptotically equal to $\sigma^2 \left[ \bI - \btheta^\top (\btheta
    \btheta^\top)^{-1} \btheta \right]$.
\end{lem}
\end{mdframed}
When the number of causes is finite, this residual covariance is nonspherical.
In contrast, the true conditional dependence is $\Cov(\bA | \bZ) = \sigma^2 \bI$.

\noindent \textit{Proof.}
  \begin{align*}
    \plim_{n \to \infty} \Cov(\bA) - \widehat\Cov(\bA)
    &= \plim_{n \to \infty} \btheta^\top \btheta + \sigma^2 \bI - \hat\btheta^\top \hat\btheta \\
    &= \btheta^\top \btheta + \sigma^2 \bI
  - \btheta^\top \bR
  \left( \bLambda_{1:k} + \sigma^2 \bI \right) \bLambda_{1:k}^{-1}
  \bR^\top \btheta \nonumber \\
  &= \sigma^2 \bI
  - \sigma^2 \btheta^\top \bR \bLambda_{1:k}^{-1} \bR^\top \btheta \nonumber \\
  &= \sigma^2 \left[ \bI - \btheta^\top (\btheta \btheta^\top)^{-1} \btheta \right]
   & \qed \\
  \intertext{Under the strong infinite confounding assumption,}
  \lim_{m \to \infty} \plim_{n \to \infty} \Cov(\bA) - \widehat\Cov(\bA)
  &= \sigma^2 \bI
  \end{align*}  

\paragraph{An Example When Strong Infinite Confounding Fails.} 
\label{a:weakinfconf}
Here we consider an example where the number of treatments increases, but strong infinite confounding does not hold. This builds on an idea found in \citet{d2019comment}.  Suppose $Z_{i} \sim \cN(0, \sigma^2)$. We will suppose that $A_{i,m} = \frac{1}{m^2} Z_{i}  + \epsilon_{i}$, where $\epsilon_{i} \sim \cN(0, \sigma^2)$.  In this example, as $\lim_{m \rightarrow \infty} \btheta \btheta^\top = \sum_{m = 1}^{\infty} \frac{1}{m^4} = \frac{\pi^4}{90}$.  Therefore, strong infinite confounding fails.  In general, in the one-dimensional case, the infinite series must diverge for strong infinite confounding to hold.

We now provide a minor result that helps characterize the behavior of the
na\"ive estimator, \eqref{e:naive}, when applied to sequences of data-generating
processes satisfying \textit{strong infinite confounding}
(Definition~\ref{def:infinite_confounding}). In Supplement~\ref{a:bias_naive}
(proof of Proposition~\ref{prop:bias_naive}), we will show the conditions for
asymptotic unbiasedness of the na\"ive estimator as $n$ and $m$ go to
infinity. Lemma~\ref{l:infinite_confounding} states that under the assumption of
strong infinite confounding, this condition is asymptotically satisfied as $m$
grows.

\subsubsection{Behavior of the Na\"ive Estimator}

\begin{mdframed}
\begin{lem}{(Na\"ive convergence under strong infinite confounding.)}
  \label{l:infinite_confounding}
  A sequence of strongly infinitely confounded data-generating processes
  satisfies
  \begin{align*}
    \lim_{m \to \infty} \btheta \left( \btheta^\top \btheta + \sigma^2 \bI
    \right)^{-1} \btheta^\top
    &= \bI
  \end{align*}

\end{lem}
\end{mdframed}

\noindent \textit{Proof.} By the Woodbury matrix identity,
\begin{align*}
\left(\btheta\btheta^\top \frac{1}{\sigma^4}\bI + \frac{1}{\sigma^2} \bI\right)^{-1} &= \sigma^2 \bI  - \sigma^4 \bI \btheta ( \sigma^4 \bI + \sigma^2 \bI \btheta^\top \btheta)^{-1} \btheta^\top \\
\left(\btheta\btheta^\top \frac{1}{\sigma^2}\bI + \bI\right)^{-1}\sigma^2& = \sigma^2 \bI  - \sigma^2 \btheta ( \sigma^2 \bI + \btheta^\top \btheta)^{-1} \btheta^\top \\
\left(\btheta\btheta^\top \frac{1}{\sigma^2}\bI + \bI\right)^{-1} & = \bI  -  \btheta ( \sigma^2 \bI + \btheta^\top \btheta)^{-1} \btheta^\top \\
\end{align*} 
for any $m$. Because both the entries and number of columns of $\btheta$ are
finite, the strong infinite confounding condition requires that the diagonal
elements of $\btheta \btheta^\top$ also tend to infinity as $m$ grows large. Therefore $ \lim_{m
  \to \infty} \left( \frac{1}{\sigma^2} \btheta \btheta^\top + \bI
\right)^{-1} = \bm{0}$, and $ \lim_{m \to \infty} \btheta \left(
\btheta^\top \btheta + \sigma^2 \bI \right)^{-1} \btheta^\top = \bI$. \qed

Supplement~\ref{a:bias_subset} (proof of Proposition~\ref{prop:bias_subset}) shows
that unbiasedness of the subset deconfounder estimator requires $\lim_{m \to
  \inty} \left( \btheta \btheta^\top \right)^{-1} = \bm{0}$, which is trivially
satisfied for strongly infinitely confounded sequences of data-generating
processes.

\subsection{Bias of the Na\"ive Estimator}
\label{a:bias_naive}

For convenience, we reiterate the data-generating process, na\"ive estimation
procedure. We will suppose, without loss of generality, that the $m$
causes, $\bA$, are divided into $m_F$ focal causes of interest, the column
subset $\bA_F$, and $m_N$ nonfocal causes, $\bA_N$. As before, we consider $n$
observations drawn i.i.d. as follows.
\begin{align*}
  \underset{n \times k}{\bZ} &\sim \cN(\bm{0}, \bI) \\
  \underset{n \times m_F}{\bnu_F} &\sim \cN(\bm{0}, \sigma^2 \bI) \\
  \underset{n \times m_N}{\bnu_N} &\sim \cN(\bm{0}, \sigma^2 \bI) \\
  \underset{n \times m_F}{\bA_F} &= \bZ \btheta_F + \bnu_F \\
  \underset{n \times m_N}{\bA_N} &= \bZ \btheta_N + \bnu_N \\
  \underset{n \times 1}{\bepsilon} &\sim \cN(\bm{0}, \omega^2) \\
  \underset{n \times 1}{\bY} &= \bA_F \bbeta_F + \bA_N \bbeta_N + \bZ \bgamma + \bepsilon
\end{align*}

The na\"ive estimator estimates the treatment effects by conducting a regression of
the outcome on both focal and nonfocal causes, producing estimates for both
focal and nonfocal effects, then discarding the latter. The full regression
coefficients are 
\begin{align} \label{e:naive}
  \left[
    \begin{array}{l}
      \hat\bbeta_F^\naive \\
      \hat\bbeta_N^\naive
    \end{array}
    \right]
&\equiv
\left(
\left[ \bA_F, \bA_N \right]^\top \left[ \bA_F, \bA_N \right]
\right)^{-1}
\left[ \bA_F, \bA_N \right]^\top \bY.
\end{align}

We will prove Proposition~\ref{prop:bias_naive_focal}, a generalization of Proposition~\ref{prop:bias_naive} that distinguishes between focal and non-focal treatment cases. The proof of Proposition~\ref{prop:bias_naive} is by reduction to the special case in which all causes are of interest, so that $\bA_F = \bA$ and $\bA_N$ is empty.

\begin{mdframed}
\begin{prop}
\label{prop:bias_naive_focal}
(Asymptotic Bias of the Na\"ive Regression under Strong Infinite Confounding.)

Suppose that the $m$ causes, $\bA$, are divided into focal causes of interest,
the column subset $\bA_F$, and nonfocal causes, $\bA_N$, without loss of
generality. The bias of the na\"ive estimator, \eqref{e:naive}, for the
corresponding focal effects, $\bbeta_F$, is given by
\begin{align*}
\plim_{n \to \infty} \hat\bbeta_F^\naive - \bbeta_F 
&= 
\left[
\btheta_F^\top \btheta_F
+ \sigma^2 \bI
- \btheta_F^\top \bOmega \btheta_F
\right]^{-1}
\left[
\btheta_F^\top
- \btheta_F^\top \bOmega
\right]
\bgamma,
\end{align*}
where $\bOmega = \btheta_N
\left( \btheta_N^\top \btheta_N + \sigma^2 \bI \right)^{-1}
\btheta_N^\top $ and $\btheta_F$ and $\btheta_N$ are the corresponding column subsets of
$\btheta$. Under the assumptions of the linear-linear model,
\begin{align*}
\lim_{m \to \infty} \plim_{n \to \infty}  \hat\bbeta_F^\naive - \bbeta_F &= \bm{0}.
\end{align*}
\end{prop}
\end{mdframed}

\noindent\textit{Proof of Proposition~\ref{prop:bias_naive_focal}.}
By the Frisch-Waugh-Lovell theorem, $\hat\bbeta_F$ can be re-expressed in terms
of the portion of $\bA_F$ not explained by $\bA_N$.  Without loss of generality we suppose that the set of focal treatments, $\bA_F$ are fixed at the outset and are finite.  We denote the residualized
focal treatments as $\tilde\bA_F^\naive = \bA_F - \hat\bA_F^\naive$, where
\begin{align*}
  \hat\bA_F^\naive
  &= \hat\E[ \bA_F | \bA_N ] = \bA_N \hat\bzeta \text{, } \\
  \hat\bzeta
  &= (\bA_N^\top \bA_N)^{-1} \bA_N^\top \bA_F \text{, and} \\
  \plim_{n \to \infty} \hat\bzeta
  &\equiv \bzeta = \left( \btheta_N^\top \btheta_N + \sigma^2 \bI \right)^{-1} \btheta_N^\top \btheta_F .
\end{align*}

The na\"ive estimator is then rewritten as follows:
\begin{align*}
  \hat{\bbeta}_F^\naive
  &= \left( \frac{1}{n} \tilde\bA_F^{\naive \top} \tilde\bA_F^\naive \right)^{-1} \frac{1}{n} \tilde\bA_F^\naive \bY
\end{align*}
We now characterize the asymptotic bias of this estimator by examining the
behavior of $\frac{1}{n} \tilde\bA_F^{\naive \top} \tilde\bA_F^\naive$ and
$\frac{1}{n} \tilde\bA_F^\naive \bY$ in turn. Beginning with the residual
variance of the focal causes,
\begin{align}
  \frac{1}{n} \tilde\bA_F^{\naive \top} \tilde\bA_F^\naive 
  &=
  \frac{1}{n}
  \left( \bA_F - \hat\bA_F^\naive \right)^\top
  \left( \bA_F - \hat\bA_F^\naive \right) 
  \nonumber \\
  &= \frac{1}{n} \left(
  \bA_F^\top \bA_F
  + \hat\bA_F^{\naive \top} \hat\bA_F^\naive
  - \bA_F^\top \hat\bA_F^\naive
  - \hat\bA_F^{\naive \top} \bA_F
  \right) 
  \nonumber \\
  &= 
  \frac{1}{n} (\bZ \btheta_F + \bnu_F)^\top (\bZ \btheta_F + \bnu_F) 
  \nonumber\\
  &\qquad + \frac{1}{n} \hat\bzeta^\top (\bZ \btheta_N + \bnu_N)^\top (\bZ \btheta_N + \bnu_N) \hat\bzeta 
  \nonumber \\
  &\qquad
  - \frac{1}{n} (\bZ \btheta_F + \bnu_F)^\top (\bZ \btheta_N + \bnu_N) \hat\bzeta 
  \nonumber \\
  &\qquad - \frac{1}{n} \hat\bzeta^\top (\bZ \btheta_N + \bnu_N)^\top (\bZ \btheta_F + \bnu_F) 
  \nonumber \\
  \plim_{n \to \infty} \tilde\bA_F^{\naive \top} \tilde\bA_F^\naive 
  &= \plim_{n \to \infty}
  \btheta_F^\top \btheta_F + \sigma^2 \bI 
  + \hat\bzeta^\top (\btheta_N^\top \btheta_N + \sigma^2 \bI) \hat\bzeta 
  - \btheta_F^\top \btheta_N \hat\bzeta 
  - \hat\bzeta^\top \btheta_N^\top \btheta_F 
  \nonumber \\
  &= 
  \btheta_F^\top \btheta_F + \sigma^2 \bI
  - \btheta_F^\top \btheta_N \left( \btheta_N^\top \btheta_N + \sigma^2 \bI \right)^{-1}
  \btheta_N^\top \btheta_F, \quad \text{and}
  \label{e:naive_denominator_large_n_small_m} \\
  \lim_{m \to \infty} \plim_{n \to \infty} \tilde\bA_F^{\naive \top} \tilde\bA_F^\naive 
  &= 
  \sigma^2 \bI \quad \text{under the infinite confounding assumption.}
  \label{e:naive_denominator_large_n_large_m}
\end{align}
Turning to the residual covariance between the focal causes and the outcome,
\begin{align}
  \frac{1}{n} \tilde\bA_F^{\naive \top} \bY  
  &= 
  \frac{1}{n} \left( \bA_F - \hat\bA_N \hat\bzeta \right)^\top
  \left( \bA_F \bbeta_F + \bA_N \bbeta_N + \bZ \bgamma + \bepsilon \right) 
  \nonumber \\
  &= 
  \frac{1}{n} \bA_F^\top \bA_F \bbeta_F
  + \frac{1}{n} \bA_F^\top \bA_N \bbeta_N
  + \frac{1}{n} \bA_F^\top \bZ \bgamma 
  + \frac{1}{n} \bA_F^\top \bepsilon
  \nonumber \\
  &\qquad
  - \frac{1}{n} \hat\bzeta^\top \bA_N^\top \bA_F \bbeta_F
  - \frac{1}{n} \hat\bzeta^\top \bA_N^\top \bA_N \bbeta_N
  - \frac{1}{n} \hat\bzeta^\top \bA_N^\top \bZ \bgamma
  - \frac{1}{n} \hat\bzeta^\top \bA_N^\top \bepsilon
  \nonumber \\
  &= 
  \frac{1}{n} (\btheta_F^\top \bZ^\top + \bnu_F^\top) (\bZ \btheta_F + \bnu_F) \bbeta_F 
  + \frac{1}{n} (\btheta_F^\top \bZ^\top + \bnu_F^\top) (\bZ \btheta_N + \bnu_N) \bbeta_N 
  \nonumber \\
  &\qquad
  + \frac{1}{n} (\btheta_F^\top \bZ^\top + \bnu_F^\top) \bZ \bgamma
  + \frac{1}{n} \bA_F^\top \bepsilon
  - \frac{1}{n} \hat\bzeta^\top (\btheta_N^\top \bZ^\top + \bnu_N^\top)
  (\bZ \btheta_F + \bnu_F)
  \bbeta_F 
  \nonumber \\
  &\qquad - \frac{1}{n} \hat\bzeta^\top (\btheta_N^\top \bZ^\top + \bnu_N^\top)
  (\bZ \btheta_N + \bnu_N)
  \bbeta_N 
  - \frac{1}{n} \hat\bzeta^\top (\btheta_N^\top \bZ^\top + \bnu_N^\top) \bZ \bgamma
  - \frac{1}{n} \hat\bzeta^\top \bA_N^\top \bepsilon
  \nonumber 
  \end{align}
  Taking limits,
  \begin{align}
  \plim_{n \to \infty} \frac{1}{n} \tilde\bA_F^{\naive \top} \bY  
  &= 
  \left[ \btheta_F^\top \btheta_F + \sigma^2 - \btheta_F^\top \btheta_N \left( \btheta_N^\top \btheta_N + \sigma^2 \bI \right)^{-1} \btheta_N^\top \btheta_F \right] \bbeta_F 
  \nonumber \\
  &\qquad
  + \left[ \btheta_F^\top - \btheta_F^\top \btheta_N \left( \btheta_N^\top \btheta_N + \sigma^2 \bI \right)^{-1} \btheta_N^\top  \right] \bgamma, \quad \text{and}
  \label{e:naive_numerator_large_n_small_m}
  \\
  \lim_{m \to \infty} \plim_{n \to \infty} \frac{1}{n} \tilde\bA_F^{\naive \top} \bY
  &= \sigma^2 \bbeta_F
  \quad \text{under the infinite confounding assumption.}
  \label{e:naive_numerator_large_n_large_m}
\end{align}

Combining
~\eqref{e:naive_denominator_large_n_small_m}~and~\eqref{e:naive_numerator_large_n_small_m} and applying Lemma 3 to $\btheta_{N}$,
\begin{align*}
  \plim_{n \to \infty} \hat\bbeta_F^\naive
  &= \bbeta_F +
  \left[
  \btheta_F^\top \btheta_F + \sigma^2 \bI
  - \btheta_F^\top \btheta_N \left( \btheta_N^\top \btheta_N + \sigma^2 \bI \right)^{-1}
  \btheta_N^\top \btheta_F
  \right]^{-1} 
  \\
  &\qquad\qquad \cdot
  \left[ \btheta_F^\top - \btheta_F^\top \btheta_N \left( \btheta_N^\top \btheta_N + \sigma^2 \bI \right)^{-1} \btheta_N^\top  \right] \bgamma
  ,
  \intertext{and under the infinite confounding assumption, \eqref{e:naive_denominator_large_n_large_m}~and~\eqref{e:naive_numerator_large_n_large_m} yield}
  \lim_{m \to \infty} \plim_{n \to \infty} \hat\bbeta_F^\naive 
  &= \bbeta_F .
\end{align*}

When all effects are of interest, the above reduces to
\begin{align*}
  \hat\bbeta^\naive 
  &\equiv
  \left( \bA^\top \bA \right)^{-1}
  \bA^\top \bY
  \\
  \plim_{n \to \infty} \frac{1}{n} \bA^\top \bA
  &=
  \btheta^\top \btheta + \sigma^2 \bI
  \\
  \plim_{n \to \infty} \frac{1}{n} \bA^\top \bY
  &=
  \plim_{n \to \infty} \frac{1}{n}
  (\btheta^\top \bZ^\top + \bnu^\top )
  (\bZ \btheta \bbeta + \bnu \bbeta + \bZ \bgamma + \bepsilon)
  \\
  &=
  \btheta^\top \btheta \bbeta +
  \btheta^\top \bgamma +
  \sigma^2 \bI \bbeta
  \\
  \plim_{n \to \infty} \hat\bbeta^\naive 
  &=
  \bbeta +
  (\btheta^\top \btheta + \sigma^2 \bI)^{-1}
  \btheta^\top \bgamma
  \\
  \lim_{m \to \infty} \plim_{n \to \infty} \hat\bbeta^\naive
  &=
  \bbeta \qed
\end{align*}

\subsection{Bias of the Penalized Deconfounder Estimator}
\label{a:bias_ridge}

For convenience, we reiterate the data-generating process and penalized deconfounder
estimation procedure here, along with identities that will be useful in the
proof of Proposition~\ref{prop:bias_ridge}. As before, we consider $n$ observations
drawn i.i.d. as follows.
\begin{align*}
  \underset{n \times k}{\bZ} &\sim \cN(\bm{0}, \bI) \\
  \underset{n \times m}{\bnu} &\sim \cN(\bm{0}, \sigma^2 \bI) \\
  \underset{n \times m}{\bA} &= \bZ \btheta + \bnu \\
  \underset{n \times 1}{\bepsilon} &\sim \cN(\bm{0}, \omega^2) \\
  \underset{n \times 1}{\bY} &= \bA \bbeta + \bZ \bgamma + \bepsilon
\end{align*}

The penalized deconfounder estimator (1) takes the singular value decomposition
$\bA = \bU \bD \bV^\top$; (2) extracts the first $k$ components, $\hat\bZ \equiv
\sqrt{n} \bU_{1:k}$; and (3) estimates the focal effects by computing
\begin{align*}
  \left[\begin{array}{l}
    \hat\bbeta^\ridge \\
    \hat\bgamma^\ridge
  \end{array} \right]
&\equiv
\left(
\left[ \bA, \hat\bZ \right]^\top \left[ \bA, \hat\bZ \right] + \lambda(n) \bI
\right)^{-1}
\left[ \bA, \hat\bZ \right]^\top \bY
\end{align*}
and discarding $\hat\bgamma^\ridge$. The $\lambda(n)$ term indicates the
strength of the ridge penalty; we allow this term to scale with $n$ for full
generality. Note that identification is purely from this term ridge
penalty---because $\hat\bZ$ is merely a linear transformation of $\bA$, the
above is non-estimable when $\lambda(n) = 0$.

We now restate Proposition~\ref{prop:bias_ridge} for convenience.

\begin{mdframed}
  \textbf{Proposition~\ref{prop:bias_ridge}.}  
  \it (Asymptotic Bias of the Penalized Full Deconfounder.)

  Consider the linear-linear data-generating process, in which $n$ observations
  are sampled i.i.d. by drawing $k$ unobserved confounders,
  $\bZ \sim \cN(\bm{0}, \bI)$; these generate $m \ge k$ observed treatments,
  $\bA \sim \cN(\bZ \btheta, \sigma^2 \bI)$; and a scalar outcome is drawn from
  $\bY \sim \cN(\bA \bbeta + \bZ \bgamma, \omega^2)$. The penalized deconfounder
  estimator, as implemented in WB,
  is
  \begin{align*} \hat\bbeta^\ridge
  &\equiv \left( \left[ \bA, \hat\bZ \right]^\top \left[ \bA, \hat\bZ \right]
  + \lambda(n) \bI \right)^{-1} \left[ \bA, \hat\bZ \right]^\top \bY,
  \end{align*}  
  where $\hat\bZ$ is obtained by taking the singular value decomposition $\bA
  = \bU \bD \bV^\top$ and extracting the first $k$ components,
  $\hat\bZ \equiv \sqrt{n} \bU_{1:k}$, and $\lambda(n)$ is a ridge penalty that
  is assumed to be sublinear in $n$. The asymptotic bias of this estimator is
  given by
\begin{align*}
\plim_{n \to \infty} \hat\bbeta^\ridge - \bbeta 
  &=
  \overbrace{
  -
  \bQ_{1:k}
  \ \mathrm{diag}_j \left( \frac{
  1
  }{
  \sigma^2 + \Lambda_j + 1
  } \right)
  \bQ_{1:k}^\top \bbeta
  }^{\text{Regularization}} \\
  &\qquad
  \underbrace{+ \bQ_{1:k}
  \ \mathrm{diag}_j \left( \frac{
    \Lambda_j
  }{
    \sigma^2 + \Lambda_j + 1
  } \right)
  \bQ_{1:k}^\top \btheta^\top (\btheta \btheta^\top)^{-1} \bgamma}_{\text{Omitted Variable Bias}},
  \intertext{where $\bQ$ and $\bLambda = [\Lambda_1, \ldots, \Lambda_k, 0, \ldots]$ are respectively eigenvectors and eigenvalues obtained from decomposition of $\btheta^\top \btheta$. Under strong infinite confounding,}
\lim_{m \to \infty} \plim_{n \to \infty} \hat\bbeta^\ridge - \bbeta &= \bm{0}.
\end{align*}
\end{mdframed}

In what follows, we prove Proposition~\ref{prop:bias_ridge} by relating the
asymptotic behavior of the penalized deconfounder to the eigendecomposition
$\btheta^\top \btheta = \bQ \bLambda \bQ^\top = \bQ_{1:k} \bLambda_{1:k}
\bQ_{1:k}^\top$. We will rely on the singular value decompositions of $\bA$ and
$[ \bA, \hat\bZ]$. To distinguish these, for this section only, we denote the
former as $\bA = \bU_A \bD_A \bV_A^\top$ and the latter as $[ \bA, \hat\bZ] =
\bU_{AZ} \bD_{AZ} \bV_{AZ}^\top $. Lemma~\ref{l:svd_az} characterizes the
relationship between these.

\begin{mdframed}
\begin{lem}
  \label{l:svd_az}
  For any $n$, the singular value decomposition $[ \bA, \hat\bZ] = \bU_{AZ}
  \bD_{AZ} \bV_{AZ}^\top $ obeys
  \begin{align*}
    \bU_{AZ}
    &= \left[ \bU_A,\ \ast \right] \\
    \bD_{AZ}
    &= \left[
      \begin{array}{lllll}
        \left( \bD_{A,1:k}^2 + n \bI \right)^{\frac{1}{2}}, &&
        \bm{0}, &&
        \bm{0} \\
        \enspace\bm{0}, &&
        \bD_{A,(k+1):m}, &&
        \bm{0} \\
        \enspace\bm{0}, &&
        \bm{0}, &&
        \bm{0}
      \end{array}
      \right]
    \\
    \bV_{AZ}^\top
    &= \left[ \begin{array}{lll}
        \left( \bD_{A,1:k}^2 + n \bI \right)^{-\frac{1}{2}} \bD_{A,1:k} \bV_{A,1:k}^\top, &&
        \sqrt{n} \left( \bD_{A,1:k}^2 + n \bI \right)^{-\frac{1}{2}} \\[1ex]
        \enspace\bV_{A,(k+1):m}^\top, &&
        \bm{0} \\[1ex]
        \enspace\ast && \ast
      \end{array} \right],
  \end{align*}
  where $\ast$ indicates irrelevant normalizing columns in $\bU_{AZ}$ and $\bV_{AZ}$.
\end{lem}
\end{mdframed}

\textit{Proof of Lemma~\ref{l:svd_az}.}
The first equality follows from the fact that the newly appended $\hat\bZ$
columns are merely linear transformations of $\bA$, so that the leading $m$ left
singular vectors remain unchanged.

Of the unchanged left singular vectors, each of the first $k$ is directly
proportional to the corresponding column of $\hat\bZ$. Because $\hat\bZ$ is
standardized by construction, the variance explained by each of the first $k$
left singular vector increases by one; the $(k+1)$-th through $m$-th left
singular vectors are orthogonal to the newly appended $\hat\bZ$ and so their
singular values remain unchanged. This yields the second equality.

The third equality can verified by $\bU_{AZ} \bD_{AZ} \bV_{AZ}^\top = [\bU_A \bD_A \bV_A^\top, \sqrt{n} \bU_{A,1:k}] = [ \bA,
  \hat\bZ]$ and $\bV_{AZ,1:m}^\top \bV_{AZ,1:m} = \bI$. \qed

We now examine the asymptotic behavior of the penalized deconfounder estimator.

\noindent\textit{Proof of Proposition~\ref{prop:bias_ridge}.}
\begin{align}
  \left[\begin{array}{l}
      \hat\bbeta^\ridge \\ \hat\bgamma^\ridge
    \end{array}\right]
  &= \left( [ \bA ,\ \hat\bZ ]^\top [ \bA ,\ \hat\bZ ] + \lambda(n) \bI \right)^{-1} [ \bA ,\ \hat\bZ ]^\top \ \bY \\
  &= \bV_{AZ} \left(
  \bD_{AZ}^{\ 2} + \lambda(n) \bI
  \right)^{-1} \bD_{AZ} \bU_{AZ}^\top \bY 
  \intertext{By Lemma~\ref{l:svd_az},}
  &= \left[\begin{array}{ll}
      \bV_A \bD_A, & 
      \ast
      \\[1ex]
      \sqrt{n} \bI, & 
      \ast
    \end{array}\right]
  \left[
    \begin{array}{ll}
      \left( \bD_A^2 + \lambda(n) \bI + n \cdot \mathrm{diag}_j 1\{ j \le k \} \right)^{-1}, &
      \bm{0} \\[1ex]
      \enspace\bm{0}, &
      \bm{0}
    \end{array}
    \right] [ \bU_A , \ \ast ]^\top \bY \nonumber \\
  \intertext{where asterisks denote irrelevant blocks, eliminated below.}
  &= \left[\begin{array}{l}
      \bV_A \bD_A \\
      \sqrt{n} \bI
    \end{array}\right]
  \left( \bD_A^2 + \lambda(n) \bI + n \cdot \mathrm{diag}_j 1\{ j \le k \} \right)^{-1}
  \bU_A^\top \left( \
  \bA \bbeta + \bZ \bgamma + \bepsilon
  \right) \nonumber \\
  \intertext{We now subset to $\hat\bbeta^\ridge$, then substitute $\bA = \bU_A \bD_A \bV_A^\top$, $\bU_A^\top = \bD_A^{-1} \bV_A^\top \bA^\top$ and $\bZ = (\bA - \bnu) \btheta^\top (\btheta \btheta^\top)^{-1}$,}
  \plim_{n \to \infty} \hat\bbeta^\ridge 
  &= \plim_{n \to \infty}
  \bV_A \bD_A 
  \left( \bD_A^2 + \lambda(n) \bI + n \cdot \mathrm{diag}_j 1\{ j \le k \} \right)^{-1}
  \bD_A \bV_A^\top \bbeta
  \label{e:ridge_goto}
  \\
  &\qquad\qquad
  + \bV_A \bD_A 
  \left( \bD_A^2 + \lambda(n) \bI + n \cdot \mathrm{diag}_j 1\{ j \le k \} \right)^{-1}
  \bD_A \bV_A^\top \btheta^\top (\btheta \btheta^\top)^{-1} \bgamma
  \nonumber \\
  &\qquad\qquad
  - \bV_A 
  \left( \bD_A^2 + \lambda(n) \bI + n \cdot \mathrm{diag}_j 1\{ j \le k \} \right)^{-1}
  \bV_A^\top \bA^\top
  \bnu \btheta^\top (\btheta \btheta^\top)^{-1} \bgamma
  \nonumber \\
  &= \plim_{n \to \infty}
  \bV_{A}
  \ \mathrm{diag}_j \left( \frac{
    \sigma^2 + 1\{ j \le k \} \Lambda_j
  }{
    \sigma^2 + \lambda(n) / n + 1\{ j \le k \} (\Lambda_j + 1)
  } \right)
  \bV_{A}^\top \bbeta
  \nonumber \\
  &\qquad\qquad
  + \bV_{A}
  \ \mathrm{diag}_j \left( \frac{
    \sigma^2 + 1\{ j \le k \} \Lambda_j
  }{
    \sigma^2 + \lambda(n) / n + 1\{ j \le k \} (\Lambda_j + 1)
  } \right)
  \bV_{A}^\top \btheta^\top (\btheta \btheta^\top)^{-1} \bgamma
  \nonumber \\
  &\qquad\qquad
  - \bV_A 
  \ \mathrm{diag}_j \left( \frac{
    \sigma^2
  }{
    \sigma^2 + \lambda(n) / n + 1\{ j \le k \} (\Lambda_j + 1)
  } \right)
  \bV_A^\top \btheta^\top (\btheta \btheta^\top)^{-1} \bgamma
  \nonumber \\
  \intertext{By properties of PPCA \citep{TipBis99},}
  &= \plim_{n \to \infty}
  \bbeta - \bV_A
  \ \mathrm{diag}_j \left( \frac{
    \lambda(n) / n +  1\{ j \le k \} 
  }{
    \sigma^2 + \lambda(n) / n + 1\{ j \le k \} (\Lambda_j + 1)
  } \right)
  \bV_A^\top \bbeta
  \nonumber \\
  &\qquad\qquad
  + \bV_{A}
  \ \mathrm{diag}_j \left( \frac{
    1\{ j \le k \} \Lambda_j
  }{
    \sigma^2 + \lambda(n) / n + 1\{ j \le k \} (\Lambda_j + 1)
  } \right)
  \bV_{A}^\top \btheta^\top (\btheta \btheta^\top)^{-1} \bgamma
  \nonumber \\  
  &= 
  \bbeta -
  \bQ_{1:k}
  \ \mathrm{diag}_j \left( \frac{
    \lambda(n) / n + 1
  }{
    \sigma^2 + \lambda(n) / n + \Lambda_j + 1
  } \right)
  \bQ_{1:k}^\top 
  \bbeta
  \nonumber \\
  & \qquad - \left( \frac{
    \lambda(n) / n
  }{
    \sigma^2 + \lambda(n) / n
  } \right)
  \bQ_{(k+1):m} \bQ_{(k+1):m}^\top
  \bbeta 
  \nonumber \\
  &\qquad
  + \bQ_{1:k}
  \ \mathrm{diag}_j \left( \frac{
    \Lambda_j
  }{
    \sigma^2 + \lambda(n) / n + \Lambda_j + 1
  } \right)
  \bQ_{1:k}^\top \btheta^\top (\btheta \btheta^\top)^{-1} \bgamma
  \nonumber \\
  \intertext{When $\lambda(n)$ is sublinear,}
  \plim_{n \to \infty} \hat\bbeta^\ridge
  &=
  \bbeta -
  \bQ_{1:k}
  \ \mathrm{diag}_j \left( \frac{
    1
  }{
    \sigma^2 + \Lambda_j + 1
  } \right)
  \bQ_{1:k}^\top \bbeta 
  \nonumber \\
  & \qquad + \bQ_{1:k}
  \ \mathrm{diag}_j \left( \frac{
    \Lambda_j
  }{
    \sigma^2 + \Lambda_j + 1
  } \right)
  \bQ_{1:k}^\top \btheta^\top (\btheta \btheta^\top)^{-1} \bgamma . \nonumber
\end{align}
Under strong infinite confounding, it can be seen that $\lim_{m \to \infty} \plim_{n \to \infty}
\hat\bbeta^\ridge = \bbeta$. This follows from 
$\lim_{m \to \infty} \bLambda_{1:k}^{-1} = \bm{0}$
and $\lim_{m \to \infty} \btheta^\top (\btheta \btheta^\top)^{-1} \bgamma = \bm{0}$. \qed

\subsection{Bias of the Penalized Deconfounder Estimator under Nonlinear Confounding}
\label{a:bias_ridge_nonlinear}

In this section, we evaluate the behavior of the deconfounder in a more general
setting. We consider $n$ observations drawn i.i.d. from the below
data-generating process.
\begin{align}
  \underset{n \times k}{\bZ} &\sim \cN(\bm{0}, \bI) \nonumber \\
  \underset{n \times m}{\bnu} &\sim \cN(\bm{0}, \sigma^2 \bI) \nonumber \\
  \underset{n \times m}{\bA} &= \bZ \btheta + \bnu \nonumber \\
  \underset{n \times 1}{\bepsilon} &\sim \cN(\bm{0}, \omega^2) \nonumber \\
  \intertext{However, we relax the outcome model to allow for arbitrary
    additively separable confounding,}
  \underset{n \times 1}{\bY} &=
  \left[ \bA_i^\top \bbeta + g_Y(\bZ_i) + \bepsilon_i \right] \label{e:outcome_nonlinear}
\end{align}
The linear-linear model defined in Section~\ref{s:asymptotic} is a special
case of the linear-separable model. Note that in empirical settings, the
specific functional form of $g_Y(\cdot)$ is rarely known except when analyzing
simple physical systems. However, any $g_Y(\cdot)$ can be approximated to
arbitrary degree $d$ as follows. First, denote the polynomial basis expansion of
$\bZ_i$ as $h(\bZ_i) \equiv \left[ \prod_{k'=1}^k Z_{i,k'}^{d'_{k'}}
  \right]_{\sum_{k'=1}^k d'_{k'} \le d}$ and collect these in rows of $h(\bZ) =
[ h(\bZ_i) ]$. Then, \eqref{e:outcome_nonlinear} can be rewritten by Taylor
expansion as
\begin{align}
  \bY &= \bA \bbeta + h(\bZ) \bxi + \bepsilon \label{e:outcome_nonlinear_expansion}
\end{align}
with approximation error that grows arbitrarily small as $d$ grows large. We
will choose $d$ sufficiently to fully capture $g_Y(\cdot)$. Let $\bW$ be the
orthogonal higher-order polynomials of $\bZ$. Then, \eqref{e:outcome_nonlinear}
can be rewritten yet again as
\begin{align}
  \bY &= \bA \bbeta + \gamma \bZ + \delta \bW + \bepsilon.
\end{align}
As before, the effects $\bbeta$ are the causal quantities of interest. Note that
the confounding, $g_Y(\bZ)$, is a nuisance term, so there is no need to
reconstruct it from its expansion. We will assume that $g_Y(\bZ)$ is zero-mean for
convenience; this assumption is trivial to relax using an added intercept.

We will derive the asymptotic behavior of the flexible penalized deconfounder,
which generalizes the penalized full deconfounder of Supplement~\ref{a:bias_ridge}
for all additively separable forms of confounding. The flexible penalized
deconfounder estimator consists of the following procedure: (1) take the
singular value decomposition $\bA = \bU \bD \bV^\top$; and (2) extract the first
$k$ components, $\hat\bZ \equiv \sqrt{n} \bU_{1:k}$. To allow for nonlinear
confounding, (3) compute $h(\hat\bZ)$ and take its QR decomposition, $h(\hat\bZ)
= \bQ_Z \bR_Z = \left[ \frac{1}{\sqrt{n}} \hat\bZ ,\ \frac{1}{\sqrt{n}} \hat\bW
  \right] \bR_Z$.\footnote{The invariance of $\hat\bZ$ follows from its
  orthonormality.} Finally, (4) estimate $\bbeta$ by a ridge regression of the
form
$$[ \hat\bbeta^{\nonlinear \top} ,\ \hat\bgamma^{\nonlinear \top}
  ,\ \hat\bdelta^{\nonlinear \top}]^\top \ = \ \left( [ \bA ,\ \hat\bZ
  ,\ \hat\bW ]^\top [ \bA ,\ \hat\bZ ,\ \hat\bW ] + \lambda(n) \bI \right)^{-1}
[ \bA ,\ \hat\bZ ,\ \hat\bW ]^\top \bY.$$ As in Supplement~\ref{a:bias_ridge}, the
$\lambda(n)$ term indicates the strength of the ridge penalty and is allowed to
scale sublinearly in $n$. Again, identification is purely from this ridge
penalty, because $\hat\bZ$ is merely a linear transformation of $\bA$ and thus
the matrix is non-invertible without regularization.

\begin{mdframed}
\begin{prop}
  \label{prop:bias_ridge_nonlinear} (Asymptotic Bias of the
  Flexible Penalized Deconfounder under Additively Separable Confounding.)

  For all data-generating processes containing a linear factor model and
  additively separable confounding, the asymptotic bias of the flexible ridge
  deconfounder is given by
\begin{align*}
\plim_{n \to \infty} \hat\bbeta^\nonlinear - \bbeta 
  &=
  -
  \bQ_{1:k}
  \ \mathrm{diag}_j \left( \frac{
  1
  }{
  \sigma^2 + \Lambda_j + 1
  } \right)
  \bQ_{1:k}^\top \bbeta
  \\
  & \qquad 
  + \bQ_{1:k}
  \ \mathrm{diag}_j \left( \frac{
    \Lambda_j
  }{
    \sigma^2 + \Lambda_j + 1
  } \right)
  \bQ_{1:k}^\top \btheta^\top (\btheta \btheta^\top)^{-1} \bgamma,
  \intertext{where $\bQ$ and $\bLambda = [\Lambda_1, \ldots, \Lambda_k, 0, \ldots]$ are
  respectively eigenvectors and eigenvalues obtained from decomposition of
  $\btheta^\top \btheta$. Under strong infinite confounding,}
\lim_{m \to \infty} \plim_{n \to \infty} \hat\bbeta^\nonlinear &= \bbeta.
\end{align*}
\end{prop}
\end{mdframed}

We briefly offer intuition for the form of this bias before proceeding to the
proof. The bias expressions in Proposition~\ref{prop:bias_ridge_nonlinear} are
identical to those of Proposition~\ref{prop:bias_ridge}, though the
interpretation diverges slightly due to the flexible nature of the confounding
function, $g_Y(\bZ)$. The term $\bgamma$ represents the portion of the confounding
due to the linear trend in $g_Y(\bZ)$, which induces bias as described above. In
contrast, $\bdelta$ represents the nonlinear portion of the confounding that
remains after eliminating the main linear trend. Because this part of $g_Y(\bZ)$
is by construction orthogonal to $\bZ$ (and therefore to $\bA$, due to the
linear nature of the factor model) it cannot induce bias in
$\hat\bbeta^\nonlinear$.

\noindent\textit{Proof of Proposition~\ref{prop:bias_ridge_nonlinear}.} In what
follows, we will relate the asymptotic behavior of the flexible penalized
deconfounder to the eigendecomposition $\btheta^\top \btheta = \bQ \bLambda
\bQ^\top = \bQ_{1:k} \bLambda_{1:k} \bQ_{1:k}^\top$. To do so, we will rely on
the singular value decompositions of $\bA$, $[ \bA, \hat\bZ]$, and $[ \bA,
  \hat\bZ, \hat\bW]$. For this section only, we respectively denote these as
$\bA = \bU_A \bD_A \bV_A^\top$, $[ \bA, \hat\bZ] = \bU_{AZ} \bD_{AZ}
\bV_{AZ}^\top $, and $[ \bA, \hat\bZ, \hat\bW] = \bU_{AZW} \bD_{AZW}
\bV_{AZW}^\top$. Lemma \ref{l:svd_az} characterizes the relationship between the
first two; we now describe the latter.

For any $n$, the singular value decomposition $[ \bA, \hat\bZ, \hat\bW] = \bU_{AZW}
\bD_{AZW} \bV_{AZW}^\top $ can be seen to obey
\begin{align*}
  \bU_{AZW}
  &= \left[
    \frac{1}{\sqrt{n}} \hat\bZ,\ \bU_{A, (k+1):m},\ \ast,\ \frac{1}{\sqrt{n}} \hat\bW
    \right]
  \\
  \bD_{AZW}
  &= \left[
    \begin{array}{lllllll}
      \left( \bD_{A,1:k}^2 + n \bI \right)^{\frac{1}{2}}, &&
      \bm{0}, &&
      \bm{0}, &&
      \bm{0}\\
      \enspace\bm{0}, &&
      \bD_{A,(k+1):m}, &&
      \bm{0}, &&
      \bm{0} \\
      \enspace\bm{0}, &&
      \bm{0}, &&
      \bm{0}, &&
      \bm{0} \\
      \bm{0}, &&
      \bm{0}, &&
      \bm{0}, &&
      \sqrt{n} \bI
    \end{array}
    \right]
  \\
  \bV_{AZW}^\top
  &= \left[ \begin{array}{lllll}
      \left( \bD_{A,1:k}^2 + n \bI \right)^{-\frac{1}{2}} \bD_{A,1:k} \bV_{A,1:k}^\top, &&
      \sqrt{n} \left( \bD_{A,1:k}^2 + n \bI \right)^{-\frac{1}{2}}, &&
      \bm{0}
      \\[1ex]
      \enspace\bV_{A,(k+1):m}^\top, &&
      \bm{0}, &&
      \bm{0}
      \\[1ex]
      \enspace\ast, &&
      \ast, &&
      \ast
      \\[1ex]
      \enspace\bm{0}, &&
      \bm{0}, &&
      \bI
    \end{array} \right],
\end{align*}
where $\ast$ indicates irrelevant normalizing columns in $\bU_{AZW}$ and
$\bV_{AZW}$. The above is due to Lemma~\ref{l:svd_az} for the first $k+m$
columns. The behavior of the trailing columns follows from the fact that
$\hat\bW$ is normalized and orthogonal to $\hat\bZ$ (and therefore to $\bA$) by
construction, and therefore remains invariant in the decomposition.

We now substitute the singular value decomposition of $[ \bA, \hat\bZ, \hat\bW]$
into the ridge estimator.
\begin{align*}
  \left[\begin{array}{l}
      \hat\bbeta^\nonlinear \\ \hat\bgamma^\nonlinear \\ \hat\bdelta^\nonlinear
    \end{array}\right]
  &= \left(
       [ \bA ,\ \hat\bZ ,\ \hat\bW ]^\top
       [ \bA ,\ \hat\bZ ,\ \hat\bW ]
       + \lambda(n) \bI \right)^{-1}
       [ \bA ,\ \hat\bZ ,\ \hat\bW ]^\top \bY
  \\
  &= \bV_{AZW} \left(
       \bD_{AZW}^{\ 2} + \lambda(n) \bI
  \right)^{-1} \bD_{AZW} \bU_{AZW}^\top \bY 
  \intertext{Eliminating dimensions with zero
    singular values and subsetting to $\hat\bbeta^\nonlinear$, we obtain}
  &= 
      \bV_A \bD_A
      \left( \bD_A^2 + \lambda(n) \bI + n \cdot \mathrm{diag}_j 1\{ j \le k \} \right)^{-1}
      \bU_A^\top \bY 
\end{align*}
and go to \eqref{e:ridge_goto} in the proof of
Proposition~\ref{prop:bias_ridge}. \qed

\subsection{Bias of the White-noised Deconfounder Estimator}
\label{a:bias_noise_white}

In one of the tutorial simulations in \citet{wang2019github}, gaussian noise is added to the substitute confounder to render it estimable. This simulation and our reanalysis is discussed in Supplement~\ref{a:logisticsim}. Here we prove properties of this general strategy.

For convenience, we reiterate the data-generating process and white-noised
deconfounder estimation procedure here. As before, we consider $n$ observations
drawn i.i.d. as follows.
\begin{align*}
  \underset{n \times k}{\bZ} &\sim \cN(\bm{0}, \bI) \\
  \underset{n \times m}{\bnu} &\sim \cN(\bm{0}, \sigma^2 \bI) \\
  \underset{n \times m}{\bA} &= \bZ \btheta + \bnu_F \\
  \underset{n \times 1}{\bepsilon} &\sim \cN(\bm{0}, \omega^2) \\
  \underset{n \times 1}{\bY} &= \bA \bbeta + \bZ \bgamma + \bepsilon
\end{align*}

The white-noised deconfounder estimator (1) takes the singular value
decomposition $\bA = \bU \bD \bV^\top$; (2) extracts the first $k$ components,
$\hat\bZ \equiv \sqrt{n} \bU_{1:k}$ and accompanying $\hat\btheta \equiv
\frac{1}{\sqrt{n}} \bD_{1:k} \bV_{1:k}^\top$; adds noise $\bS \sim \cN(\bm{0},
\psi^2 \bI)$ to $\hat\bZ$ to break perfect collinearity with $\bA$; and (4)
estimates effects by computing
\begin{align}
  \left[\begin{array}{l}
    \hat\bbeta^\wndf \\
    \hat\bgamma^\wndf    
  \end{array} \right]
&\equiv
\left(
\left[ \bA, \hat\bZ + \bS \right]^\top \left[ \bA, \hat\bZ + \bS \right]
\right)^{-1}
\left[ \bA, \hat\bZ + \bS \right]^\top \bY.
\label{e:noise_posterior_white_estimator}
\end{align}

We now restate Proposition~\ref{prop:bias_noise_white} before proceeding to the proof.
\begin{mdframed}
\textbf{Proposition~\ref{prop:bias_noise_white}.}
\label{prop:bias_noise_white}
 \it (Asymptotic Bias of the White-noised Deconfounder.)

  Consider $n$ observations drawn from a data-generating process with $k$
  unobserved confounders, $\bZ \sim \cN(\bm{0}, \bI)$; $m \ge k$ observed
  treatments, $\bA \sim \cN(\bZ \btheta, \sigma^2 \bI)$; and outcome $\bY \sim
  \cN(\bA \bbeta + \bZ \bgamma, \omega^2)$. The white-noised
  deconfounder estimator, is
  \begin{align*}
    \left[\begin{array}{l}
    \hat\bbeta^\wndf \\ \hat\bgamma^\wndf
    \end{array}\right]
    &\equiv
    \left(
    \left[ \bA, \hat\bZ + \bS \right]^\top \left[ \bA, \hat\bZ + \bS \right]
    \right)^{-1}
    \left[ \bA, \hat\bZ + \bS \right] \bY,
  \end{align*}
  where $\hat\bZ$ is obtained by taking the singular value decomposition $\bA =
  \bU \bD \bV^\top$ and extracting the first $k$ components, $\hat\bZ \equiv
  \sqrt{n} \bU_{1:k}$; the addition of white noise, $\bS \sim \cN(0, \psi^2
  \bI)$, makes this regression estimable. The asymptotic bias of this estimator
  is given by
  \begin{align*}
\plim_{n \to \infty} \hat\bbeta^\wndf - \bbeta
  &=
  \left\{ \btheta^\top
  \left[\bI - \frac{\sigma^2}{\psi^2} ( \btheta \btheta^\top )^{-1} \right]
  \btheta +
  \frac{\sigma^2}{\psi^2} (1 + \psi^2) \bI
  \right\}^{-1}
  \btheta^\top \bgamma , \\
 \intertext{and under strong infinite confounding,}
  \lim_{m \to \infty} \plim_{n \to \infty} \hat\bbeta^\wndf - \bbeta
  &=
  \left[ \btheta^\top \btheta +
    \frac{\sigma^2}{\psi^2} (1 + \psi^2) \bI
    \right]^{-1}
  \btheta^\top \bgamma 
  \end{align*}
\end{mdframed}

\noindent\textit{Proof of Proposition~\ref{prop:bias_noise_white}.}

After subsetting \eqref{e:noise_posterior_white_estimator} to the treatment effects, the estimator can be rewritten as
\begin{align*}
  \hat\bbeta^\wndf
  &= (\bA^\top
\bM^\wndf \bA)^{-1} \bA^\top \bM^\wndf \bY ,\quad\text{where} \\
\bM^\wndf
&\equiv
\bI - (\hat\bZ + \bS) \left[ (\hat\bZ + \bS)^\top (\hat\bZ + \bS)
  \right]^{-1} (\hat\bZ + \bS)^\top.
\end{align*}
Note that
\begin{align*}
  \plim_{n \to \infty} \hat\bbeta
  &= \plim_{n \to \infty} (\bA^\top \bM^\wndf \bA)^{-1} \bA^\top \bM^\wndf \bY \\
  &= \plim_{n \to \infty} (\bA^\top \bM^\wndf \bA)^{-1} \bA^\top \bM^\wndf (\bA \bbeta + \bZ \bgamma + \bepsilon) \\
  &= \bbeta + \plim_{n \to \infty} (\bA^\top \bM^\wndf \bA)^{-1} \bA^\top \bM^\wndf \bZ \bgamma.
\end{align*}
We will proceed by first examining $\plim_{n \to \infty} \frac{1}{n} \bA^\top
\bM^\wndf \bA$ and then $\plim_{n \to \infty} \frac{1}{n} \bA^\top \bM^\wndf
\bZ$.
\begin{align*}
  \plim_{n \to \infty} \frac{1}{n} \bA^\top \bM^\wndf \bA
  &=
  \plim_{n \to \infty} \frac{1}{n} \bA^\top \left\{
  \bI - (\hat\bZ + \bS) \left[ (\hat\bZ +
  \bS)^\top (\hat\bZ + \bS) \right]^{-1} (\hat\bZ +
  \bS)^\top
  \right\} \bA
  \\
  &= \plim_{n \to \infty} \frac{1}{n} \bA^\top \left[
    \bI -
    \frac{1}{n (1 + \psi^2 )} (\hat\bZ + \bS)
    (\hat\bZ + \bS)^\top
    \right] \bA
  \\
   &= \plim_{n \to \infty} \frac{1}{n} \bA^\top \bA -
   \frac{1}{1 + \psi^2} \left(
     \frac{1}{n} \bA^\top
     \hat\bZ
     \right)
   \left(
   \frac{1}{n} \hat\bZ^\top
   \bA
   \right)
   \\
   &= \plim_{n \to \infty} \frac{1}{n} \bA^\top \bA -
   \frac{1}{1 + \psi^2}
   \hat\btheta^\top
   \hat\btheta
   \\
   &= \frac{\psi^2}{1 + \psi^2} \btheta^\top \btheta +
   \sigma^2 \bI - \frac{\sigma^2}{1 + \psi^2} \btheta^\top \left(\btheta \btheta^\top\right)^{-1} \btheta 
   \quad\text{by Lemma~\ref{l:residual_dependence}}
   \\
   &= \frac{\psi^2 }{1 + \psi^2} \btheta^\top
   \left[\bI - \frac{\sigma^2}{\psi^2} ( \btheta \btheta^\top )^{-1} \right]
   \btheta +
   \sigma^2 \bI
   \\
   \plim_{n \to \infty} \frac{1}{n} \bA^\top \bM^{\wndf\ast} \bZ
   &= \plim_{n \to \infty} \frac{1}{n} \bA^\top \left\{
   \bI -
   (\hat\bZ + \bS)
   \left[ (\hat\bZ + \bS)^\top (\hat\bZ + \bS) \right]^{-1}
   (\hat\bZ + \bS)^\top
   \right\} \bZ
   \\
   &= \plim_{n \to \infty} \btheta^\top -
   \frac{1}{1 + \psi^2} \left( \frac{1}{n} \bA^\top \hat\bZ \right)
   \left( \frac{1}{n} \hat\bZ^\top \bZ \right)
   \\
   &= \plim_{n \to \infty} \btheta^\top -
   \frac{1}{1 + \psi^2} \btheta^\top
   (\hat\btheta \hat\btheta^\top)^{-1} \hat\btheta \left[
     \frac{1}{n} \bA^\top (\bA - \bnu)
     \right] \btheta^\top (\btheta \btheta^\top)^{-1} 
   \\
   &= \plim_{n \to \infty} \btheta^\top -
   \frac{1}{1 + \psi^2} \btheta^\top
   (\hat\btheta \hat\btheta^\top)^{-1} \hat\btheta 
   \btheta^\top \btheta \btheta^\top (\btheta \btheta^\top)^{-1} 
   \\
\intertext{By Lemma~\ref{l:theta_hat},}
   &= \plim_{n \to \infty} \btheta^\top -
   \frac{1}{1 + \psi^2} \btheta^\top \bR \bLambda_{1:k}^{-\frac{1}{2}} \left( \bLambda_{1:k} + \sigma^2 \bI \right)^{\frac{1}{2}}
    (\hat\btheta \hat\btheta^\top)^{-1}
    \nonumber \\
    & \qquad \qquad \qquad \qquad \qquad \cdot \hat\btheta \hat\btheta^\top 
   \left( \bLambda_{1:k} + \sigma^2 \bI \right)^{-\frac{1}{2}} \bLambda_{1:k}^{\frac{1}{2}} \bR^\top
   \\
   &=
   \frac{\psi^2}{1 + \psi^2} \btheta^\top
   \\
   \plim_{n \to \infty} \hat\bbeta^\wndf - \bbeta
   &=
   \plim_{n \to \infty} (\bA^\top \bM^\wndf \bA)^{-1} \bA^\top \bM^\wndf \bZ \bgamma
   \\
   &=
   \left\{ \btheta^\top
   \left[\bI - \frac{\sigma^2}{\psi^2} ( \btheta \btheta^\top )^{-1} \right]
   \btheta +
   \frac{\sigma^2}{\psi^2} (1 + \psi^2) \bI
   \right\}^{-1}
   \btheta^\top \bgamma \quad\text{and}
   \\
   \lim_{m \to \infty} \plim_{n \to \infty} \hat\bbeta^\wndf - \bbeta
   &=
   \left[ \btheta^\top \btheta +
     \frac{\sigma^2}{\psi^2} (1 + \psi^2) \bI
     \right]^{-1}
   \btheta^\top \bgamma \qed
\end{align*}

\subsection{Bias of the Posterior-mean Deconfounder Estimator}
\label{a:bias_noise_posterior}

For convenience, we reiterate the data-generating process and posterior-mean
deconfounder estimation procedure here. As before, we consider $n$ observations
drawn i.i.d. as follows.
\begin{align*}
  \underset{n \times k}{\bZ} &\sim \cN(\bm{0}, \bI) \\
  \underset{n \times m}{\bnu} &\sim \cN(\bm{0}, \sigma^2 \bI) \\
  \underset{n \times m}{\bA} &= \bZ \btheta + \bnu_F \\
  \underset{n \times 1}{\bepsilon} &\sim \cN(\bm{0}, \omega^2) \\
  \underset{n \times 1}{\bY} &= \bA \bbeta + \bZ \bgamma + \bepsilon
\end{align*}

The posterior-mean deconfounder estimator is an approximate Bayesian procedure,
in the sense that it obtains an estimate for the effects $\bbeta$ by integrating
over an approximation to the full joint posterior, $f(\bbeta, \bgamma,
\bz | \bY, \bA)$, as follows.

First, the full posterior is factorized as $f(\bbeta, \bgamma | \bY,
\bA, \bz) f(\bz | \bY, \bA)$. Then, $f(\bz | \bA)$ is obtained by a
Bayesian principal components analysis of $\bA$ alone---i.e., ignoring
information from $\bY$---and used as an approximation to $f(\bz | \bY,
\bA)$. A Bayesian linear regression of $\bY$ on $\bA$ and $\bz$ is used to
obtain the conditional posterior $f(\bbeta, \bgamma | \bY, \bA,
\bz)$, and finally $\bz$ is integrated out. The posterior-mean
deconfounder estimator is thus
\begin{align*}
  \left[ \hat\bbeta^{\pndf \top}, \hat\bgamma^{\pndf \top} \right]^\top
  &\equiv
  \int f(\bz | \bA) \ 
  \E\left\{
       \left[ \bbeta^\top, \bgamma^\top \right]^\top \big| \ 
       \bY, \bA, \bz
       \right\}
       \dd{\bz}.
\end{align*}
We leave priors for all parameters unspecified; by the Bernstein-von Mises
theorem, our results are invariant to the choice of any prior with positive
density on the true parameters.

We now restate Proposition~\ref{prop:bias_noise_posterior} before proceeding to the proof.

\begin{mdframed}
\textbf{Proposition \ref{prop:bias_noise_posterior}.}
\it (Asymptotic Bias of the Posterior-Mean Deconfounder.) 

Consider $n$ observations drawn from a data-generating process with $k$
    unobserved confounders, $\bZ \sim \cN(\bm{0}, \bI)$; $m \ge k$ observed
    treatments, $\bA \sim \cN(\bZ \btheta, \sigma^2 \bI)$; and outcome $\bY \sim
    \cN(\bA \bbeta + \bZ \bgamma, \omega^2)$, following WB. The posterior-mean
    deconfounder estimator is
  \begin{align*}
    \left[\begin{array}{l}
    \hat\bbeta^\pndf \\ \hat\bgamma^\pndf
    \end{array}\right]
    &\equiv
    \int
    \left(
    \left[ \bA, \bz \right]^\top \left[ \bA, \bz \right]
    \right)^{-1}
    \left[ \bA, \bz \right] \bY \ 
    f(\bz | \bA) \dd{\bz}
    ,
  \end{align*}
  where $f(\bz | \bA)$ is a posterior obtained from Bayesian principal component
  analysis.\footnote{While the regression cannot be estimated when $\bz = \E[\bZ
      | \bA]$, it is almost surely estimable for samples $\bz^\ast \sim f(\bz |
    \bA)$ due to posterior uncertainty, which eliminates perfect collinearity
    with $\bA$. The posterior-mean implementation of WB evaluates the integral by Monte Carlo methods and thus is able to compute the regression coefficients for each sample.} The asymptotic bias of this
  estimator is given by
  \begin{align*}
    \plim_{n \to \infty} \hat\bbeta^\pndf - \bbeta &=
 (\btheta^\top \btheta + \sigma^2 \bI)^{-1} \btheta^\top \bgamma,
 \intertext{and under strong infinite confounding,}
     \lim_{m \to \infty} \plim_{n \to \infty} \hat\bbeta^\pndf - \bbeta &=
    \bm{0}
  \end{align*}
\end{mdframed}

\noindent\textit{Proof of Proposition~\ref{prop:bias_noise_posterior}.}

Under the Bayesian principal components generative model,
\begin{align*}
  \left[ \begin{array}{l}
      \bZ_i \\ \bA_i
    \end{array} \right]
  &\sim \cN\left(
  \bm{0},
  \left[ \begin{array}{ll}
      \bI, & \btheta \\
      \btheta^\top, & \btheta^\top \btheta + \sigma^2 \bI
    \end{array} \right]
  \right)\\
  \intertext{and by properties of the multivariate normal,}
  f(\bz_i | \bA_i, \btheta, \sigma^2) 
  &= \phi\left( \bz_i ; ~
  \btheta (\btheta^\top \btheta + \sigma^2 \bI)^{-1} \bA_i, ~
  \bI - \btheta (\btheta^\top \btheta + \sigma^2 \bI)^{-1} \btheta^\top
  \right).
\end{align*}

We will decompose the conditional posterior over confounders as $f(\bz | \bA) = \\
f(\bz | \btheta, \sigma^2, \bA) f(\btheta, \sigma^2 | \bA)$. A sample $\bz_i^\ast$
can be drawn from the Bayesian principal component posterior by first sampling
$\btheta^\ast$ and $\sigma^{\ast 2}$ from $f(\btheta, \sigma^2 | \bA)$,
deterministically constructing $\hat\bZ_i^\ast \equiv \E[\bZ_i | \btheta^\ast,
  \sigma^{\ast 2}, \bA_i] = \btheta^\ast (\btheta^{\ast \top} \btheta^\ast +
\sigma^{\ast 2} \bI)^{-1} \bA_i$, sampling $\bs_i^\ast$ from $f(\bs_i |
\btheta^\ast, \sigma^{\ast 2}) = \phi\left( \bs_i ; ~ \bm{0}, ~ \bI -
\btheta^\ast (\btheta^{\ast \top} \btheta^\ast + \sigma^{\ast 2} \bI)^{-1}
\btheta^{\ast \top} \right) $, and deterministically taking $\bz_i^\ast =
\hat\bZ_i^\ast + \bs_i^\ast$.

We can then rewrite
\begin{align*}
  \left[ \hat\bbeta^{\pndf \top}, \hat\bgamma^{\pndf \top} \right]^\top
  &=
  \int f(\btheta^\ast, \sigma^{\ast 2}, \bs^\ast | \bA) \ 
  \E\left\{
  \left[ \bbeta^\top, \bgamma^\top \right]^\top \big| \ 
  \bY, \bA, \btheta^\ast, \sigma^{\ast 2}, \bs^\ast
  \right\}
  \dd{\btheta^\ast}
  \dd{\sigma^{\ast 2}}
  \dd{\bs^\ast}
\end{align*}
where
\begin{align}
  &\E\left\{
  \left[ \bbeta^{\ast \top}, \bgamma^{\ast \top} \right]^\top \big| \ 
  \bY, \bA, \btheta^\ast, \sigma^{\ast 2}, \bs^\ast
  \right\} \nonumber \\
  &\qquad= \left(
  \left[ \bA,\ (\hat\bZ^\ast + \bs^\ast) \right]^\top
  \left[ \bA,\ (\hat\bZ^\ast + \bs^\ast) \right]
  \right)^{-1}
  \left[ \bA,\ (\hat\bZ^\ast + \bs^\ast) \right]^\top
  \bY.
  \label{e:noise_posterior_deconfounder_conditional_estimate}
\end{align}

Note that the posterior $f(\btheta, \sigma^2 | \bA)$ concentrates on true
$\sigma^2$ and $\btheta$ (up to a rotation).
Thus, candidate $\btheta^\ast$ and $\sigma^{\ast 2}$ values that fail to satisfy
$\btheta^{\ast \top} \btheta^\ast = \btheta^\top \btheta$ and $\sigma^{\ast 2} =
\sigma^2$ grow vanishingly unlikely as $n$ grows large. We examine the
asymptotic behavior of the conditional estimator,
\eqref{e:noise_posterior_deconfounder_conditional_estimate}, in this region and
show that the bias is constant. Thus, the estimator remains asymptotically
biased after integrating over all possible rotations of $\btheta^\ast$.

After subsetting \eqref{e:noise_posterior_deconfounder_conditional_estimate} to
the treatment effects, the conditional estimator can be rewritten as $
\hat\bbeta^\ast \equiv \E\left[ \bbeta^\ast | \ \bY, \bA, \btheta^\ast,
  \sigma^{\ast 2}, \bs^\ast \right] = (\bA^\top \bM^{\pndf\ast} \bA)^{-1}
\bA^\top \bM^{\pndf\ast} \bY $, where $\bM^{\pndf\ast}$ denotes the conditional
annihilator, $\bI - (\hat\bZ^\ast + \bs^\ast) \left[ (\hat\bZ^\ast +
  \bs^\ast)^\top (\hat\bZ^\ast + \bs^\ast) \right]^{-1} (\hat\bZ^\ast +
\bs^\ast)^\top$. Note that
\begin{align*}
  \plim_{n \to \infty} \hat\bbeta^\ast
  &= \plim_{n \to \infty} (\bA^\top \bM^{\pndf\ast} \bA)^{-1} \bA^\top \bM^{\pndf\ast} \bY \\
  &= \plim_{n \to \infty} (\bA^\top \bM^{\pndf\ast} \bA)^{-1} \bA^\top \bM^{\pndf\ast} (\bA \bbeta + \bZ \bgamma + \bepsilon) \\
  &= \bbeta + \plim_{n \to \infty} (\bA^\top \bM^{\pndf\ast} \bA)^{-1} \bA^\top \bM^{\pndf\ast} \bZ \bgamma .
\end{align*}
for any $\btheta^\ast$ and $\sigma^{\ast 2}$. We will proceed by first examining
$\plim_{n \to \infty} \frac{1}{n} \bA^\top \bM^{\pndf\ast} \bA$ and then $\plim_{n
  \to \infty} \frac{1}{n} \bA^\top \bM^{\pndf\ast} \bZ$. %
\begin{align*}
  \plim_{n \to \infty} \frac{1}{n} \bA^\top \bM^{\pndf\ast} \bA
  &=
  \plim_{n \to \infty} \frac{1}{n} \bA^\top \left\{
  \bI - (\hat\bZ^\ast + \bs^\ast) \left[ (\hat\bZ^\ast +
  \bs^\ast)^\top (\hat\bZ^\ast + \bs^\ast) \right]^{-1} (\hat\bZ^\ast +
  \bs^\ast)^\top
  \right\} \bA
  \\
 &= \plim_{n \to \infty} \frac{1}{n} \bA^\top \left[
  \bI -
  \frac{1}{n} (\hat\bZ^\ast + \bs^\ast)
  (\hat\bZ^\ast + \bs^\ast)^\top
  \right] \bA
  \\
  &= \plim_{n \to \infty} \frac{1}{n} \bA^\top \bA -
  \left(
    \frac{1}{n} \bA^\top
    \hat\bZ^\ast
    \right)
  \left(
  \frac{1}{n} \hat\bZ^{\ast \top}
  \bA
  \right)
  \\
  &= \plim_{n \to \infty} \frac{1}{n} \bA^\top \bA -
  (\btheta^\top \btheta + \sigma^2 \bI)
  (\btheta^{\ast \top} \btheta^\ast + \sigma^{\ast 2} \bI)^{-1}
  \btheta^{\ast \top}
  \btheta^\ast
  \\
  & \qquad \qquad \qquad \qquad \cdot (\btheta^{\ast \top} \btheta^\ast + \sigma^{\ast 2} \bI)^{-1}
  (\btheta^\top \btheta + \sigma^2 \bI)
  \\
  &= \btheta^\top \btheta + \sigma^2 \bI - \btheta^{\ast\top} \btheta^\ast
  \\
  &= \sigma^2 \bI
  \\
  \plim_{n \to \infty} \frac{1}{n} \bA^\top \bM^{\pndf\ast} \bZ
  &= \plim_{n \to \infty} \frac{1}{n} \bA^\top \left\{
  \bI -
  (\hat\bZ^\ast + \bs^\ast)
  \left[ (\hat\bZ^\ast + \bs^\ast)^\top (\hat\bZ^\ast + \bs^\ast) \right]^{-1}
  (\hat\bZ^\ast + \bs^\ast)^\top
  \right\} \bZ
  \\
  &= \plim_{n \to \infty} \frac{1}{n} ( \btheta^\top \bZ^\top + \bnu^\top ) \bZ -
  (\btheta^\top \btheta + \sigma^2 \bI)
  (\btheta^{\ast \top} \btheta^\ast + \sigma^{\ast 2} \bI)^{-1}
  \btheta^{\ast \top} 
  \left( \frac{1}{n} \hat\bZ^{\ast\top} \bZ \right)
  \\
  &= \plim_{n \to \infty} \btheta^\top -
  \btheta^{\ast \top} 
  \left( \frac{1}{n} \btheta^\ast
  (\btheta^{\ast \top} \btheta^\ast + \sigma^{\ast 2} \bI)^{-1}
  \bA^\top \bZ \right)
  \\
  &= \sigma^2
  (\btheta^\top \btheta + \sigma^2 \bI)^{-1} \btheta^\top
\end{align*}
Note that both expressions depend only on $\btheta^{\ast \top} \btheta^\ast$,
not $\btheta^\ast$ alone. Thus, the bias is constant over the entire asymptotic
posterior (i.e., all rotations) of $\btheta^\ast$.
\begin{align*}
  \plim_{n \to \infty} \hat\bbeta^\pndf - \bbeta
  &=
  \plim_{n \to \infty} \int f(\btheta^\ast, \sigma^{\ast 2}, \bs^\ast | \bA) \
  \E\left[ \bbeta^\ast - \bbeta | \ \bY, \bA, \btheta^\ast,
  \sigma^{\ast 2}, \bs^\ast \right]
  \dd{\btheta^\ast}
  \dd{\sigma^{\ast 2}}
  \dd{\bs^\ast}
  \\
  &=
  (\btheta^\top \btheta + \sigma^2 \bI)^{-1} \btheta^\top \bgamma
  \intertext{and under strong infinite confounding,}
  \lim_{m \to \infty} \plim_{n \to \infty} \hat\bbeta^\pndf - \bbeta
  &= \bm{0}
  \qed
\end{align*}

\subsection{Bias of the Subset Deconfounder}
\label{a:bias_subset}

For convenience, we reiterate the data-generating process and subset
deconfounder estimation procedure here. We will suppose, without loss of
generality, that the $m$ causes, $\bA$, are divided into $m_F$ focal causes of
interest, the column subset and $m_N$ nonfocal causes, $\bA_N$. As before, we
consider $n$ observations drawn i.i.d. as follows.
\begin{align*}
  \underset{n \times k}{\bZ} &\sim \cN(\bm{0}, \bI) \\
  \underset{n \times m_F}{\bnu_F} &\sim \cN(\bm{0}, \sigma^2 \bI) \\
  \underset{n \times m_N}{\bnu_N} &\sim \cN(\bm{0}, \sigma^2 \bI) \\
  \underset{n \times m_F}{\bA_F} &= \bZ \btheta_F + \bnu_F \\
  \underset{n \times m_N}{\bA_N} &= \bZ \btheta_N + \bnu_N \\
  \underset{n \times 1}{\bepsilon} &\sim \cN(\bm{0}, \omega^2) \\
  \underset{n \times 1}{\bY} &= \bA_F \bbeta_F + \bA_N \bbeta_N + \bZ \bgamma + \bepsilon
\end{align*}

The subset deconfounder estimator (1) takes the singular value decomposition
$\bA = \bU \bD \bV^\top$; (2) extracts the first $k$ components, $\hat\bZ \equiv
\sqrt{n} \bU_{1:k}$ and accompanying $\hat\btheta \equiv \frac{1}{\sqrt{n}}
\bD_{1:k} \bV_{1:k}^\top$; and (3) estimates the focal effects by computing
\begin{align*}
  \left[\begin{array}{l}
    \hat\bbeta_F^\sdf \\
    \hat\bgamma^\sdf
  \end{array} \right]
&\equiv
\left(
\left[ \bA_F, \hat\bZ \right]^\top \left[ \bA_F, \hat\bZ \right]
\right)^{-1}
\left[ \bA_F, \hat\bZ \right]^\top \bY
\end{align*}
and discarding $\hat\bgamma^\sdf$.

We now restate Proposition~\ref{prop:bias_subset} before proceeding to the proof.
\begin{mdframed}
\textbf{Proposition~\ref{prop:bias_subset}.}
(Asymptotic Bias of the Subset Deconfounder.)  
 
  The subset deconfounder
  estimator, based on Theorem 7 from WB, is
  \begin{align}
    \left[\begin{array}{l}
        \hat\bbeta_F^\sdf \\ \hat\bgamma^\sdf
      \end{array}\right]
    &\equiv
    \left(
    \left[ \bA_F, \hat\bZ \right]^\top \left[ \bA_F, \hat\bZ \right]
    \right)^{-1}
    \left[ \bA_F, \hat\bZ \right]^\top \bY.
  \end{align}
  where the column subsets $\bA_F$ and $\bA_N$ respectively partition $\bA$ into a finite number of focal causes of interest and non-focal causes.
  The substitute confounder, $\hat\bZ$, is obtained by taking the singular value decomposition $\bA =
  \bU \bD \bV^\top$ and extracting the first $k$ components, $\hat\bZ \equiv
  \sqrt{n} \bU_{1:k}$. Under the linear-linear model, the asymptotic bias of this estimator is given by
  \begin{align*}
    \plim_{n \to \infty} \hat\bbeta_F^\sdf - \bbeta_F 
    &=
    \left(
    \bI - \btheta_F^\top (\btheta \btheta^\top)^{-1} \btheta_F
    \right)^{-1}
    \btheta_F^\top (\btheta \btheta^\top)^{-1} \btheta_N \bbeta_N, \\
    \intertext{
  with $\btheta_F$ and $\btheta_N$ indicating the column subsets of $\btheta$
  corresponding to $\bA_F$ and $\bA_N$, respectively. The subset deconfounder is unbiased for $\bbeta_F$ (i) if $\btheta_F = \bm{0}$, (ii) if $\lim_{m \rightarrow \infty} \btheta_N \bbeta_N = \bm{0}$ and $\lim_{m \rightarrow \infty} \left[
  \bI - \btheta_F^\top (\btheta \btheta^\top)^{-1} \btheta_F
  \right]^{-1}$ is convergent or (iii) if both strong infinite confounding holds and $(\btheta \btheta^\top)^{-1} \btheta_{N} \bbeta_{N}$ goes to $\bm{0}$, as $m \rightarrow \infty$.  If one of these additional conditions hold, 
}
  \lim_{m \to \infty} \plim_{n \to \infty} \hat\bbeta_F^\sdf - \bbeta_F 
    &= \bm{0}
  \end{align*}
\end{mdframed}

\textit{Proof of Proposition~\ref{prop:bias_subset}.}
By the Frisch-Waugh-Lovell theorem, $\hat\bbeta_F^\sdf$ can be re-expressed in
terms of the portion of $\bA_F$ not explained by $\hat\bZ$. We denote the
residualized focal treatments as $\tilde\bA_F^\sdf = \bA_F - \hat\bA_F^\sdf =
\bA_F^\sdf - \hat\bZ \hat\btheta_F = \bU_{(k+1):m} \bD_{(k+1):m}
\bV_{(k+1):m}^\top$. The subset deconfounder estimator is then rewritten as
follows:
\begin{align*}
  \hat{\bbeta}_F^\sdf
  &= \left( \frac{1}{n} \tilde\bA_F^{\sdf \top} \tilde\bA_F^\sdf \right)^{-1} \frac{1}{n} \tilde\bA_F^\sdf \bY
\end{align*}
We now characterize the asymptotic bias of this estimator by examining the
behavior of $\frac{1}{n} \tilde\bA_F^{\sdf \top} \tilde\bA_F^\sdf$ and
$\frac{1}{n} \tilde\bA_F^\sdf \bY$ in turn. Beginning with the residual variance
of the focal causes,
\begin{align}
  \frac{1}{n} \tilde\bA_F^{\sdf \top} \tilde\bA_F^\sdf 
  &= \frac{1}{n}
  \left( \bA_F^\top - \hat\bA_F^{\sdf \top} \right) \left( \bA_F - \hat\bA_F^\sdf \right) \nonumber \\ 
  &= \frac{1}{n} \left(
  \bA_F^\top \bA_F
  + \hat\bA_F^{\sdf \top} \hat\bA_F^\sdf
  - \bA_F^\top \hat\bA_F^\sdf
  - \hat\bA_F^{\sdf \top} \bA_F  \right) \nonumber \\ 
  &= 
  \frac{1}{n} \left( \btheta_F^\top \bZ^\top + \bnu_F^\top \right)
  \left( \bZ \btheta_F + \bnu_F \right)
  + \frac{1}{n} \hat\btheta_F^\top \hat\bZ^\top \hat\bZ \hat\btheta_F \nonumber \\ 
  &\qquad
  - \frac{1}{n} \left( \bV_{1:k,F} \bD_{1:k} \bU_{1:k}^\top
  + \bV_{(k+1):m,F} \bD_{(k+1):m} \bU_{(k+1):m}^\top \right)
  \bU_{1:k} \bD_{1:k} \bV_{1:k,F}^\top \nonumber \\ 
  &\qquad
  - \frac{1}{n} \bV_{1:k,F} \bD_{1:k} \bU_{1:k}^\top
  \left( \bU_{1:k} \bD_{1:k} \bV_{1:k,F}^\top
  + \bU_{(k+1):m} \bD_{(k+1):m} \bV_{(k+1):m,F}^\top \right) \nonumber \\ 
  &= 
  \frac{1}{n} \left( \btheta_F^\top \bZ^\top + \bnu_F^\top \right)
  \left( \bZ \btheta_F + \bnu_F \right) 
  + \hat\btheta_F^\top \hat\btheta_F 
  - \frac{2}{n} \bV_{1:k,F} \bD_{1:k}^2 \bV_{1:k,F}^\top
  \end{align}
  Taking limits,
  \begin{align}
  \plim_{n \to \infty} \frac{1}{n} \tilde\bA_F^{\sdf \top} \tilde\bA_F^\sdf
  &=
  \btheta_F^\top \btheta_F + \sigma^2 \bI - \hat\btheta_F^\top \hat\btheta_F \nonumber \\
  &= \sigma^2 \bI
  - \sigma^2 \btheta_F^\top (\btheta \btheta^\top)^{-1} \btheta_F \quad \text{by Lemma~\ref{l:residual_dependence}, and}
  \label{e:subset_deconfounder_denominator_large_n_small_m} \\
  \lim_{m \to \infty} \plim_{n \to \infty} \frac{1}{n} \tilde\bA_F^{\sdf \top} \tilde\bA_F^\sdf
  &= \sigma^2 \bI \quad \text{under strong infinite confounding.}
  \label{e:subset_deconfounder_denominator_large_n_large_m} 
\end{align}
Turning to the residual covariance between the focal causes and the outcome,
\begin{align}
  \frac{1}{n} \tilde\bA_F^{\sdf \top} \bY 
  &= \frac{1}{n} 
  \tilde\bA_F^{\sdf \top} \left(
  \bA_F \bbeta_F + \bA_N \bbeta_N  + \bZ \bgamma + \bepsilon
  \right) \nonumber \\
  &= \frac{1}{n} \left[
    \left(\bA_F^\top - \hat\bA_F^{\sdf \top}\right) \bA_F \bbeta_F
    + \tilde\bA_F^{\sdf \top} \bA_N \bbeta_N
    + \tilde\bA_F^{\sdf \top} \bZ \bgamma
    + \tilde\bA_F^{\sdf \top} \bepsilon
    \right] \nonumber \\
  &= \frac{1}{n}
  (\bA_F^\top - \hat\bA_F^{\sdf \top}) \bA_F \bbeta_F %
  + \frac{1}{n}
  \bV_{(k+1):m,F} \bD_{(k+1):m} \bU_{(k+1):m}^\top
  \bU \bD \bV_{N}^\top \bbeta_N \nonumber \\
  &\qquad
  + \frac{1}{\sqrt{n}}
    \bV_{(k+1):m,F} \bD_{(k+1):m} \bU_{(k+1):m}^\top
      \bU_{1:k} \bgamma%
  + \frac{1}{n}
  \tilde\bA_F^{\sdf \top} \bZ \bgamma 
  + \frac{1}{n}
  \tilde\bA_F^{\sdf \top} \bepsilon \nonumber \\
    &= \frac{1}{n}
  (\bA_F^\top - \hat\bA_F^{\sdf \top}) \bA_F \bbeta_F %
  + \frac{1}{n}
  \bV_{(k+1):m,F} \bD_{(k+1):m}^2 \bV_{(k+1):m, N}^\top \bbeta_N \nonumber \\
  &\qquad
  + \frac{1}{n}
  ( \btheta_F^\top \bZ^\top + \bnu_F^\top - \hat\btheta_F^\top \hat\bZ^\top) \bZ \bgamma
  + \frac{1}{n}
  \tilde\bA_F^{\sdf \top} \bepsilon \nonumber 
  \end{align}
  We will proceed by reducing $(\bA_F^\top - \hat\bA_F^{\sdf \top}) \bA_F$ as above, substituting 
  \begin{align*}
\hat\bZ^\top \bZ = \big[ (\hat\btheta \hat\btheta^\top)^{-1} \hat\btheta \bA^\top \big] \big[ (\bA - \bnu) \btheta^\top (\btheta \btheta^\top)^{-1} \big]      
  \end{align*} and invoking Lemmas~\ref{l:theta_hat}~and~\ref{l:residual_dependence}.

  \begin{align}
  \plim_{n \to \infty} \frac{1}{n} \tilde\bA_F^{\sdf \top} \bY 
  &= 
  \sigma^2 \left[
    \bI - \btheta_F^\top (\btheta \btheta^\top)^{-1} \btheta_F
    \right] \bbeta_F
  - \sigma^2 \btheta_F^\top (\btheta \btheta^\top)^{-1} \btheta_N \bbeta_N
  \nonumber \\
  &\qquad
  + \btheta_F^\top \bgamma - \plim_{n \to \infty} \frac{1}{n} \hat\btheta_F^\top (\hat\btheta \hat\btheta^\top)^{-1} \hat\btheta \bA^\top (\bA - \bnu) \btheta^\top (\btheta \btheta^\top)^{-1} \bgamma
  \nonumber \\
  &= \sigma^2 \left[
    \bI - \btheta_F^\top (\btheta \btheta^\top)^{-1} \btheta_F
    \right] \bbeta_F
  - \sigma^2 \btheta_F^\top (\btheta \btheta^\top)^{-1} \btheta_N \bbeta_N
  \nonumber \\
  &\qquad
  + \btheta_F^\top \bgamma - \hat\btheta_F^\top (\hat\btheta \hat\btheta^\top)^{-1} \hat\btheta \left( \btheta^\top \btheta + \sigma^2 \bI - \sigma^2 \bI \right) \btheta^\top (\btheta \btheta^\top)^{-1} \bgamma
  \nonumber \\
  &= \sigma^2 \left[
    \bI - \btheta_F^\top (\btheta \btheta^\top)^{-1} \btheta_F
    \right] \bbeta_F
  - \sigma^2 \btheta_F^\top (\btheta \btheta^\top)^{-1} \btheta_N \bbeta_N
  \nonumber \\
  &\qquad
  + \btheta_F^\top \bgamma - \hat\btheta_F^\top (\hat\btheta \hat\btheta^\top)^{-1}
  \hat\btheta \btheta^\top \bgamma
  \nonumber \\
  &= \sigma^2 \left[
    \bI - \btheta_F^\top (\btheta \btheta^\top)^{-1} \btheta_F
    \right] \bbeta_F
  - \sigma^2 \btheta_F^\top (\btheta \btheta^\top)^{-1} \btheta_N \bbeta_N
  \nonumber \\
  &\qquad
  + \btheta_F^\top \bgamma - \hat\btheta_F^\top (\hat\btheta \hat\btheta^\top)^{-1}
  \hat\btheta \hat\btheta^\top
  \left( \bLambda_{1:k} + \sigma^2 \bI \right)^{-\frac{1}{2}}
  \bLambda_{1:k}^{\frac{1}{2}} \bR^\top
  \bgamma
  \nonumber \\
  &= \sigma^2 \left[
    \bI - \btheta_F^\top (\btheta \btheta^\top)^{-1} \btheta_F
    \right] \bbeta_F
  - \sigma^2 \btheta_F^\top (\btheta \btheta^\top)^{-1} \btheta_N \bbeta_N
  \nonumber \\
  &\qquad
  + \btheta_F^\top \bgamma
  - \btheta_F^\top \bR
  \left( \bLambda_{1:k} + \sigma^2 \bI \right)^{\frac{1}{2}} \bLambda_{1:k}^{-\frac{1}{2}}
  \left( \bLambda_{1:k} + \sigma^2 \bI \right)^{-\frac{1}{2}}
  \bLambda_{1:k}^{\frac{1}{2}} \bR^\top
  \bgamma
  \nonumber \\
  &= \sigma^2 \left[
    \bI - \btheta_F^\top (\btheta \btheta^\top)^{-1} \btheta_F
    \right] \bbeta_F
  - \sigma^2 \btheta_F^\top (\btheta \btheta^\top)^{-1} \btheta_N \bbeta_N 
  \label{e:subset_deconfounder_numerator_large_n_small_m} \\ 
  \intertext{and under strong infinite confounding,}
  \lim_{m \to \infty} \plim_{n \to \infty} \frac{1}{n} \tilde\bA_F^{\sdf \top} \bY
  &= \sigma^2 \bbeta_F .
  \label{e:subset_deconfounder_numerator_large_n_large_m}
\end{align}

Combining \eqref{e:subset_deconfounder_denominator_large_n_small_m}~and~\eqref{e:subset_deconfounder_numerator_large_n_small_m},
\begin{align*}
  \plim_{n \to \infty} \hat\bbeta_F^\sdf
  &= \bbeta_F +
  \left[
  \bI - \btheta_F^\top (\btheta \btheta^\top)^{-1} \btheta_F
  \right]^{-1}
  \btheta_F^\top (\btheta \btheta^\top)^{-1} \btheta_N \bbeta_N ,
  \intertext{Consider the additional conditions. (i) If $\btheta_{F} = \bm{0}$ then convergence is immediate.  (ii) If $\lim_{m \rightarrow \infty} \btheta_{N} \bbeta_{N}$ and $\lim_{m \rightarrow \infty} \left[
  \bI - \btheta_F^\top (\btheta \btheta^\top)^{-1} \btheta_F
  \right]^{-1}$ is convergent then $\lim_{m \rightarrow \infty} (\btheta \btheta^\top)^{-1} \btheta_N \bbeta_N = \bm{0}$ so the product of the limits is $\bm{0}$. (iii) If $\lim_{m \rightarrow \infty} \left(\btheta \btheta^{\top} \right)^{-1} \btheta_{N} \bbeta_{N} = \bm{0} $ and strong infinite confounding holds then $\lim_{m \rightarrow \infty} \left[
  \bI - \btheta_F^\top (\btheta \btheta^\top)^{-1} \btheta_F
  \right]^{-1} = \bI$  so the bias term also goes to zero.  Therefore, combining Equations \eqref{e:subset_deconfounder_denominator_large_n_large_m}~and~\eqref{e:subset_deconfounder_numerator_large_n_large_m} if one of these conditions holds yields}
  \lim_{m \to \infty} \plim_{n \to \infty} \hat\bbeta_F^\sdf 
  &= \bbeta_F .
\qed
\end{align*} 

Proposition \ref{prop:bias_naive_focal} demonstrates that the na\"ive regression estimator is an unbiased estimator for focal treatments; it is a generalization of Proposition~\ref{prop:bias_naive}.  The proof is given in Section~\ref{a:bias_naive}.

\subsection{Subset Deconfounder Requires Assumptions about Treatment Effects} \label{a:subset_add} 
To see why strong infinite confounding is insufficient, consider a simple example.  Using the linear-linear data-generating process, consider the case of $k = 1$. For the sake of this example, we will suppose that for each treatment $j$ that $\theta_{j} = \bar{\theta}$ and that $\beta_{j} = \bar{\beta}$.  This clearly satisfies strong infinite confounding, PCA will be a consistent estimator of the substitute confounder, and na\"ive regression will be a consistent estimator of the treatment effects. But this is not the case for the subset deconfounder.  Using Proposition 4, the bias for arbitrary focal treatment $j$ as $m \rightarrow \infty$ is given by:  
\begin{align}
\lim_{m \rightarrow \infty} \E[\hat\beta_j] - \beta_j 
&= \lim_{m \rightarrow \infty}  \left(1 - \frac{\bar\theta^2}{m \bar\theta^2}\right)^{-1} \frac{(m-1) \bar{\theta}^2 \bar{\beta}}{m \theta^2} \nonumber \\
&=  \bar{\beta} \nonumber 
\end{align} 
So long as $\bar{\beta} \neq 0$ then the bias is non-zero, regardless of how many treatments are present. The intuition is that as we add more treatments we are having two countervailing effects on our estimator.  Additional treatments are giving us a better estimate of the substitute confounder, which is reducing the correlation between the focal and non-focal treatments, which on its own would reduce bias. But at the same time, we're adding additional omitted variable bias.  That additional omitted variable bias causes the subset deconfounder's bias to not decrease as the treatments are added.

\paragraph{Subset Deconfounder as Regularization} Connections between the subset deconfounder and na\"ive regression can be seen through the following straightforward argument.  Consider first na\"ive regression. By the Frisch-Waugh-Lovell theorem, for any treatment $\bA_j$, its estimated effect is related to the residualized treatment, $\tilde{\bA}^\naive_j = \bA_j - \hat{\bA}^\naive_j $, where $\hat{\bA}^\naive_j  = \left( \bA_{\setminus j}^{\top} \bA_{\setminus j} \right)^{-1} \bA_{\setminus j}^{\top} \bA_j$ is the part of $\bA_j$ that can be predicted from the other treatments, $\bA_{\setminus j}$. Specifically, $\hat\beta_j^{\text{na\"ive}}  = \frac{\Cov(\bY, \tilde{\bA}^\naive_j) }{ \Var(\tilde{\bA}^\naive_j )}$. Denoting the SVD of $\bA_{\setminus j}$ as $\bm{U}_{\setminus j}\bm{D}_{\setminus j} \bm{V}_{\setminus j}^{\top}$, then  $\hat{\bA}^\naive_j  = \bm{U}_{\setminus j} \bm{U}_{\setminus j}^{\top} \bA_j $.  As $m , n\rightarrow \infty$ then under linear-linear confounding this approaches $\bm{U} \bm{U}^{\top} \bA_j$.

Now consider the subset deconfounder. Also by Frisch-Waugh-Lovell, for any single treatment $\bA_j$, its estimated effect is $\hat\beta_j^{\text{subset}}  = \frac{\Cov(\bY, \tilde{\bA}_j^{\text{subset}}) }{ \Var(\tilde{\bA}_j^{\text{subset}} )}$ and $\tilde{\bA}_j^{\text{subset}} = \bA_j^{\text{subset}} - \hat{\bA}_j^{\text{subset}} $. Denoting the SVD of $\bA$ as $\bA = \bm{U}\bm{D} \bm{V}^{\top}$, it can be seen that $\hat{\bA}_j^{\text{subset}} = \bm{U}_{1:k} \bm{U}_{1:k}^{\top} \bA_j$.  

This makes clear that the na\"ive regression is adjusting along the same eigenvectors as the subset deconfounder.  Further, it shows that the subset deconfounder is performing the same regression as the na\"ive regression but with a particular form of regularization.  Specifically, the deconfounder imposes no penalty on the first $k$ eigenvectors, but then regularizes by suppressing the influence of dimensions $k+1$ to $m$.

\subsection{Convergence of Deconfounder and Na\"ive Estimators}
\label{a:deconfounder_naive_equality}

In this section, we extend our previous results on asymptotic equivalence
between the deconfounder and na\"ive analyses to a broad class of
nonlinear-nonlinear data-generating processes. We consider factor models in
which the distributions of $\bZ_i$ and $\bA_i | \bZ_i$ have continuous
density. Following the deconfounder papers, we analyze the case in which pinpointedness holds and this requires strong infinite confounding and infinite $m$.

We also follow the deconfounder papers in restricting attention to outcome models with constant
treatment effects and finite variance. That is, we study outcome models
satisfying $\E[ Y_i | \bA, \bZ ] = \bA_i^\top \bbeta + g_Y(\bZ_i)$, allowing for
arbitrarily complex nonlinear confounding. This family is more restrictive than
the outcome models considered in
Section~\ref{a:deconfounder_inconsistency_nonlinear} but nevertheless nests all
data-generating processes studied in prior deconfounder papers and our paper.

For convenience, we restate Theorem~\ref{thm:deconfounder_naive_equality} here.
\begin{mdframed}
  \textbf{Theorem~\ref{thm:deconfounder_naive_equality}.}
  \it (Deconfounder-Na\"ive Convergence under Strong Infinite Confounding.)

Consider all data-generating processes in which (i) treatments are drawn from a
factor model with continuous density that is a function $\bZ$, (ii) $\bZ$ is pinpointed, and (iii) the
outcome model contains constant treatment effects and treatment assignment is ignorable nonparametrically given $\bZ$. Any consistent
deconfounder converges to a na\"ive estimator for any finite subset of
treatment effects.
\end{mdframed}

\noindent \textit{Proof.} We begin with some preliminaries. As in
Section~\ref{a:bias_naive}, we partition the $m$ causes, $\bA_i$, into finite
$m_F$ focal causes of interest, $\bA_{i,F}$, and $m_N$ nonfocal causes,
$\bA_{i,N}$. Then the conditional expectation function of the outcome can be
rewritten $\E[ Y_i | \bA, \bZ ] = \bA_{i,F}^\top \bbeta_F + \bA_{i,N}^\top
\bbeta_N + g_Y(\bZ_i)$. In what follows, we will also use the conditional
expectation function $g_{\bA}(\bz) \equiv \E[ \bA_i | \bZ_i=\bz ]$, as well as
its partitioned counterparts, $g_{\bA_F}(\bz) \equiv \E[ \bA_{i,F} | \bZ_i=\bz
]$ and $g_{\bA_N}(\bz) \equiv \E[ \bA_{i,N} | \bZ_i=\bz ]$.

Pinpointedness implies that $g_{\bA}(\bz)$ must be invertible and
consistently estimable; it also requires that both $n$ and $m$ go to infinity. When these
conditions hold, $\plim_{n \to \infty} \hat{g}_{\bA}^{-1}(\bA_i) =
g_{\bA}^{-1}(\bA_i) = \bZ_i$, up to symmetries such as rotation invariance. The
deconfounder estimator then reduces to the partially linear regression
\begin{align*}
  \left( \hat\bbeta_F^\deconf, \hat\bbeta_N^\deconf, \hat{g}_Y^\deconf \right)
  &= \argmin_{\bbeta_F^\ast, \bbeta_N^\ast, g_Y^\ast} \ \sum_{i=1}^n \left(
  Y_i - \bA_{i,F}^\top \bbeta_F^\ast - \bA_{i,N}^\top \bbeta_N^\ast - g_Y^\ast(\hat{g}_{\bA}^{-1}(\bA_i))
  \right)^2
\end{align*}
which is consistent for $\bbeta_F$ \citep{robinson1988}.

Now consider the following alternative estimator, the partially linear na\"ive regression
\begin{align}
  \hat{h}_{\bA_F}^\naive
  &= \argmin_{h_{\bA_F}^\ast} \sum_{i=1}^n \left\| \bA_{i,F} - h_{\bA_F}^\ast(\bA_{i,N}) \right\|_F^2
  \\
  \hat{h}_Y^\naive
  &= \argmin_{h_Y^\ast} \sum_{i=1}^n \left( Y_i - h_Y^\ast(\bA_{i,N}) \right)^2 
  \\
  \hat\bbeta_F^\naive
  &=
  \left( \tilde\bA_F^{\naive \top} \tilde\bA_F^\naive \right)^{-1}
  \tilde\bA_F^{\naive \top} \tilde\bY^\naive
\end{align}
where $\tilde\bA_F^\naive$ collects
$\bA_{i,F} - \hat{h}_{\bA_F}^\naive(\bA_{i,N})$
and $\tilde\bY^\naive$ collects
$Y_i - \hat{h}_Y^\naive(\bA_{i,N})$.  Like its linear-linear analogue,
\eqref{e:naive}, this generalized na\"ive estimator models the outcome in terms
of the treatments only, ignoring the existence of confounding. It can be seen
that \eqref{e:naive_nonlinear_af} and \eqref{e:naive_nonlinear_y} flexibly
estimate the conditional means of $\bA_{i,F}$ and $Y_i$, respectively, given
$\bA_{i,N}$.

We now note that because pinpointedness only holds with infinite $m$, it must
also hold with $m - m_F$ treatments. That is, if $g_{\bA}^{-1}(\bA_i) = \bZ_i$,
then also $g_{\bA_N}^{-1}(\bA_{i,N}) = \bZ_i$. Next, because $g_{\bA_F}(\cdot)$
by definition minimizes the conditional variance in the focal treatments,
$g_{\bA_F}( g_{\bA_N}^{-1}(\bA_{i,N}) )$ is asymptotically the minimizer of
\eqref{e:naive_nonlinear_af}. Similarly, it is easy to see that $\bA_{i,N}^\top
\bbeta_N + g_Y(g_{\bA_N}^{-1}(\bA_{i,N}))$ asymptotically solves
\eqref{e:naive_nonlinear_y}. As a result \eqref{e:naive_nonlinear} identifies
using $\bA_{i,F} - \E[\bA_{i,F} | \bZ_i]$, the component of the focal treatments
which is uncontaminated by confounding. Thus, $\hat\bbeta_F^\deconf$ and
$\hat\bbeta_F^\naive$ both converge in probability to $\bbeta_F$, and hence to
each other. \qed

\subsection{Inconsistency of the Deconfounder in Nonlinear Settings}
\label{a:deconfounder_inconsistency_nonlinear}

We now generalize our previous results to a broad class of nonlinear-nonlinear
data-generating processes. We consider all factor models in which the
distributions of $\bZ_i$ and $\bA_i | \bZ_i$ have continuous density. We
restrict our attention to the class of outcome models with additively separable
confounding, a family that nests all data-generating processes studied in our paper and in applications of the deconfounder. That is, we study outcome models satisfying $\E[ Y_i |
  \bA, \bZ ] = f(\bA_i) + g_Y(\bZ_i)$, allowing for arbitrarily complex
confounding and arbitrarily complex, interactive treatment effects.

For convenience, we restate Theorem~\ref{thm:deconfounder_inconsistency_nonlinear} here.

\begin{mdframed}
  \textbf{Theorem~\ref{thm:deconfounder_inconsistency_nonlinear}.}
  \it (Inconsistency of the Deconfounder in Nonlinear Settings.)

Consider all data-generating processes in which finite treatments are drawn from
a factor model with continuous density. The deconfounder is inconsistent for
any outcome model with additively separable confounding.

\end{mdframed}

This result follows from a relatively simple premise, building on the intuition behind Lemma~\ref{l:pinpointing}. Because there is always
noise in $\bA_i$, the analyst can never recover the exact value of the $\bZ_i$
when there are a finite number of treatments. The error in $\hat\bZ_i$ depends
on $\bA_i$, and the outcome $Y_i$ is in part dependent on this error (that is,
$Y_i$ is dependent on $\bZ_i$, which is only partially accounted for by
$\hat\bZ_i$). Therefore, an outcome analysis that neglects the unobserved
mismeasurement, $\hat\bZ_i - \bZ_i$, will necessarily be affected by omitted
variable bias.

We proceed by cases, focusing primarily on a highly implausible best-case
scenario. As we discuss below, alternate cases either (i) asymptotically almost
never occur or (ii) lead trivially to inconsistency of the deconfounder. Thus,
because the deconfounder is inconsistent even in this ideal setting, it is
inconsistent in all cases.

\noindent \textit{Proof.} First, consider the case in which $g_{\bA}(\bz) \equiv
\E[ \bA_i | \bZ_i=\bz]$ is invertible. (Invertibility of the conditional
expectation function is a necessary condition for pinpointedness; in the case
where it does not hold, $\bZ_i$ cannot be learned, and inconsistency follows
immediately.) In the invertible case, it is easy to see that the unobserved
confounder could be recovered with $ g_{\bA}^{-1}( \E[ \bA_i | \bZ_i ] ) =
\bZ_i$, if both $g_{\bA}(\cdot)$ and the conditional expectation of the
treatments (i.e., without noise) was known. However, the fundamental problem is
that the analyst only observes the treatments for unit $i$, $\bA_i$, including
noise---its conditional expectation given the latent confounder,
$\E[\bA_i | \bZ_i]$, is unknown because the true $\bZ_i$ is unknown.

Let $h_{\bA}(\bA_i, \bA_i - \E[ \bA_i | \bZ_i ])$ represent an unobservable
clean-up function that corrects the resulting error when $\bA_i$ is used instead
of $\E[ \bA_i | \bZ_i ]$, so that $ h_{\bA}(\bA_i, \bA_i - \E[ \bA_i | \bZ_i ])
+ g_{\bA}^{-1}( \bA_i ) \ = \ \bZ_i $. For concreteness, consider the
linear-linear setting as an example: here, $ g_{\bA}(\bZ_i) = \bZ_i^\top \btheta
$, $ g_{\bA}^{-1}( \bA_i ) = \bA_i^\top \btheta^\top (\btheta \btheta^\top
)^{-1}$, and $h_{\bA}(\bA_i, \bnu_i) = -\bnu_i^\top \btheta^\top (\btheta
\btheta^\top )^{-1}$.

We next assume that factor-model parameters can be consistently estimated,
setting aside likelihood invariance and other issues, so that $\plim_{n \to
  \infty} \hat{g}_{\bA}(\cdot) = g_{\bA}(\cdot)$. (In cases where these
conditions do not hold, $\bZ_i$ is again unidentified, and the deconfounder is
again inconsistent.)  The deconfounder procedure asks the analyst to compute
$\hat\bZ_i \equiv \hat{g}_{\bA}^{-1}( \bA_i )$. As $n$ grows large, the analyst
is left with the unobserved error $\plim_{n \to \infty} \hat\bZ_i - \bZ_i =
-h_{\bA}(\bA_i, \bA_i - \E[ \bA_i | \bZ_i ])$. In other words, consistency of
$\hat{g}_{\bA}$ does not imply consistency of $\hat\bZ_i$---even when factor
model parameters are known perfectly, noise in $\bA_i$ will almost surely
deceive the factor model about the true value of $\bZ_i$.

We now turn to the outcome model, $\E[Y_i | \bA_i, \bZ_i] = f(\bA_i) +
g_Y(\bZ_i)$. The deconfounder requires analysts to correct for confounding by
estimating $g_Y(\cdot)$. Here, we consider three cases. The first is the
knife-edge case in which errors in $\hat{g}_Y(\cdot)$ and $\hat\bZ_i$ exactly
offset one another, so that $\hat{g}_Y(\hat\bZ_i) = g_Y(\bZ_i)$. This occurs on a set of densities with measure zero, because as $n$ grows large the probability that all
noise-induced errors in $\hat\bZ$ can be rectified by a deterministic correction
goes to zero. The second is where the confounding function is not perfectly
recovered and errors do not cancel. Note that in practice, this scenario
dominates; the confounding function generally cannot be estimated reliably
because its inputs are only observed with measurement error. Here, deconfounder
inconsistency again follows trivially. The third is yet another unrealistic,
best-case scenario: the confounding function, $g_Y(\cdot)$, is perfectly
recovered.

Even in this third, ideal scenario, the deconfounder procedure introduces
further unobservable error when substituting $\hat{g}_Y(\hat\bZ_i) =
g_Y(\hat\bZ_i)$ instead of $g_Y(\bZ_i)$. We now define a second clean-up
function, $h_Y(\bZ_i, \hat\bZ_i - \bZ_i)$, that ensures $\E[Y_i | \bA_i, \bZ_i]
= f(\bA_i) + g_Y(\hat\bZ_i) + h_Y(\bZ_i, \hat\bZ_i - \bZ_i)$. As a concrete
example, in the linear-linear case, $g_Y(\hat\bZ_i) = \hat\bZ_i^\top \btheta$,
and $h_Y(\bZ_i, \hat\bZ_i - \bZ_i) = (\bZ_i - \hat\bZ_i)^\top \btheta$ corrects
for mismeasurement of $\bZ$.

Finally, the deconfounder asks analysts to compute a regression after adjustment
for the confounding function. Specifically, the deconfounder learns the
relationship between $Y_i$ and $f(\bA_i)$, seeking to correct for the estimated
confounding, $\hat{g}_Y(\hat\bZ_i)$. Even in this best-case
scenario, which grants $\hat{g}_Y(\cdot) = g_Y(\cdot)$, this
regression fails to account for the unobserved term $h_Y(\bZ_i, h_{\bA}(\bA_i,
\bA_i - \E[\bA_i | \bZ_i]))$, which is associated with both $\bA_i$ and
$Y_i$. Thus, omitted-variable bias arises everywhere in the model space except
when knife-edge conditions are satisfied. As a concrete example of such a
knife-edge condition, consider the partially linear constant-effects model $\E[
  Y_i | \bA, \bZ ] = \bA_i^\top \bbeta + g_Y(\bZ_i)$. Here, the deconfounder is
inconsistent except in the special cases where either $\mathrm{Cov}\left(Y_i,
h_Y(\bZ_i, h_{\bA}(\bA_i, \bA_i - \E[\bA_i | \bZ_i])) \right)$ or
$\mathrm{Cov}\left(\bA_i, h_Y(\bZ_i, h_{\bA}(\bA_i, \bA_i - \E[\bA_i | \bZ_i]))
\right)$ equal zero. \qed

\section{Simulation Details}
\label{app:sim}
Our simulation code is based on replication materials for \citet{WanBle19} that were provided to us by Wang and Blei in December of 2019 and two simulations that were publicly posted tutorials on the \texttt{blei-lab} github page from September 20, 2018 \citep{wang2019github} until its removal on March 22nd, 2020, after we downloaded and analyzed them. After reading our working paper and discussing our analyses with them, Wang, Zhang and Blei posted reference implementations on July 2nd, 2020 at \url{https://github.com/blei-lab/deconfounder_public} and \url{https://github.com/zhangly811/Medical_deconfounder_simulation}. Because the reference implementation was produced in response to our analyses, all references are with respect to the December 2019 code and the details provided in the published papers. 

This appendix provides additional details on the simulations and attempts to explain why our results deviate from the previously published results. In this introduction to the appendix, we define a few terms that we will use repeatedly.  We then detail common deviations between our simulations and the original designs and provide discussion of why our results differ. We then run through each simulation in turn.

\noindent \textbf{Common Terms:}
\begin{itemize}
\item Na\"ive Regression\\
When not otherwise specified the na\"ive regression is an OLS regression of the outcome on all the treatments.  When creating confidence intervals they are always $95\%$ intervals calculated using the classical covariance matrix estimated under homoskedasticity.  
\item Oracle Regression\\
This follows the same setup as the na\"ive regression but also controls for the unmeasured confounder.
\item PCA\\
 When we compute a principal components analysis we always center and scale the treatments.  We take the top $k$ principal components where $k$ is set to the true number of unobserved components.  This is obviously not feasible in real world settings, but is as generous as possible to the deconfounder.
\item pPCA \\
To compute probabilistic Principal Components Analysis we follow the \texttt{rstan} code provided to us by Wang and Blei for replicating the smoking example in \citet{WanBle19}.  We remove the computation of heldout samples for reasons that will be explained below. We estimate the model using automatic differentiation variational inference using their convergence settings. When unspecified we use the posterior mean as our estimate of $\hat{Z_i}$.
\item Deconfounder \\
When not otherwise specified we are using the substitute confounder version of the deconfounder not the reconstructed causes.
\end{itemize}

\subsection{Common Deviations in Simulation Setups}
\label{a:common_deviations}
We have tried to remain as faithful as possible to the original simulation setups, but we have made changes where necessary to assess the deconfounder's performance.  In each simulation and application, we detail all deviations in procedures---here we summarize the most common such deviations. 

\paragraph{Stabilizing Estimation}
There are two areas where estimation issues become a concern in the deconfounder: the estimation of the factor model and the estimation of the outcome model.

Instability in the factor model arises for a number of reasons.  In at least two of the simulations (Medical Deconfounder 2 and Smoking), the original design calls for simulating factor models which have more latent dimensions than observed data dimensions ($m$).  In these settings the extra factors are just noise as $n$ factors would be sufficient to exactly reconstruct the data under a frequentist model. In these settings, we report results for the models with $k > m$ but caution against overinterpreting the results.  See the smoking simulation for an examination of factor model instability in this setting.  In other settings (GWAS, Actors and Breast Cancer), the code uses a procedure that analyzes a subset of the data when fitting factor models and updates in batches.  In Breast Cancer and Actors we replace their procedure with code that analyzes the entire dataset at once. In many of the applications we replace their PPCA with standard PCA where possible to avoid the noise from the posterior approximation.

Many of the original simulations estimate the na\"ive and oracle regressions with bayesian linear regression with normal priors fit with black-box varitional estimation in \texttt{Stan}. We always use the computationally cheaper and more stable OLS which explains why in several simulations (e.g. Medical Deconfounder) the original papers report oracle coverage rates that are below the nominal levels. Instability in the deconfounder outcome model is particularly high because of near-perfect collinearity between the treatments and the substitute confounder.  This problem is in turn exacerbated by the black box variational estimation in \texttt{Stan}. This drives our development of the PCA+CV-Ridge deconfounder variant we deploy in the Medical Deconfounder 1 simulation. 

\paragraph{Probing the Simulations}
We extend many of the simulations to more thoroughly probe the properties of the estimator.  A number of the original simulations (Quadratic, Logistic, Medical Deconfounder 1) only report results from a single realization of the data generating process, while we repeat hundreds or thousands of times.  We also extend our search over reasonable parameter spaces of the original data generating process by examining different numbers of treatments and levels of confounding (Quadratic), different levels of noise added to the substitute confounder (Logistic) and different coefficients in the outcome model (Medical Deconfounder 1).  In other cases, we use the same simulation designs, but explore the results in different ways (e.g. breaking out GWAS results by causal and non-causal coefficients).

\paragraph{Removing the Heldout Procedure}
Many of the original simulations describe holding out $20\%$ of the data for predictive checks.  We generally assume the correct latent specification and are interested in comparing all models that are presented in the original paper so we skip this step.  See also our discussion in Supplement~\ref{app:smokingsim} for an explanation of why the heldout procedure implemented in the smoking example yields different results.

\subsection{Medical Deconfounder Simulations}
The simulations are both taken from Section 3 of \citet{Zha19} on the medical deconfounder.  They are used in conjunction with two case studies to establish the performance of the medical deconfounder. We recreate these simulations from details provided in the paper (reusing \texttt{Stan} code provided to us by Wang and Blei provided for the smoking example).

\subsubsection{Simulation Study 1: One Real Cause}
\label{a:med1}
\textbf{Summary:} We replicate and extend a simulation design designed to support the medical deconfounder from \citet{Zha19} which uses penalized regression to be estimable.

The true data generating process for each of 1000 patients indexed by $i$ with true scalar confounder $Z_i$ is as follows.
\begin{align*}
Z_i &\sim \text{Normal}(0,1) \\
A_{1,i} &\sim \text{Normal}(0.3Z_i,1) \\
A_{2,i} &\sim \text{Normal}(0.4Z_i,1) \\
Y_i &\sim \text{Normal}(0.5Z_i + 0.3A_{2,i},1)
\end{align*}
In the actual paper, the notation writes the error term as $\epsilon_i$ for that is shared across $A_1,A_2,Y$, but we take this to be a typo.

The original simulation fits a probabilistic PCA model with $k=1$ using black box variational inference \citep{ranganath2014black} in Edward and fit the outcome model using ADVI \citep{kucukelbir2017automatic}.  To keep the entire workflow in R, we fit the probabilistic PCA model with $k=1$ in \texttt{rstan} using ADVI based on the probabilistic PCA model provided to us for replicating the smoking example.  We modified the code to remove the heldout procedure designed for posterior predictive checks but otherwise kept the model with same, with the same priors.  We use the default settings in \texttt{rstanarm} for the outcome regression. We increased the maximum number of iterations on both the factor model and the outcome model by a factor of 10 to try to stabilize the estimates.

We introduce the PCA + CV-Ridge estimator.  We estimate PCA with $k=1$ and then estimate a ridge regression using \texttt{glmnet} \citep{glmnet} and choosing the penalty parameter by cross-validation. In the paper, \citet{Zha19} present one realization of this data-generating process and we repeat this process 1000 times to create a simulation.  Table 2 of \citet{Zha19} contains their results.  The results are not directly comparable in that they are showing a single realization of a set of estimators and are showing more systematic results.  

Finally, we show an additional simulation where the true treatment effects are $(-.3,.3)$ instead of $(0,.3)$ to demonstrate the results are sensitive to the settings of the true coefficients, consistent with our results about the subset deconfounder.

\begin{framed}
\noindent \textbf{Deviations}
\begin{itemize}
\item we take more than one draw from the simulation and report aggregate quantities
\item we correct a typo in the manuscript and do not share errors across treatments
\item we use \texttt{rstan} instead of \texttt{Edward} for pPCA
\item we increase the maximum iterations of the bayesian procedures
\item we add the PCA + CV-Ridge Estimator.
\item we assess in terms of bias, standard deviation and RMSE rather than with coverage or $p$-values.
\item we don't holdout $20\%$ of the data or do predictive checks
\item we use OLS rather than bayesian linear regression for our na\"ive and oracle estimators
\end{itemize}

\noindent \textbf{Their Results:}  They argue based on one draw from their simulated process that the deconfounder leads to the right conclusions and the na\"ive leads to the wrong conclusions.\\
\noindent \textbf{Our Results:} We demonstrate that even with the PCA + CV-Ridge Estimator which provides large improvements over the medical deconfounder, the performance of the deconfounder is in practice no better than the na\"ive regression.
\end{framed}

\subsubsection{Simulation Study 2: A multi-medication simulation}
\label{a:med2}
\textbf{Summary: }We replicate and extend a simulation design created to support the medical deconfounder from \citet{Zha19} which uses a nonlinear functional form to be estimable.

\citet{Zha19} simulate 50 total treatments of which only 10 have a non-zero effect on the outcome.  They use the data generating process,
\begin{align*}
Z_{i,k} &\sim \text{Normal}(0,1), &k=1\dots10\\
A_{i,j} &\sim \text{Bernoulli}\left(\sigma\left(\sum_{k=1}^{10}\lambda_{k,j}Z_{i,k} \right) \right), &j=1\dots50\\
Y_i &\sim \text{Normal}\left(\sum_{j=1}^{10} \beta_jA_{ij} + \sum_{k=1}^{10} \gamma_kZ_{i,k},1\right)\\
\lambda_{k,j} &\sim \text{Normal}(0,.5^2) \\
\gamma_k,\beta_j &\sim \text{Normal}(0,.25^2)
\end{align*}
where $\sigma$ is the sigmoid function.  Only the first 10 treatments have non-zero coefficients---these are the medications that work.

\citet{Zha19} report results for the oracle, the na\"ive estimator, a poisson matrix factorization (PMF) with $K=450$ and a deep exponential family---the latter two implemented in \texttt{Edward}.  We estimate the Poisson matrix factorization using the \texttt{R} package \texttt{poismf} with $L_2$ regularization (\texttt{l2\_reg=.01}) and $K=450$ run for 100 iterations. All outcome regressions are computed using standard linear regression with homoskedastic standard errors.  Coverage is computed with respect to $95\%$ confidence intervals.

\begin{table}[ht]
\centering
\begin{tabular}{|r|c|ccc|}
\hline \hline 
  & & \multicolumn{3}{c|}{Coverage Proportion}\\
  \hline 
 & RMSE & All & Non-Zero & Zero \\ 
  \hline
  Oracle & 0.03 & 0.95 & 0.95 & 0.95 \\ 
  Na\"ive  & 0.13 & 0.42 & 0.41 & 0.42 \\ 
  Deconfounder & 0.13 & 0.43 & 0.43 & 0.44 \\ 
   \hline
   \hline
\end{tabular}
\caption{\textbf{Replication of Simulation Study 2 of \citet{Zha19}.} The deconfounder is estimated using a 450-dimensional poisson matrix factorization. Coverage rates for nominal 95\% intervals are reported separately for zero coefficients (no causal effect) and non-zero coefficients.}
\label{tab:med2}
\end{table}

We calculate RMSE by calculating all the squared errors $(\hat{\beta_j} - \beta)^2$ and taking the square root of the mean over all coefficients in all simulations. We simulate the entire data generating process each time and conduct 1000 simulations.

We note that original results in \citet{Zha19} do not show the oracle achieving $95\%$ coverage---in fact, they show only $50\%$ coverage for the non-zero coefficients whereas we achieve the nominal rate.  The results of the deconfounder and the na\"ive estimator are essentially indistinguishable in our setting, but we note that they aren't particularly different as reported in the original paper, either. Because the factor models are fit with $k=450$ to a 50-dimensional the exact form of the prior specification will have a huge impact on the values of the substitute confounder.  In an unpenalized setting, $k=50$ should be sufficient to exactly reconstruct the observed data.

\begin{framed}
\noindent \textbf{Deviations}
\begin{itemize}
\item we report only the poisson matrix factorization (not the deep exponential family) and use \texttt{poismf} instead of \texttt{Edward}.
\item we don't holdout $20\%$ of the data or do predictive checks
\item we use OLS rather than bayesian linear regression for all outcome regressions
\end{itemize}

\noindent \textbf{Their Results:}  They argue that the deconfounder produces better results than the na\"ive.\\
\noindent \textbf{Our Results:} We cannot compare to the deep exponential family, but the poisson matrix factorization does not show the gains in improvement over na\"ive that they claim.  We show better RMSE for all estimators as well as better coverage. 
\end{framed}

\subsection{Tutorial Simulations}
The simulations were IPython notebooks in a folder marked \texttt{toy simulations}.  Each shows a simulated data generating process and then walks through a single draw and compares the na\"ive regression (labeled ``noncausal estimation'') with deconfounder estimates based on reconstructed causes and the substitute confounder.  We understand that public tutorials need to be fast to run and thus may often be less nuanced than authors would prefer.  That said, we think these tutorials are important to replicate because they are they way many potential users would be exposed to the deconfounder and would come to understand its properties.

\subsubsection{Logistic Simulation}
\label{a:logisticsim}
\textbf{Summary:} This simulation adds noise to the substitute confounder to make the model estimable---we explore how variations on the amount of noise affect simulation results.

The clearest application of the noised deconfounder is in the logistic tutorial simulation \citep{wang2019github}.  The simulation uses the following data generating process to create 10,000 observations:
\begin{eqnarray}
(X_{1}, X_{2}, Z) & \sim & \text{Multivariate Normal}\left((0,0,0), \begin{pmatrix} 1 & 0.4 & 0.4 \\
               0.4 & 1 & 0.4 \\
               0.4 & 0.4 & 1 \\
\end{pmatrix}\right) \nonumber  \\
y & \sim & \text{Bernoulli}\left( \frac{1}{1 + \text{exp}\left(-(.4 + .2X_1 + 1X_2 + .9Z)\right)}    \right) \nonumber
\end{eqnarray}          
The substitute confounder is found by using PCA of ($X_1, X_2$) and then adding random noise, such that
\begin{align*}
\hat{Z} &= \text{Normal}(\underbrace{\eta_{1} X_{1} + \eta_{2} X_{2}}_{\text{PCA}}, .1^2) 
\end{align*} 
The simulation then estimates a logistic regression with a linear predictor ($X_1,X_2,\hat{Z}$).  The random noise is introduced to break the perfect collinearity between the treatments and the substitute confounder.  For a single draw of the data generating process, the tutorial simulation claims that with the na\"ive regression ``none of the confidence intervals include the truth", but with the deconfounder ``both of the confidence intervals (for $X1, X2$) include the truth." The implication is that the deconfounder improves upon the na\"ive regression estimates.  We repeat the simulation 100,000 times to summarize properties more generally and vary the standard deviation of the noise added to the deconfounder from $.1$ to $(1,.1,.01,.001)$.

We extend the tutorial to assess bias, standard deviation, coverage and RMSE all at different values of the noise parameter with results shown in Table~\ref{tab:logistic}.  At no point does the overall performance of the deconfounder exceed that of the na\"ive estimator.  For large noise, the deconfounder approaches the performance of the na\"ive estimator; as the noise grows small and collinearity increases, estimator variance and RMSE get large very quickly.

\begin{table}[ht]
\centering
\begin{tabular}{|lc|cccc|}
\hline \hline
{} & {} & \multicolumn{2}{c}{Treatment 1}&  \multicolumn{2}{c|}{Treatment 2} \\ 
{} & Noise S.D. & Deconfounder & Na\"ive & Deconfounder & Na\"ive \\ 
\hline
\multirow{4}{*}{Bias} & $10^{-3}$ & 0.200	& \multirow{4}{*}{0.210} & 0.116 & \multirow{4}{*}{0.126} \\ 
								& $10^{-2}$ & 0.202 	&  								 & 0.118 &  \\ 
 								& \textbf{$10^{-1}$} & 0.210 	& 									 & 0.127 &  \\ 
 								& $10^{0}$ & 0.210 	&  								 & 0.126 & \\ 
 \hline
\multirow{4}{*}{Std. Dev.} 	& $10^{-3}$& 16.566 	& \multirow{4}{*}{0.026} & 16.567 &   \multirow{4}{*}{0.030}\\ 
										 	& $10^{-2}$ & 1.659 		& & 1.659 &  \\ 
  											&  \textbf{$10^{-1}$} & 0.167 		&  & 0.168 &  \\ 
  											& $10^{0}$ & 0.031 		&  & 0.035 &  \\ 
  \hline
\multirow{4}{*}{Coverage} 	& $10^{-3}$ & 0.949 		& \multirow{4}{*}{0.000} & 0.949 & \multirow{4}{*}{0.012} \\ 
  											& $10^{-2}$ & 0.948 		&  & 0.949 & \\ 
 											& \textbf{$10^{-1}$}& 0.759 		&  & 0.884 &  \\ 
 											& $10^{0}$ & 0.000 		&  & 0.042 &  \\ 
  \hline
\multirow{4}{*}{RMSE} 			& $10^{-3}$ & 16.568 	&   \multirow{4}{*}{0.211} & 16.567 &   \multirow{4}{*}{0.130} \\ 
											& $10^{-2}$ & 1.671 		&  & 1.663 & \\ 
											& \textbf{$10^{-1}$} & 0.269 		&  & 0.211 & \\ 
											& $10^{0}$ & 0.212 		&  & 0.131 &  \\ 
   \hline\hline
\end{tabular}
\caption{\textbf{Logistic Tutorial Simulation.} 100,000 simulations are summarized in terms of bias, standard deviation of the sampling distribution, coverage of $95\%$ confidence intervals, and root mean squared-error for various levels of simulated noise.  As the noise gets small, the standard deviation and the RMSE of the deconfounder explode (the estimator approaches perfect collinearity). As the noise increases, the deconfounder collapses on the performance of the na\"ive estimator. \label{tab:logistic}}
\end{table}

To help explain the discrepancy in our results, we note that the single draw shown in the workbook is unusual in terms of its error.  The argument made in the writeup is that the confidence intervals of the deconfounder contain the truth while the na\"ive estimator doesn't.  This is a fairly common occurrence---approximately $75\%$ of the simulations because the na\"ive estimator has coverage close to zero.  However, in only $42\%$ of the simulations did the deconfounder produce answers closer to the truth than the na\"ive (along both dimensions).  The deconfounder is unusually close to (and the the na\"ive estimator unusually far from) the truth in terms of mean absolute error across the two coefficients.  The deconfounder only performs as well as reported $8\%$ of the time and na\"ive only performs as poorly as reported $16\%$ of the time. Thus the reported draw is not a representative indicator of performance.

\begin{framed}
\noindent \textbf{Deviations}
\begin{itemize}
\item we repeat the process to create a simulation
\item we examine only substitute confounder and not reconstructed causes
\item we explore different noise levels
\end{itemize}

\noindent \textbf{Their Results:}  The simulation shows an example where the confidence interval for the deconfounder covers the truth and the na\"ive estimator does not. \\
\noindent \textbf{Our Results:} We show that the coverage result is relatively typical but the one draw shown is abnormally accurate for the deconfounder. By evaluating across levels of noise added to the substitute confounder we demonstrate that the results are highly sensitive to the noise level and are at no level better than na\"ive on bias, variance or RMSE.
\end{framed}

\subsubsection{Quadratic Simulation}
\label{app:quadsim}
\textbf{Summary:} This simulation uses a transformation of PCA to make the model estimable---we explore how variations in simulation parameters affect results.

10,000 observations are simulated from the following data-generating process,
\begin{align*}
\left[\begin{array}{l} A_{i,1} \\ A_{i,2} \\ Z_i \end{array} \right]
&\sim \cN\left(
\left[\begin{array}{l} 0 \\ 0 \\ 0 \end{array} \right]
,
\left[\begin{array}{lll}
 1 & \rho & \rho \\
 \rho & 1 & \rho \\
 \rho & \rho & 1
\end{array} \right]
\right)
\nonumber \\[1ex]
Y_i &\sim
\cN(0.4 + 0.2 A_{i,1}^2 + 1 A_{i,2}^2 + 0.9 Z_i^2,\ 1) \label{e:quadratic_outcome}
\end{align*}           
for $\rho=.4$. The substitute confounder, $\hat{Z}$, is based on a PCA of $(A_{1}, A_{2})$, $\hat{Z} = \eta_{1} A_{1} + \eta_{2} A_{2}$.  The outcome regression is $Y = \tau_{0} + \tau_{1} A_{1}^{2} + \tau_{2}A_{2}^{2} + \gamma \hat{Z}^2 + \epsilon_{i}$.

We modify this simulation by adding noise to the outcome drawn from Normal(0,1).  We also evaluate the performance of the parametric specification $\hat{Z}_{(\text{alt})} = A_{1}^2 + A_{2}^2 + 2 A_{1} A_{2}$ and show that it has superior performance. We also demonstrate that the deconfounder breaks down for negative values of the correlation coefficient.

In a footnote of the main text, we noted that the proposed inference procedure uses PCA on $A_1,A_2$ but estimates treatment effects on their squares.  This either presumes that we are interested in the treatment effect of the squared treatment or that we have access to the square root of our real treatments of interest.  However, in that case, we lose the true sign of $A_1, A_2$.  To approximate this scenario, we ran a version of the analysis where we estimate PCA on the absolute value of the variables ($|A_1|$), $|A_2|$) with results shown in Table~\ref{tab:sqrt}.  This leads to an approximately five-fold increase in the RMSE for the deconfounder and roughly a doubling in RMSE for the parametric specification. The relatively worse performance of the deconfounder is due to centering before performing PCA, but the knowledge to not center can only be leveraged if we are sure that the underlying $\bA$ was centered which requires more knowledge that we cannot have.

\begin{table}[ht]
\centering
\begin{tabular}{r|cc}
 & \multicolumn{2}{c}{Treatment Number} \\ 
model & 1 & 2 \\ 
  \hline
Na\"ive & 0.1248 & 0.1252 \\ 
Oracle & 0.0072 & 0.0074 \\ 
Parametric & 0.0425 & 0.0431 \\ 
Deconfounder & 0.1028 & 0.1033 \\ 
\end{tabular}
\caption{RMSE of Quadratic Simulation with Original Settings ($\rho=.4$ and $m=2$) with PCA of Absolute Value of $\bA_1,\bA_2$.}
\label{tab:sqrt}
\end{table}

\begin{framed}
\noindent \textbf{Deviations}
\begin{itemize}
\item we simulate the outcome with error
\item we repeat the process to create a simulation
\item we examine only substitute confounder and not reconstructed causes
\end{itemize}

\noindent \textbf{Their Results:}  The original simulation shows for a single draw that the deconfounder is closer to the truth than the na\"ive and they claim that the confidence intervals contain the truth. \\
\noindent \textbf{Our Results:} We extend the simulation and while performance is in fact strong at the given settings, changing the correlation between $\bA$'s to moderately negative causes the deconfounder to perform much worse than the na\"ive.  As our theory predicts, when the number of treatments is large, the difference between the deconfounder and the na\"ive regression disappears. 
\end{framed}

\subsection{GWAS}\label{a:gwas}
\textbf{Summary:} \citet{WanBle19} applies the deconfounder to study the effects of genes on traits.  We use replication provided to us by Wang and Blei to perform a simulation under similar conditions and we find that the deconfounder and the na\"ive regression perform almost identically.  On closer inspection this is expected.  Figure 4 of \citet{WanBle19} shows nearly identical performance between the deconfounder and the na\"ive regression.  

\subsubsection{Overview}
As a motivating example of the deconfounder, \citet{WanBle19} evoke the use of methods similar to the deconfounder in the genetics literature to explain the effect of genes on the expression of traits.  The genetics database that was used to produce the simulations in the original paper could not be shared with us because of restrictions in data access.  Instead, Wang and Blei shared a purely synthetic simulation procedure intended to replicate the characteristics of the simulations in the original paper as well as code for several factor models applied to this data set.  In this section we describe our results using this synthetic example of the GWAS simulation.  We find that the deconfounder offers essentially identical performance to the na\"ive regression.  This is not surprising, because Figure 4 of \citet{WanBle19} shows that RMSE from the deconfounder and the na\"ive regression are nearly identical.  

\subsubsection{Simulation Procedure} We follow the description of the data generating process in \citet{WanBle19} for the high SNR setting, using a synthetic genetic simulation provided to us to generate data under the Balding-Nichols procedure.  We generate data with 5000 individuals and 5000 genetic markers, with genetic and environmental variation set to 0.4, and with 10\% of the genes assumed to have a causal effect on the outcome. Note, that because of this assumption, any method that shrinks coefficient estimates towards zero will obtain better performance on the non-causal genes, so we divide our results into causal and non-causal genes.  We use the provided code to estimate substitute confounders for deep exponential families (with a 100-dimensional substitute confounder), pca (10-dimensional), poisson matrix factorization (10-dimensional), linear factor analysis (10-dimensional), and probabilistic principal components (5-dimensional).  We avoid the use of the holdout procedure described in the code because it incorrectly sets all held out values to be zero. The posterior predictive checks, as implemented in the code, suggest that the factor models have unrealistic model fits.   

We then generate a single set of effects on the causal genes $\bbeta \sim \mathcal{N}(0, 1)$ and confounding variables $\boldsymbol{\lambda}$ using a slightly modified function from WB, where the modification enabled us to draw the coefficients only once.  Following the original simulation design, we set all non-causal coefficients to zero.  Using the draws of $\bbeta$ and $\boldsymbol{\lambda}$ we simulated the outcome vector, $\bY$, 100 times.

As in the original simulation, we use a ridge regression and nonlinear functional form to render the deconfounder estimable. In each simulation we estimate a ridge regression, cross validating to obtain the penalty parameter. For the na\"ive regression we condition on all 5,000 genes. For each of the factor models we also include the corresponding estimated substitute confounders.  We write our own code to estimate the average root mean squared error, which we display in Table \ref{t:genes}.

Table \ref{t:genes} shows that the na\"ive regression outperforms the deconfounder on this simulation on the genes that have a causal effect on the outcome. On the non-causal genes the other models perform slightly better, but all models offer a nearly identical improvement over the na\"ive regression of the outcomme on one gene at a time. %

\begin{table}[hbt!]
\caption{Using the Synthetic Genetic Data Set, The Deconfounder Offers No Improvement Over Na\"ive Regression} \label{t:genes}
\centering
\begin{tabular}{l|ccc|}
\hline \hline 
    & \multicolumn{3}{c|}{RMSE}  \\
    \hline
    & Causal & Non-Causal & Overall  \\
    \hline 
Na\"ive & 0.737 & 0.127 & 0.263 \\ 
Oracle & 0.742 & 0.125 & 0.263 \\ 
Deconfounder (DEF) & 0.746 & 0.123 & 0.263 \\ 
Deconfounder (PCA) & 0.745 & 0.123 & 0.263 \\ 
Deconfounder (PMF) & 0.745 & 0.123 & 0.263 \\ 
Deconfounder (LFA) & 0.746 & 0.123 & 0.263 \\ 
Bivariate Naive & 1.576 & 1.607 & 1.604 \\ 
\hline\hline
\end{tabular}
\end{table}

\begin{framed}
\noindent \textbf{Deviations}
\begin{itemize}
	\item We use a synthetic simulation created to approximate the simulation in the original paper
	\item We draw the genes and $\lambda$ once.
	\item We evaluate bias and RMSE (see discussion in Supplement~\ref{a:definitions}.
	\item We use a ridge regression with a cross validation-selected penalty, using mean squard error as a cross validation statistic
	\item Following the genetics literature, we examine performance differences on causal and non-causal genes
\end{itemize}
\noindent \textbf{Takeaways}	
\begin{itemize}
\item The deconfounder offers marginal improvements on the non-causal genes and performs worse than the na\"ive estimator on the causal genes. 
\item \citet{WanBle19} uses this simulation as evidence that DEFs are useful. While DEFs do provide a better estimate of the non-causal genes, they are worse on the causal genes and offer---at best---marginal improvements in effect estimation.  
\end{itemize}
\end{framed}
\subsection{Subset Deconfounder}
\label{a:subsetsim}
We use a new simulation design to examine the finite sample properties of the subset deconfounder under the linear-linear data generating process. We create a single-dimensional confounder, $\bZ$, and allow this confounder to satisfy strong infinite confounding.

Our simulation is designed to demonstrate how the average RMSE of the subset deconfounder depends on the underlying treatment effect sizes.  In all of our settings we set each $\theta_{m} = 10$, ensuring strong infinite confounding is satisfied.  We suppose $A_{i} \sim \theta_{m} Z_{i} + \nu_{m}$ where $\nu_{m} \sim \cN(0, 0.01)$.  We then generate outcome data using the linear outcome model using the following coefficient values: 
\begin{enumerate}
    \item $\beta_{m} = 10$ 
    \item $\beta_{m} = 100$ 
    \item $\beta_{m} \sim \cN(1, 4)$
    \item $\beta_{m} = \frac{1}{m} $
\end{enumerate}

We suppose $\gamma= 10$ for all simulations and that $Y_{i} = A_{i} \bbeta + Z_{i} \gamma  + \epsilon_{i}$ where $\epsilon_{i} \sim \cN(0, 0.01)$.  

The results from this simulation align exactly with the predictions from Proposition 4.  Specifically, from Proposition 4 we predict for $\beta_{m} = 10$ a bias of magnitude 10, $\beta_{m} = 100$ a bias of magnitude 100, $\beta_{m} \sim \cN(1, 4)$ a bias of magnitude 1, and for $\beta_{m} = \frac{1}{m} $ the bias at $M$ treatments will bias$_M = \frac{ \sum_{m=1}^{M} \frac{1}{M} }{ M}$ or the average value of $\beta_{m}$.  

\section{Smoking Simulation}
\label{app:smokingsim}
\textbf{Summary:} We replicate the first empirical case study in \citet{WanBle19}, a semi-synthetic dataset about the causes of smoking. We argue that the simulation design is not informative about the performance of the deconfounder because (1) the factor models often have $k \geq m$, and (2) the controls used to compare with a strategy of measuring confounders are themselves uninformative about the confounding. We first quickly review the details of the original design based on the paper and replication code provided by Wang and Blei in December 2019.  We then briefly detail our argument along with some additional results.

\subsection{Original Design}

The smoking simulation is a semi-synthetic study that uses data from the 1987 National Medical Expenditures Survey (NMES) to generate a real joint distribution of three variables which are combined with a linear model to create a synthetic outcome.  

\subsubsection{Data Generating Process in \citet{WanBle19}}
The original simulation in \citet{WanBle19} selects two observed treatments from the NMES: the individual's martial status, $A_{\text{mar}}$, and the exposure to smoking measured as the number of cigarettes per day divided by 20 times number of years smoked,$A_{\text{exp}}$.  The last age at which the person smoked, $A_{\text{age}}$, is designated as the unobserved confounder.  All variables are centered and scaled.  In equations 23 of \cite{WanBle19}, the data generating process for the synthetic outcome is laid out as,
\begin{align*}
Y_i &= \text{Normal}\left(\beta_0 + \beta_{\text{mar}}A_{\text{mar},i} + \beta_{\text{exp}}A_{\text{exp},i} + \beta_{\text{age}}A_{\text{age},i},1\right)
\end{align*}
where the intercept $\beta_0$ is included in the replication code provided to us but not the paper.

In \citet{WanBle19} equation 24, the data generating process of the coefficients is described as, 
\begin{align*}
\beta_{\text{mar}} &\sim \text{Normal}(0,1) \\
\beta_{\text{exp}} &\sim \text{Normal}(0,1) \\
\beta_{\text{age}} &\sim \text{Normal}(0,1) 
\end{align*}
although in the provided replication code the coefficient for the last variable (which will be used as the unobserved confounder) is multiplied by 2.5, leading to,
\begin{align*}
\beta_{\text{age}} &\sim \text{Normal}(0,2.5^2) 
\end{align*}

One of the two treatments, $A_{\text{mar}}$, is a factor variable with 5 levels.  While the factor variable is unlabeled, by examining earlier sources\footnote{The data comes from \citet{imai2004causal} which in turn gets it from \citet{johnson2003disease} which obtains the data from the original source.}, we are confident that level 1 corresponds to married and level 5 corresponds to never married.  We think levels 2-4 correspond to widowed, divorced and separated respectively.  This treatment is treated as a numeric variable in \texttt{R} although factor levels 2-4 ($22\%$ of the data) aren't meaningfully ordered. 

The original simulation treats the first two variables as observed and the final($A_{\text{age},i}$) as the unobserved confounder.  The paper reports the results for 12 configurations of models: na\"ive regression, oracle, linear factor with one dimension (substitute confounder and reconstructed cause), quadratic factor with one, two and three dimensions (substitute confounder and reconstructed cause) and the one dimensional quadratic factor model with additional covariates.

\subsubsection{Factor Model Inference in \citet{WanBle19}}
For each simulation in \citet{WanBle19}, factor models are fit using automatic variational bayes as implemented in \texttt{rstan}.  The model for the quadratic factor analysis (the linear is analogous) as implemented in the provided replication code is,
\begin{align*}
\alpha &\sim \text{Gamma}(1,1) \\
\theta^{(0)} &\sim \text{Normal}(0, 1/\alpha) \\
\theta^{(1)}_k &\sim \text{Normal}(0, 1/\alpha) \\
Z_{i,k} &\sim \text{Normal}(0, 2^2) \\
\bA_{i} &\sim \text{Normal}\left(\theta^{(0)} + \sum_{k=1}^K\theta_k^{(1)} Z_{i,k} + \sum_{k=1}^K \theta^{(2)}_k Z_{i,k}^2, .1^2\right)
\end{align*}
The Normal variances are held fixed for $\bZ$ and $\bA$ and in the equations above we have set them to the values given in the code.  The model is fit with the default settings for ADVI (fully factorized gaussian approximation) except with a fixed step size of .25.  Before beginning the variational approximation, the initial values are set by optimizing the joint posterior with LBFGS for a maximum of 1000 iterations.

For the substitute confounder the $\bZ$ variables are used directly.  For the reconstructed causes, the model outputs:
\begin{align*}
\hat{\bA}_{\text{WB}} &\sim \text{Normal}\left(\theta^{(0)} + \sum_{k=1}^K\theta_k^{(1)} Z_{i,k} + \sum_{k=1}^K \theta^{(2)}_k Z_{i,k}^2, .1^2\right)
\end{align*}  
This differs from \citet{WanBle19} on page 1582 which defines the reconstructed causes as the posterior predictive mean.

\subsubsection{A Note on the Holdout Procedure.}
In order to calculate the posterior predictive checks, the code holds back approximately $5\%$ of the individual cells of the matrix $\bA$ sampled at random.  The holdout percentage is approximate because the sampling procedure allows duplicates which are then removed.  The heldout values are replaced by zero (which due to centering is also the mean of the data).  These values are not resampled in the inference program and so they are effectively treated as mean single imputations of the missing values.  This presumably has both an effect on the fit of the factor model and the posterior predictive checks themselves (which are now conducted exclusively on data that the model is trained believing are exact zeroes). This procedure was corrected in the reference implementation released in July 2020.

\subsubsection{Evaluation Procedure in \citet{WanBle19}}
After estimating the factor model the code from \citet{WanBle19} fits one of the following adjustment strategies:
\begin{enumerate}
\itemsep0em 
\item  Substitute Confounder \\
control for $\hat{\bZ}$
\item Reconstructed Causes \\
replacing the treatment with $\bA - \hat{\bA}$.  This is the version described in the paper
\item Reconstructed Causes 2 \\
the two-parameter version where they control for $\hat{A}$ 
\item Substitute Confounder with Controls \\
controlling for $\hat{\bZ}$ and five controls (see below)
 \item Reconstructed Causes with Controls \\
 replacing treatment with $\bA - \hat{\bA}$ and five controls
 \item Oracle \\
 controlling for the true confounder
 \item Na\"ive \\
regression of $Y$ on all treatments only
\end{enumerate}
The controls include the following variables: age started smoking, binary sex indicator, 3-level factor variable for race, 3-level factor variable for seatbelt use (rarely/sometimes/(always or almost always), 4-level factor variable for education (college graduate/some college/high school/other).\footnote{We pieced these together from \citet{johnson2003disease} and \citet{imai2004causal} but we can't be sure without definitions. These definitions due line up approximately with the summary statistics reported in \citet{johnson2003disease}.}  Unlike the treatments and confounders, these control variables are not standardized or centered.  The factor variables are entered as scalars (rather than contrast coded factors).  For some variables like education that are ordered this produces a linear approximation to the factor model but for the race value there is no guarantee it produces anything in particular.Each of these models is estimated with one of two different outcome regressions: bayesian linear regression estimated using ADVI in \texttt{rstanarm} and OLS.

\subsubsection{Outcome Regression: Bayesian Linear Regression:}
The ultimate goal for the simulations from \citet{WanBle19} is to study properties of the joint posterior distribution $f(\bbeta,\bz | \bY, \bA)$.  Samples from this joint distribution are obtained by factorizing as $f(\bbeta | \bY,\bA, \bz) f(\bz | \bY, \bA)$.  Samples from $f(\bz | \bA)$ are taken from the factor model's posterior---ignoring information from $\bY$---and used as an approximation to $f(\bz | \bY,\bA)$. Then a Bayesian linear regression of $\bY$ on $\bA$ and $\bz$ is used to sample from the conditional posterior $f(\bbeta| \bY, \bA,\bz)$. 

Let $\tilde{\beta}_{j, s, f, d}$ be a draw from this approximate posterior distribution for treatment $j$ in simulation $s$ where:
\begin{itemize}
\item $j \in 1 \dots 2$ indexes the two observed treatments
\item $s \in 1\dots S$ indexes the simulation (i.e. one dataset drawn from the semi-synthetic data generating process)
\item $f \in 1\dots F$ indexes samples from the factor model's posterior distribution $f(\bz | \bA)$
\item $d \in 1\dots D$ indexes the sample from the outcome regression's  posterior distribution $f(\bbeta| \bY, \bA,\bz)$.
\end{itemize}
and we use $\tilde{\cdot}$ to emphasize that it is a sample from a posterior distribution.  In the replication code $f(\bz | \bA)$ (the factor model posterior) is approximated with five samples and so we will set $F=5$. The code then approximates $f(\bbeta| \bY, \bA,\bz)$ (outcome regression conditional posterior) with a single sample and so we will set $D=1$. Let $\beta_{j,s}$ indicate the true treatment effect for treatment $j$ in simulation $s$. 

The \citet{WanBle19} code computes three quantities which effectively treat the sample from the posterior as the estimator and compute properties of that estimator within a given simulation (treating the posterior draws as independent realizations of that estimator) and then average over simulations and treatments. We explore each below.
\label{a:definitions}

\paragraph{Bias Calculation.} The following quantity is computed: 
\begin{align*}
\text{Bias}^2_{\text{WB}} = \frac{1}{2}\sum_{j=1}^2 \left( 
\frac{1}{S} \sum_{s=1}^S 
\left( \left(\frac{1}{5}\sum_{f=1}^5 \underbrace{\tilde{\beta}_{j,s,f,1}}_{\text{estimator}}\right)
- \underbrace{\beta_{j,s}}_{\text{truth}}
 \right)^2
 \right)
\end{align*}
The quantity marked ``estimator'' is a draw from the posterior distribution and the expectation for the bias is taken with respect to that posterior distribution. The metric can also be interpreted as the mean-squared error of the posterior mean estimator approximated with five samples from the posterior distribution. 

\paragraph{Variance Calculation.} Let the function $\widehat{\text{Var}}(\cdot)$ be the sample variance of its arguments. The code defines,
\begin{align*}
\text{Var}_{\text{WB}} = \frac{1}{2} \sum_{j=1}^2 \left( 
\frac{1}{S} \sum_{s=1}^S  \underbrace{\widehat{\text{Var}}\left(\tilde{\beta}_{j,s,1,1}\dots\tilde{\beta}_{j,s,5,1}\right)}_{\text{posterior var}} \right)
\end{align*}
Thus, this reports the per-simulation average posterior variance summed over treatments. 

\paragraph{Mean Squared Error.}
Finally the code computes,
\begin{align*}
\text{MSE}_{\text{WB}} = \frac{1}{2}\sum_{j=1}^2 \left( 
\frac{1}{S} \sum_{s=1}^S 
\left(
\frac{1}{5}\sum_{f=1}^5 
\left(\tilde{\beta}_{j,s,f,1} - \beta_{j,s}\right)^2
\right)
 \right)
\end{align*}
This is the per-simulation, average squared posterior deviation summed over the treatments.

\paragraph{Na\"ive and Oracle Regressions.}
In both the oracle and na\"ive regression there are, of course, no samples from the factor model.  The oracle simply averages over all 1000 samples from the outcome regression's posterior and the na\"ive regression averages over 5 samples from the outcome regression's posterior.  This will make estimates for the na\"ive regression much noisier.

\paragraph{Priors.}
In the original replication code, the Bayesian outcome regression uses different priors depending on the specification. The substitute confounder uses \texttt{rstanarm}'s default Normal(0,1) prior.  The reconstructed causes, oracle and na\"ive regressions use the default \texttt{hs\_plus()} prior which is hierarchical shrinkage prior which is a Normal centered at 0 with a standard deviation that is the product of two independent half Cauchy parameters. The latter has much more mass at 0 and correspondingly fatter tails.

\subsubsection{Outcome Regression: OLS}
The original replication code employs a corresponding set of definitions when using ordinary least squares. Denote $\hat{\beta}_{j,s,f}$ to be the coefficient for treatment $j$ fit on simulation $s$ conditional on draw $f$ from the factor model.  

 \paragraph{Reported Bias Calculation.} The following quantities are computed:
\begin{align*}
\text{Bias}^2_{\text{WB}} = \frac{1}{2}\sum_{j=1}^2 \left( 
\frac{1}{S} \sum_{s=1}^S 
\left( \underbrace{\left(\frac{1}{5}\sum_{f=1}^5 \hat{\beta}_{j,s,f}\right)}_{\text{estimator}}
- \underbrace{\beta_{j,s}}_{\text{truth}}
 \right)^2
 \right)
\end{align*}
This corresponds to the calculation in the Bayesian case but plugging in the coefficient estimates for the sample from the posterior.  

\paragraph{Reported Variance Calculation.} Let the function $\widehat{\text{Var}}(\cdot)$ be the same variance of its arguments and $\widehat{SE}(\cdot)$ be the estimated standard error of its argument.  The code defines:
\begin{align*}
\text{Var}_{\text{WB}} = \frac{1}{2}\sum_{j=1}^2 \left( 
\frac{1}{S} \sum_{s=1}^S  \underbrace{\widehat{\text{Var}}\left(\hat{\beta}_{j,s,1}\dots\hat{\beta}_{j,s,5}\right)}_{\text{var of coefs}} + \underbrace{\frac{1}{5}\sum_{f=1}^5\widehat{SE}(\hat{\beta_{j,s,f}})^2}_{\text{avg of vars}}  \right)
\end{align*}
This uses the law of total variance to provide a more efficient estimator of the sum of the average variance of $f(\bbeta,\bz | \bY, \bA)$.  

\paragraph{Reported Mean Squared Error.}
The code computes,
\begin{align*}
\text{MSE}_{\text{WB}} = \text{Bias}^2_{\text{WB}} + \text{Var}_{\text{WB}} 
\end{align*}

\subsubsection{Reported Results}
Table 3 of \citet{WanBle19} presents the findings.  The discussion highlights the improved performance of the one-dimensional and two-dimensional quadratic models over the na\"ive regression although no estimator is particularly close to the oracle.  In the corresponding discussion (1585-1586), the results are used to emphasize three points: (1) the value of the posterior predictive check for signaling whether results are biased, (2) controlling for observed confounders increases variance but does not decrease bias, and (3) the deconfounder outperforms na\"ive regression.

\subsection{New Results}

In this section, we briefly present a conceptual argument about the design before providing some broader results to contextualize the findings.

\subsubsection{Factor models where $k \geq m$}
In four of the eight original factor model specifications including two of the three highlighted in \citet{WanBle19} for performance, the dimensionality of the latent factors exceeds the dimensionality of the data. In a frequentist setting, these models would exactly reconstruct the observed treatments (they all nest 2-dimensional PCA as a special case). The models are fit with bayesian methods but using broad priors and so it would appear that they only reason they don't perfectly reconstruct the treatments is noise in the posterior approximation. This renders the models estimable, but uninformative about performance of the deconfounder. 

\subsubsection{Deviations in Our Procedure}
In re-implementing the simulation we tried to strike a balance between remaining comparable to the original design and making changes that we felt were essential to being able to interpret the simulation.  In total, we estimate three baseline specifications: na\"ive regression, the oracle model and a regression controlling for WB's controls as well as five sets of deconfounder models based on specifications by WB (Linear Factor Model with 1 dimension, Quadratic Factor Model with 1-3 dimensions and Quadratic Factor Model with 1 dimension and controls).  For each set of models we report three variants: the substitute confounder (controlling for $\hat{\bZ}$), reconstructed causes as stated in the paper (replacing the treatment with $\bA - \hat{\bA}$) and reconstructed causes as implemented in their code (controlling for $\hat{\bA}$).  This adds the specification of the controls alone and includes both versions of the reconstructed causes. We outline the other changes we make here along with our rationale.

\paragraph{Deviation 1: OLS Outcome Regression.}
We use an OLS outcome regression and average results over 1000 draws of the factor model's posterior distribution (rather than 5).  This ensures the computation of the approximate joint posterior mean is not too noisy.  The 1000 regressions can be done computationally efficiently by noting that the design matrix stays fixed.  Denoting the design matrix $\bX$ we precompute $(\bX'\bX)^{-1}\bX'$ which means the full set of 1000 regressions can be computed with a single additional matrix multiply.

\paragraph{Deviation 2: Fixed Simulation Coefficients}
It is difficult to evaluate properties like bias when the coefficients are varying across runs.  To see why this would be complicated, consider an estimator that was biased towards zero, and imagine that the calculation applied was $E[\hat{\beta}_i - \beta_i]$ where $\beta_i$ was sampled from a normal distribution.  The upward and downward biases would cancel each other out over draws, leading to an estimate of approximately 0. We avoid this problem by fixing the coefficients to arbitrarily chosen values (1,1,1) and then assessing the bias, variance and mean-squared-error of the posterior mean. The results are driven primarily by bias so we report only the root mean squared error.

\paragraph{Deviation 3: Removed Holdout Procedure}
We were concerned about the effects of the mean imputation so we removed the holdout procedure entirely. Once this was done, the treatments do not change across simulations.   

\paragraph{Deviation 4: Change in the Reconstructed Causes.}
We replace the reconstructed causes with 
\begin{align*}
\hat{\bA}_{i} = \theta^{(0)} + \sum_{k=1}^K\theta_k^{(1)} Z_{i,k} + \sum_{k=1}^K \theta^{(2)}_k Z_{i,k}^2
\end{align*}
in each sample from the posterior.  Due to the sampling of $\bZ$ and $\theta$, the reconstructed cause is almost surely not collinear with $\bA$.

\paragraph{Deviation 5: Remove the intercept.}
We remove the intercept from the true model to be consistent with the equations in \citet{WanBle19}.

\paragraph{Deviation 6: Make control variables factors}
We treat the control variables which are categorical as factors rather than numerics as done in the original simulation design.

\paragraph{Concerns We Do Not Address}
There were several issues in the simulation design that we did not address because we felt that to do so would too fundamentally alter the simulation.  
\begin{itemize}
\item for the treatments, we continue treating the factor variables as continuous variables because otherwise the dimension of $\bA$ changes
\item $A_{\text{exp}}$ is skewed (it is logged in \citet{imai2004causal}) which affects the normality assumptions
\item many of the models considered here involve many more dimensions than the two in the observed data. It is unclear why the model isn't fitting the data perfectly in these settings or why we would use latent variable modes with many more latent variables than observed dimensions.
\end{itemize}

\subsubsection{New Results: Instability in Factor Models}
\label{a:smoking_extra}

We first demonstrate that a substantial amount of variation in the original results is due to instability in factor model estimation. Table~\ref{tab:smokingmultimodal} shows results across four different factor model fits, labeled F1--4. Because we removed the model checking using held-out data, the inputs across all four models are identical, so only the seed changes across the four iterations. The differences in the learned factor models induce substantial differences in the RMSE of the resulting estimates. For example, the two-dimensional quadratic model with substitute confounder (one of the preferred specifications in \citet{WanBle19} and row 11 of our Table~\ref{tab:smokingmultimodal}) ranges from $12\%$ better to $30\%$ worse than the na\"ive estimate for the effect of Treatment 1. 

\begin{table}[ht!!]
\centering
\begin{tabular}{rl|cccc|cccc}
\hline \hline 
 & & \multicolumn{4}{c|}{Treatment 1} &  \multicolumn{4}{c}{Treatment 2} \\ 
 \hline
   & & F1&F2&F3&F4&F1&F2&F3&F4\\
  \hline
\multirow{3}{*}{Baseline}  & Na\"ive & \multicolumn{4}{c}{1.00}   & \multicolumn{4}{c}{1.00} \\ 
  & Oracle & \multicolumn{4}{c}{0.043} & \multicolumn{4}{c}{0.026} \\ 
  & Controls & \multicolumn{4}{c}{0.900} & \multicolumn{4}{c}{1.094} \\ 
  \hline
\multirow{3}{*}{Lin. (1 dim.)}  &Sub.			& 1.020 & 1.176 & 1.065 & 0.980 & 1.011 & 1.093 & 1.033 & 0.987 \\ 
  											 & Rec. 		& 0.253 & 0.181 & 0.843 & 0.665 & 0.617 & 0.529 & 0.911 & 0.816 \\
  & Rec. 2 									    & 1.020 & 1.176 & 1.065 & 0.980 & 1.011 & 1.093 & 1.033 & 0.987 \\  
  \hline
\multirow{3}{*}{Quad. (1 dim.)}  &Sub.		& 0.657 & 1.054 & 0.759 & 1.097 & 0.861 & 0.989 & 1.022 & 0.908 \\ 
  &Rec.													& 0.533 & 1.298 & 1.135 & 0.375 & 1.733 & 1.278 & 2.243 & 0.481 \\ 
  &Rec. 2 										& 1.299 & 1.503 & 1.026 & 0.136 & 2.186 & 1.503 & 1.702 & 0.545 \\  
  \hline
\multirow{3}{*}{Quad. (2 dim.)}  &Sub. 		&  0.882 & 1.028 & 1.317 & 1.012 & 0.834 & 0.941 & 0.569 & 0.921 \\ 
  &Rec.  												& 3.399 & 1.814 & 1.094 & 6.526 & 1.315 & 0.872 & 2.431 & 0.701 \\ 
  &Rec. 2 										& 0.640 & 0.711 & 0.498 & 0.538 & 1.848 & 1.117 & 0.915 & 1.329 \\ 
  \hline
 \multirow{3}{*}{Quad. (3 dim.)} &Sub.		& 1.638 & 1.856 & 1.003 & 1.398 & 1.060 & 0.964 & 1.003 & 0.440 \\ 
  &Rec. 												& 6.870 & 5.692 & 1.441 & 4.737 & 0.113 & 3.327 & 3.864 & 2.089 \\ 
  &Rec. 2 										& 0.502 & 1.777 & 0.850 & 0.445 & 1.859 & 1.420 & 1.124 & 1.116 \\ 
  \hline
\multirow{3}{*}{\begin{tabular}{r}
Quad.(1 dim.) \\w/ Controls
\end{tabular}}  &Sub. 
																			& 0.586 & 0.945 & 0.665 & 0.990 & 0.966 & 1.085 & 1.116 & 1.009 \\ 
  &Rec. 												& 0.636 & 1.112 & 1.033 & 0.446 & 1.744 & 1.179 & 2.146 & 0.515 \\ 
  &Rec. 2										& 1.221 & 1.334 & 0.876 & 0.173 & 2.179 & 1.528 & 1.653 & 0.659 \\  
   \hline\hline
\end{tabular}
\caption{\textbf{Smoking Simulation Results Vary Substantially By Factor Model Run}: This table shows the ratio of root mean squared-error to the na\"ive regression for 18 different specifications and four different runs of each factor model.  Values above 1 indicate that the model is performing worse than the na\"ive regression and models below 1 indicate it is performing better. The left column provides the factor model and the second column provides the adjustment strategy. ``Sub.'' is the substitute confounder; ``Rec.'' is the reconstructed causes approach stated in the paper; and ``Rec. 2'' is the two-parameter reconstructed causes approach implemented in code.  Models do not consistently perform better than na\"ive.}
\label{tab:smokingmultimodal}
\end{table}

We note that different adjustment strategies (Sub., Rec., Rec. 2) used with the same factor model can yield substantially different results.  Because the PPC is specific to the factor model and not the adjustment strategy, it cannot provide information about which would provide better performance.

As in \citet{WanBle19} we do not observe substantial benefits from including covariates with the deconfounder.  However, line three of the table makes clear that this is because the covariates alone are not sufficient to improve over the Na\"ive regression.  In practice this is because they are essentially uncorrelated with the variable chosen to be the unobserved confounder.  Thus, we should not draw conclusions from this study about the role of measured confounders.

We have only shown one set of simulated coefficients here. However, because the RMSE is driven almost entirely by the bias term, the results here are extremely well predicted by the standard omitted variable bias formula.  Define $\bX = (A_{\text{mar}}, A_{\text{exp}},\hat{\bZ})$ then,
\begin{align*}
\text{bias}(\beta_{\text{mar}},\beta_{\text{exp}}) = (\bX^T\bX)^{-1}\bX^T\bZ\beta_{\text{age}}
\end{align*}
For a fixed factor model, and thus a fixed $\hat{\bZ}$, we can calculate the bias for any setting of the true coefficient $\beta_{\text{age}}$. 

\subsubsection{New Results: Max ELBO of the Factor Models}
The variational inference procedure comes with a natural mechanism for choosing among the factor models.  We run each model twenty times and choose the one that maximizes the evidence lower bound.  In \texttt{rstan} this has to be parsed from a log that collects the material printed to the screen as it is not included in the returned output.  The results are in Table~\ref{tab:smokingelbo}.
\begin{table}[ht!!]
\centering
\begin{tabular}{rl|c|c}
\hline \hline 
 & & Treatment 1 &  Treatment 2 \\ 
 \hline
\multirow{3}{*}{Baseline}  & Na\"ive & 1.000  & 1.000 \\ 
  & Oracle & 0.044 & 0.025 \\ 
  & Controls & 0.900 & 1.094 \\ 
  \hline
\multirow{3}{*}{Lin. (1 dim.)}  &Sub.	& 1.062 & 1.033 \\
  						        & Rec. 	& 1.325 & 1.151 \\ 
                                & Rec. 2 & 1.061 & 1.033 \\ 
  \hline
\multirow{3}{*}{Quad. (1 dim.)}  &Sub.	& 0.648 & 0.844 \\ 
  &Rec.									 & 1.027 & 1.865 \\
  &Rec. 2 								& 1.748 & 2.437 \\  
  \hline
\multirow{3}{*}{Quad. (2 dim.)}  &Sub. 	& 2.648 & 0.308 \\  
  &Rec.  								& 4.073 & 1.214 \\ 
  &Rec. 2 								& 0.718 & 1.353 \\ 
  \hline
 \multirow{3}{*}{Quad. (3 dim.)} &Sub.	& 2.192 & 0.840 \\ 
  &Rec. 								& 3.290 & 1.362 \\
  &Rec. 2 								& 0.605 & 0.924  \\
  \hline
\multirow{3}{*}{\begin{tabular}{r}
Quad.(1 dim.) \\w/ Controls
\end{tabular}}  &Sub. 
											& 0.566 & 0.947 \\ 
  &Rec. 									& 1.166 & 1.889 \\ 
  &Rec. 2									& 1.670 & 2.421 \\ 
   \hline\hline
\end{tabular}
\caption{\textbf{Deconfounder Does Not Outperform the Na\"ive Regression In the Smoking Simulation}: This table shows the ratio of root mean squared-error to the na\"ive regression for 18 different specifications using the factor model which maximized the ELBO over twenty runs.  Values above 1 indicate that the model is performing worse than the na\"ive regression and models below 1 indicate it is performing better. The left column provides the factor model and the second column provides the adjustment strategy. ``Sub.'' is the substitute confounder; ``Rec.'' is the reconstructed causes approach stated in the paper; and ``Rec. 2'' is the two-parameter reconstructed causes approach implemented in code.  Models do not consistently perform better than na\"ive. See cautionary note in main text, results are very unstable. \label{tab:smokingelbo}}
\end{table}

In practice we observe that all the linear factor model fits are very similar, but the quadratic factor models vary substantially.  Those where $k\geq m$ vary the most.  Thus while we maximized the ELBO over twenty different fits, we would expect that results would be unstable under replication. We present these results simply to demonstrate that the ELBO does not provide a way to resolve the problem demonstrated in the previous subsection.

\subsection{Conclusions on Smoking}
The smoking simulation in \citet{WanBle19} seeks to use a semisynthetic design to justify a number of conclusions about the deconfounder's performance.  Unfortunately, as we have shown, these conclusions do not hold under reasonable adjustments and extensions to the simulation design.

\section{Breast Cancer Tutorial}\label{a:breast}
\textbf{Summary:}The github tutorial examines the effect of various tumor features on the diagnosis of breast cancer tumors.  The tutorial uses approximate inference to fit a probabilistic principal components model to estimate the substitute confounder and then assert this provides valid causal estimates.  This assertion is based on a non-standard assessments of whether a model is causal or not.  We show that the full deconfounder is only estimable because the approximate inference leads to considerable noise in the estimated substitute confounder.  When a more standard estimation procedure for the substitute confounder is deployed the full deconfounder is only estimable with a penalized regression. And the coefficient estimate that we obtain is entirely dependent on the amount of penalization.  This demonstrates that the deconfounder is not particularly helpful for causal inference in this setting.  

\begin{figure}
\scalebox{0.5}{\includegraphics{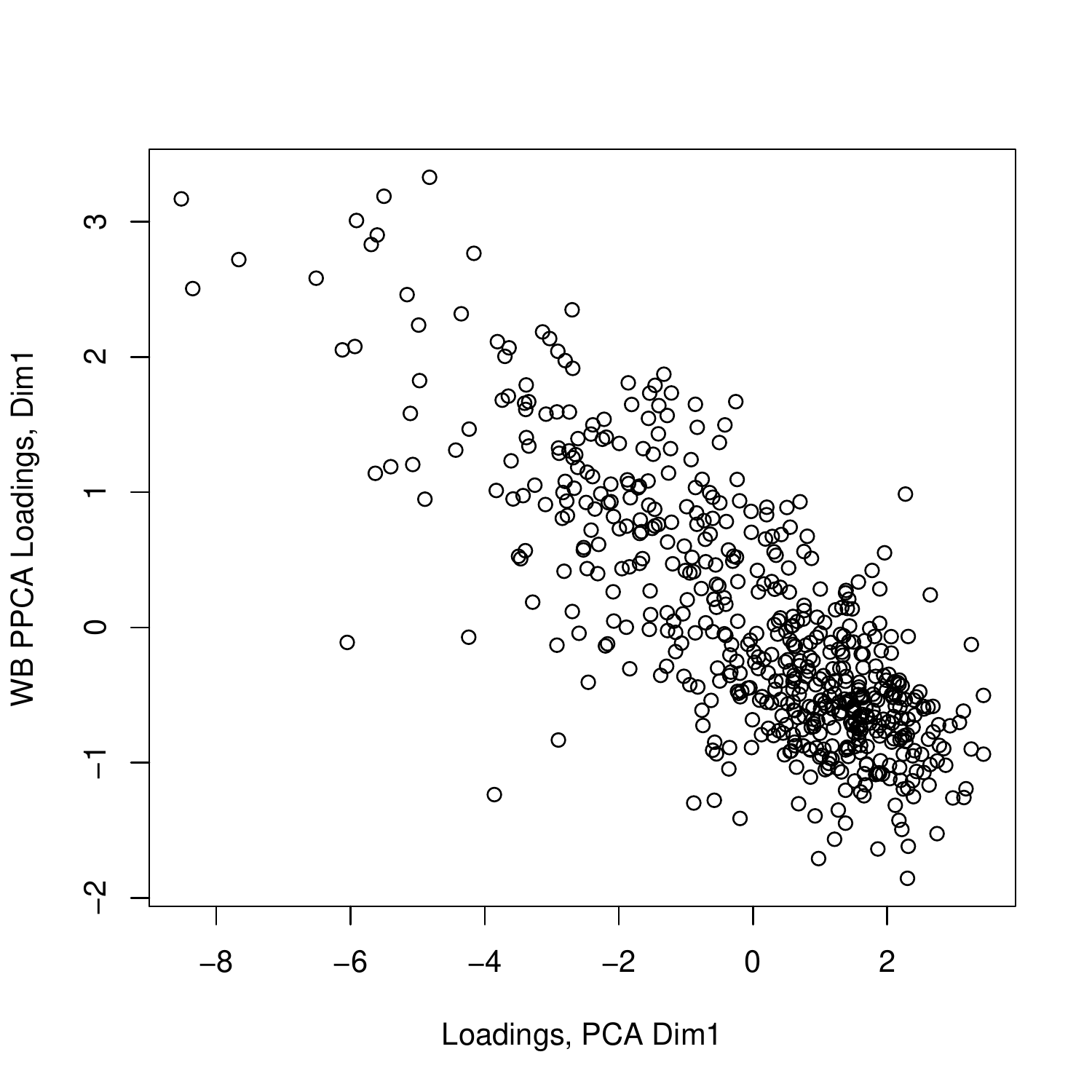}}
\scalebox{0.5}{\includegraphics{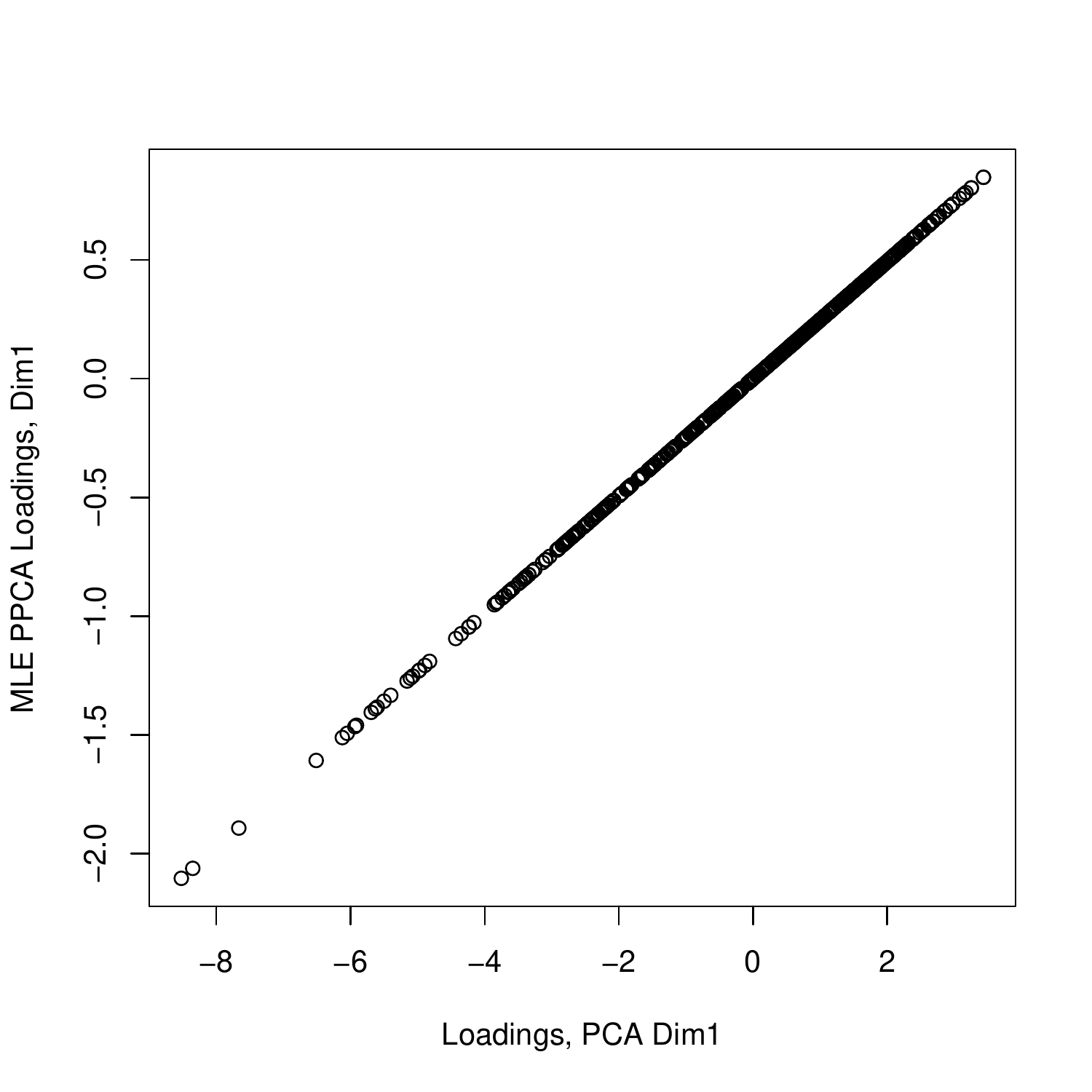}}
\caption{\textbf{The tutorial's procedure for estimating the substitute confounder adds unncessary noise.} The left-hand plots shows the relationship between the substitute confounder estimated using approximate inference and the substitute confounder estimated using traditional PCA. The approximate inference adds considerable noise. The right-hand plot shows that if we use well known MLE routines for estimating PPCA there is no disagreement between the loadings.  Approximate inference, then, leads to considerable and unnecessary error.}
\label{f:pca_ppca}
\end{figure}

Using a breast cancer data set that is distributed with SciKit learn, the tutorial estimates a substitute confounder using black box variational inference. The tutorial argues that approximate inference is completely acceptable and can be ignored.
We show this is not the case---approximate inference adds considerable noise to the estimated substitute confounder.  Consider the left-hand facet of Figure~\ref{f:pca_ppca} which compares the first dimension of the estimated substitute confounder from approximate inference (vertical axis) to the substitute confounder estimated using PCA.  This shows that the estimated substitute confounder is a noisy version of PCA (we have not rotated the loadings, which explains the negative relationship).  But the right-hand facet compares the estimate of the substitute confounder estimated from PPCA using maximum likelihood estimation against PCA. The estimated loadings using maximum likelihood to fit a PPCA model are effectively identical to the loadings from PCA, just scaled differently.  In short, the approximate inference procedure leads to a poorly estimated model. 

The result of this poorly estimated factor model is that including the substitute confounder has little effect on the actual coefficient estimates, yielding estimates that are nearly identical to a na\"ive regression.  Column 1 of Table \ref{t:breast} provides the coefficient estimates of features on breast cancer diagnosis using the original estimates of the substitute confounder in a logistic model.  Rather than make the non-standard step of subsetting to only 80\% of the data, we fit the model to the entire data set. We want to emphasize that this model is estimable \textit{solely} because the approximate inference procedure yields a poor estimate of the parameters of the PPCA model.  As we might expect give this fact, the second column shows that a na\"ive logistic regression that simply ignores those confounders yields nearly identical results.    In the third column we obtain coefficient estimates, but now we estimate the substitute confounder using standard PCA.  Without the errors in the estimated substitute confounder from approximate inference, the model is no longer estimable using a standard logistic regression.  Instead we rendered the model estimable with a ridge regression, with the penalty selected using cross validation.  This yields dramatically different results.  Many of the coefficients have effect sizes that are orders of magnitude smaller.    The final column shows the estimates from the one-at-a-time deconfounder, fit using a logistic regression.  This reveals strikingly different results from those reported in the online tutorial: the sign flips on half of the coefficients.

\begin{table}[ht]
\centering
\begin{tabular}{l|cccc}
  \hline
 & Full Deconfounder & Na\"ive & Full Deconfounder & One-at-a-Time \\ 
 & WB PCA            &        & True PCA         & True PCA \\
  \hline
mean radius & -3.10 & -3.48 & -0.81 & -0.56 \\ 
mean texture & -1.38 & -1.64 & -0.51 & -0.07 \\ 
mean smoothness & -0.87 & -1.07 & -0.38 & -1.31 \\ 
mean compactness & 1.28 & 0.88 & -0.29 & 2.59 \\ 
mean concavity & -0.77 & -1.19 & -0.50 & 1.37 \\ 
mean concave points & -1.78 & -2.27 & -0.75 & -0.97 \\ 
mean symmetry & -0.24 & -0.50 & -0.24 & -0.34 \\ 
mean fractal dimension & 0.27 & 0.19 & 0.29 & 0.98 \\ 
   \hline
\end{tabular}
\caption{\textbf{Correctly estimating PCA leads to dramatically different results}  The first column replicates the exact procedure from the github tutorial, but estimate the coefficients on the entire sample.  The second column merely drops the substitute confounder and yields very similar results.  The third column estimates the full deconfounder using the true PCA estimates, using a penalized ridge regression to fit the model.  This yields dramatically different results, consistent with our theoretical results. The fourth column provides the results from a one-at-a-time deconfounder. This shows even more deviations, with half of the coefficients changing signs, and many coefficients exhibiting orders of magnitude effect estimate changes.  } \label{t:breast}
\end{table}

\begin{framed}
\textbf{Deviations}
\begin{itemize}
	\item We fit the outcome regression on the entire data set, rather than using an 80\% held out data set. 
	\item Given the severe errors in the approximate inference procedure, we use standard PCA estimation routines
\end{itemize}
\textbf{Their Results}: The tutorial claims this model provides robust causal effect estimates. 
\textbf{Our Results}: We show that the model in WB's tutorial is estimable solely because of errors in the estimation of the PCA model.  Once corrected, we obtain different coefficient results that vary substantially depending on how we apply the deconfounder. 
\end{framed}

\section{Posterior Predictive Model Checking Does Not Reliably Identify When the Deconfounder Works}
\label{a:ppc}
We have shown that it is impossible to know when the deconfounder improves over the na\"ive regression in practice. Throughout their papers, \citet{WanBle19} and \citet{Zha19} use a framework of posterior predictive checks (PPCs) to ``greenlight'' their use of the deconfounder in practice and adjudicate between alternative estimators. \citet{WanBle19} explain,
\begin{quote}
We consider predictive scores with predictive scores larger than 0.1 to be satisfactory; we do not have enough evidence to conclude significant mismatch of the assignment model. Note that the threshold of 0.1 is a subjective design choice. We find such assignment models that pass this threshold often yield satisfactory causal estimates in practice \citep[][p. 1581]{WanBle19}  
\end{quote}
If PPCs could be used in this way, it would allow highly flexible density estimation models to be used, even when the true parametric form was unknown---as is always the case in practice.  The proof of the subset deconfounder establishes that this is impossible in that setting because the performance of the subset deconfounder depends on untestable assumptions about the treatment effects. For the full deconfounder, the check in WB are not well-suited to evaluating the conditional independence of $\bA$ given $\hat{\bZ}$ which is perhaps the most relevant observable property \citep{imai2019discussion}.  %

We evaluate the performance of PPCs on a quadratic-poisson factor model with $n=10000$ observations and $m=100$ treatments, where 
\begin{align*}
    Z_{i} &\sim \mathcal{N}(0,0.2) \\
    A_{ij} &\sim \text{Poisson}\left(\exp\left(\theta_{j1} Z_{i} + \theta_{j2} Z_{i}^2\right) \right) \\
    Y_i &\sim \mathcal{N}\left(\bA_i\bbeta + Z_i\gamma, 0.1\right)
\end{align*}, 

where $\btheta_{j1}, \btheta_{j2} \sim \cN(0, 1)$ and $\bbeta$ is set equal to $(0.8, -0.6, 0.4, -1.2)$ repeated 25 times. We compare the na\"ive regression and the oracle to the two versions of the subset deconfounder: (i) using a Deep Exponential Family (DEF) \citep{ranganath2015deep} with (5,3,1) layers to estimate the substitute confounder, and (ii) using a two-dimensional PCA to estimate the substitute confounder. 
\begin{figure}[hbt!]
\centering
\includegraphics[width=.75\textwidth]{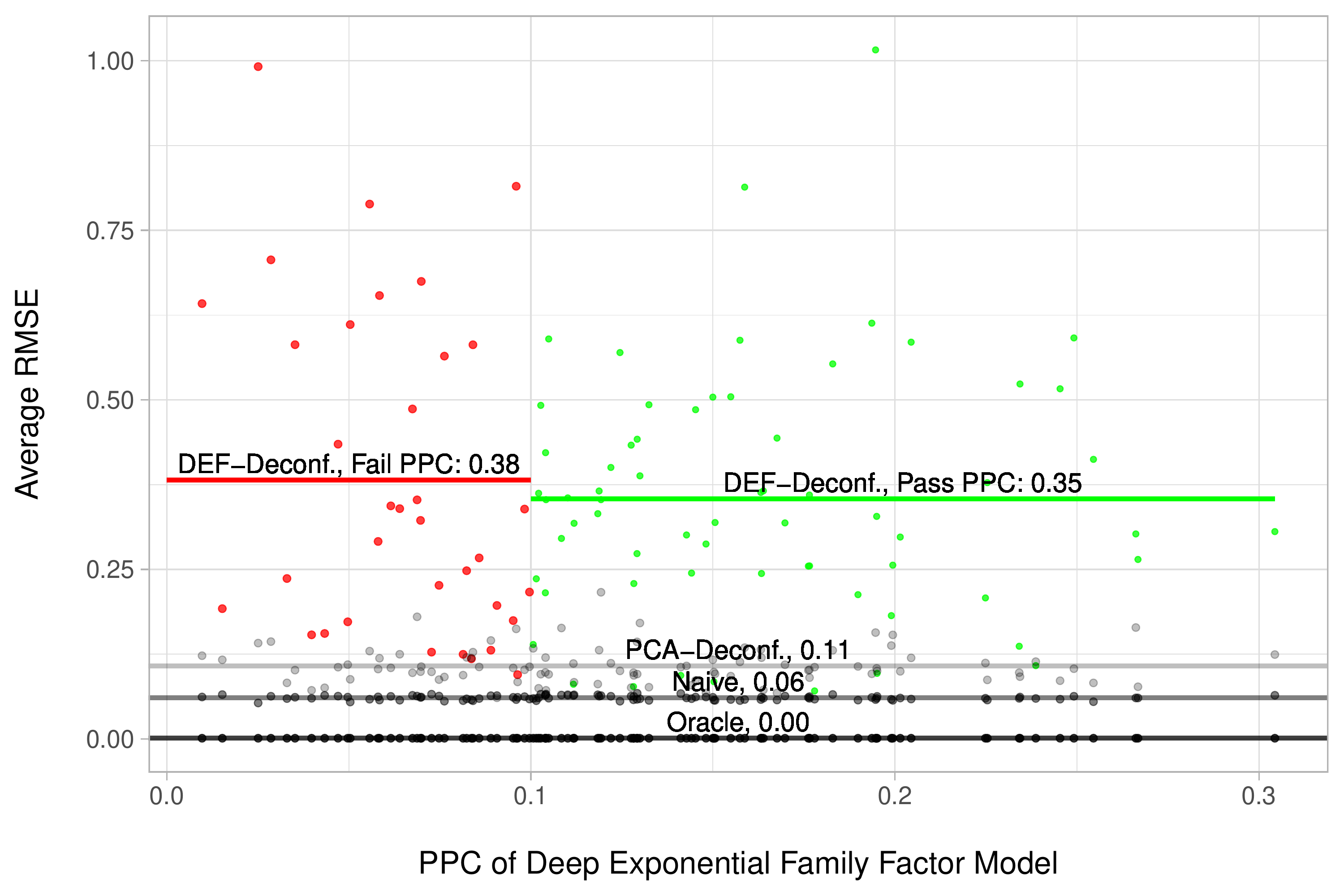}
\caption{\textbf{Posterior Predictive Checks Do Not Reliably Identify When the Deconfounder Outperforms the Na\"ive Estimator.} The horizontal axis is the predictive score from the posterior predictive check of the DEF and the vertical axis is the average RMSE for the treatment effect estimates.  The red points are the average RMSE from applications of the DEF that failed the predictive score check, while the green points are the average RMSE from applications of the DEF that passed the predictive score check.} 
\label{fig:def}
\end{figure}

The results in Figure~\ref{fig:def} show the average RMSE for each adjustment strategy plotted at the corresponding PPC for the DEF.  The average RMSE for the DEF-deconfounder is approximately equal, whether the model passes the PPC or not.  Further, we see that there is considerable variation.  We see that there can be extremely large RMSEs from estimation when the deconfounder passes the PPC and quite small when it fails the PPC.  We also find that more complex models do not outperform simple alternatives.  The average RMSE of the deconfounder using DEF is over three-times larger than the average RMSE when using PCA---even though the true underlying model that generated the treatments is nonlinear in the substitute confounder. But both implementations of the deconfounder perform considerably worse than the naive regression. The DEF deconfounder that pass the PPC has an average RMSE 5.8 times the average RMSE of the naive regression, while the PCA deconfounder has an average RMSE 1.8 times the average RMSE of the naive estimator. In every simulation, the average RMSE from the naive regression is better than either implementation of the deconfounder.

This suggests that PPCs cannot help us distinguish when the deconfounder will improve over alternatives. There is considerable noise in the RMSE of models that pass or do not pass PPCs, so the safest conclusion is that there is no difference between the RMSE of applications of the deconfounder that do and do not pass the PPC.  The simulation also demonstrates that if the functional form of the factor model is unknown, a flexible factor model can perform considerably worse than simpler models, even when the true data generating process is nonlinear and the flexible model passes model-fit checks. Most importantly, the deconfounder's estimates can be considerably worse than a na\"ive estimator which simply ignores confounding. In short, model checking and flexible nonlinear factor models cannot solve the deconfounder's problems.

\section{Additional Results from Actor Application}
\label{a:actors}
We offer several variants of the deconfounder results below.  ``WB Deconfounder'' is the cached results of the Poisson Matrix Factorization run in \citet{WanBle19}.  The ``Deconfounder, Full'' is our re-estimated model using the same code.  In \citet{WanBle19}, the analyses did not condition on any observed covariates. After we shared our draft with Wang and Blei in July 2020, they posted a reference implementation of the deconfounder that provided an illustration of conditioning on budget and runtime.
\subsection{Top 25 Actors According to Each Estimator, Ranked by Multiplicative Effect $e^{\hat\beta_j}$}

\paragraph{Naive, Full}
Stan Lee ($\times$9.31), John Ratzenberger ($\times$9.26), Sacha Baron Cohen ($\times$7.09), Leonardo DiCaprio ($\times$5.50), Josh Hutcherson ($\times$5.19), Corey Burton ($\times$4.80), Brian Doyle Murray ($\times$4.79), Tom Hanks ($\times$4.76), Julie Andrews ($\times$4.64), Desmond Llewelyn ($\times$4.55), Will Smith ($\times$4.36), Ava Acres ($\times$4.24), Crispin Glover ($\times$4.12), Warwick Davis ($\times$4.10), Andy Serkis ($\times$4.02), Eddie Murphy ($\times$3.88), Shia LaBeouf ($\times$3.81), Tomas Arana ($\times$3.79), Judi Dench ($\times$3.75), Penélope Cruz ($\times$3.66), Tom Cruise ($\times$3.49), Blythe Danner ($\times$3.46), Jay Baruchel ($\times$3.46), Harrison Ford ($\times$3.42), Michael Douglas ($\times$3.37)

\paragraph{WB Deconfounder, Full}
Stan Lee ($\times$9.29), John Ratzenberger ($\times$8.29), Sacha Baron Cohen ($\times$8.12), Josh Hutcherson ($\times$5.02), Corey Burton ($\times$4.91), Leonardo DiCaprio ($\times$4.87), Tom Hanks ($\times$4.71), Ava Acres ($\times$4.55), Julie Andrews ($\times$4.38), Will Smith ($\times$4.37), Desmond Llewelyn ($\times$4.19), Crispin Glover ($\times$4.19), Eddie Murphy ($\times$4.18), Andy Serkis ($\times$4.14), Judi Dench ($\times$3.99), Shia LaBeouf ($\times$3.92), Brian Doyle Murray ($\times$3.91), Tobin Bell ($\times$3.63), Tomas Arana ($\times$3.61), George Lopez ($\times$3.56), Warwick Davis ($\times$3.55), Blythe Danner ($\times$3.52), Mary Elizabeth Winstead ($\times$3.52), Tom Cruise ($\times$3.46), Penélope Cruz ($\times$3.45)

\paragraph{Deconfounder, Full}
Courteney Cox ($\times$15.32), Tom Cruise ($\times$14.74), John Ratzenberger ($\times$11.13), Vera Farmiga ($\times$9.86), Sacha Baron Cohen ($\times$9.83), Kim Coates ($\times$9.76), Charlize Theron ($\times$8.74), James Badge Dale ($\times$8.65), Patrick Wilson ($\times$8.58), Penélope Cruz ($\times$8.41), Rafe Spall ($\times$8.09), Tom Hanks ($\times$8.02), Jeffrey Tambor ($\times$7.63), Albert Finney ($\times$7.17), Crispin Glover ($\times$6.94), Garrett Hedlund ($\times$6.86), Octavia Spencer ($\times$6.78), Emile Hirsch ($\times$6.75), Leonardo DiCaprio ($\times$6.61), Tim Robbins ($\times$6.48), Chiwetel Ejiofor ($\times$6.45), Ewen Bremner ($\times$6.39), John Heard ($\times$6.20), Johnny Knoxville ($\times$6.01), George Clooney ($\times$6.00)

\paragraph{Budget Adjustment, Full}
Sacha Baron Cohen ($\times$6.93), Brian Doyle Murray ($\times$4.08), Conrad Vernon ($\times$4.04), Julie Andrews ($\times$3.84), Tomas Arana ($\times$3.83), Crispin Glover ($\times$3.73), John Ratzenberger ($\times$3.55), Tobin Bell ($\times$3.46), Leonardo DiCaprio ($\times$3.41), Keira Knightley ($\times$3.29), Reese Witherspoon ($\times$3.27), Josh Hutcherson ($\times$3.10), Desmond Llewelyn ($\times$3.09), Tom Hanks ($\times$2.92), George Lopez ($\times$2.87), James Rebhorn ($\times$2.85), Clea DuVall ($\times$2.85), Jason Sudeikis ($\times$2.83), Allen Covert ($\times$2.79), Alan Rickman ($\times$2.76), Anna Kendrick ($\times$2.75), Jodie Foster ($\times$2.73), Andy Samberg ($\times$2.69), John Heard ($\times$2.64), Michael Kelly ($\times$2.64)

\paragraph{Naive, One at a Time}
Jess Harnell ($\times$12.28), Ava Acres ($\times$10.16), Warwick Davis ($\times$10.09), Stan Lee ($\times$9.85), Orlando Bloom ($\times$9.50), Hugo Weaving ($\times$8.42), Chris Miller ($\times$8.35), Conrad Vernon ($\times$7.92), John Ratzenberger ($\times$7.91), Mickie McGowan ($\times$7.67), Julie Walters ($\times$7.11), Christopher Lee ($\times$7.03), Danny Mann ($\times$6.98), Lasco Atkins ($\times$6.50), Ian McKellen ($\times$6.44), Tom Felton ($\times$6.26), Daniel Radcliffe ($\times$6.23), Andy Serkis ($\times$6.19), Sacha Baron Cohen ($\times$6.10), Timothy Spall ($\times$6.03), Frank Oz ($\times$6.01), Emma Watson ($\times$5.82), Pat Kiernan ($\times$5.69), Bonnie Hunt ($\times$5.68), Denis Leary ($\times$5.26)

\paragraph{WB Deconfounder, One at a Time}
Jess Harnell ($\times$13.49), Ava Acres ($\times$10.49), Chris Miller ($\times$9.19), Orlando Bloom ($\times$9.02), Stan Lee ($\times$8.77), Conrad Vernon ($\times$8.61), Hugo Weaving ($\times$7.72), Warwick Davis ($\times$7.39), John Ratzenberger ($\times$7.35), Mickie McGowan ($\times$7.09), Christopher Lee ($\times$6.86), Danny Mann ($\times$6.50), Julie Walters ($\times$6.27), Ian McKellen ($\times$6.21), Lasco Atkins ($\times$6.16), Denis Leary ($\times$5.68), Sacha Baron Cohen ($\times$5.67), Tom Felton ($\times$5.57), John DiMaggio ($\times$5.56), Chris Ellis ($\times$5.43), Andy Serkis ($\times$5.41), Julie Andrews ($\times$5.37), Frank Oz ($\times$5.22), Bonnie Hunt ($\times$5.08), Timothy Spall ($\times$5.06)

\paragraph{Deconfounder, One at a Time}
Lasco Atkins ($\times$5.64), Sacha Baron Cohen ($\times$4.24), John Ratzenberger ($\times$4.01), Desmond Llewelyn ($\times$3.91), Will Smith ($\times$3.64), Tom Cruise ($\times$3.63), Brad Garrett ($\times$3.56), Hugo Weaving ($\times$3.46), Ving Rhames ($\times$3.38), Ian McKellen ($\times$3.30), Naomie Harris ($\times$3.27), Jeffrey Tambor ($\times$3.13), Warwick Davis ($\times$3.10), Ava Acres ($\times$3.06), Orlando Bloom ($\times$3.06), Jet Li ($\times$2.94), Julie Walters ($\times$2.92), Brent Spiner ($\times$2.91), Stan Lee ($\times$2.91), Lois Maxwell ($\times$2.90), Corey Burton ($\times$2.79), Christoph Waltz ($\times$2.78), Jess Harnell ($\times$2.77), Michelle Rodriguez ($\times$2.72), Judi Dench ($\times$2.66)

\subsection{Top 25 Actors According to Each Estimator, Ranked by Appearance-weighted Log-scale Coefficients $n_j \hat\beta_j$}

\paragraph{Naive, Full}
Stan Lee (58.01), John Ratzenberger (48.98), Tom Hanks (46.81), Tom Cruise (41.21), Harrison Ford (36.87), Arnold Schwarzenegger (36.84), Morgan Freeman (36.02), Will Smith (35.36), Eddie Murphy (32.55), Leonardo DiCaprio (32.41), Brad Pitt (31.35), Bruce Willis (30.70), Judi Dench (30.42), Robert De Niro (29.12), John Travolta (28.56), Denzel Washington (27.26), Jim Carrey (26.84), Jack Black (26.83), Robin Williams (26.06), Desmond Llewelyn (25.75), John Leguizamo (23.72), Scarlett Johansson (22.62), Octavia Spencer (21.71), Leslie Mann (21.57), Matt Damon (21.46)

\paragraph{WB Deconfounder, Full}
Stan Lee (57.96), John Ratzenberger (46.53), Tom Hanks (46.48), Tom Cruise (40.95), Harrison Ford (37.05), Arnold Schwarzenegger (36.59), Morgan Freeman (35.86), Will Smith (35.39), Eddie Murphy (34.33), Brad Pitt (32.93), Judi Dench (31.80), Leonardo DiCaprio (30.07), Bruce Willis (29.15), Robin Williams (28.20), Jack Black (28.01), Robert De Niro (27.60), Jim Carrey (27.26), John Travolta (26.68), John Leguizamo (25.68), Liam Neeson (25.24), Denzel Washington (24.91), Desmond Llewelyn (24.36), Matt Damon (24.22), Scarlett Johansson (24.21), Octavia Spencer (24.11)

\paragraph{Deconfounder, Full}
Tom Cruise (88.80), Tom Hanks (62.45), George Clooney (55.52), John Ratzenberger (53.01), Octavia Spencer (44.04), Charlize Theron (43.36), Mark Wahlberg (42.11), Bruce Willis (40.40), Morgan Freeman (38.59), Ryan Reynolds (38.07), Harrison Ford (37.17), Leonardo DiCaprio (35.88), Will Smith (35.71), Jim Broadbent (34.99), Jason Statham (34.81), Stellan Skarsgård (32.83), Jeffrey Tambor (32.52), Patrick Wilson (32.25), Jason Flemyng (31.98), Penélope Cruz (31.94), Zoe Saldana (30.97), Laurence Fishburne (30.86), Sylvester Stallone (30.23), James Remar (30.12), Mickey Rourke (29.07)

\paragraph{Budget Adjustment, Full}
Tom Hanks (32.14), John Ratzenberger (27.85), Harrison Ford (27.61), Morgan Freeman (23.66), Leonardo DiCaprio (23.30), Octavia Spencer (21.69), Reese Witherspoon (21.35), Robert De Niro (21.35), Tom Cruise (21.23), Denzel Washington (20.95), Scarlett Johansson (20.85), Eddie Murphy (20.59), Keira Knightley (20.27), Ralph Fiennes (20.20), Sacha Baron Cohen (19.36), Alan Rickman (19.27), Desmond Llewelyn (19.17), John Leguizamo (19.00), Jim Carrey (18.88), George Clooney (18.87), John Travolta (18.86), Robin Williams (18.82), Philip Seymour Hoffman (18.16), Patricia Clarkson (18.02), Brad Pitt (17.34)

\paragraph{Naive, One at a Time}
Stan Lee (59.49), John Ratzenberger (45.50), Tom Cruise (42.97), Morgan Freeman (42.31), Hugo Weaving (40.47), Tom Hanks (37.46), Samuel L  Jackson (36.98), Frank Welker (36.69), Will Smith (35.78), Jess Harnell (35.11), Ian McKellen (33.52), Christopher Lee (31.19), Bill Hader (30.36), Liam Neeson (29.64), Bruce Willis (29.30), Orlando Bloom (29.26), Cameron Diaz (29.09), Brad Pitt (28.29), Judi Dench (27.80), Warwick Davis (27.74), Danny Mann (27.20), Cate Blanchett (27.14), Stellan Skarsgård (27.05), Alan Tudyk (26.92), Harrison Ford (26.72)

\paragraph{WB Deconfounder, One at a Time}
Stan Lee (56.45), John Ratzenberger (43.90), Tom Cruise (40.47), Morgan Freeman (39.43), Hugo Weaving (38.83), Tom Hanks (38.07), Jess Harnell (36.42), Frank Welker (35.07), Will Smith (34.04), Samuel L  Jackson (32.97), Ian McKellen (32.86), Liam Neeson (32.22), Bill Hader (31.23), Christopher Lee (30.82), Cameron Diaz (30.40), Judi Dench (29.17), Bruce Willis (28.90), Cate Blanchett (28.68), Orlando Bloom (28.59), Brad Pitt (28.00), Jonah Hill (27.59), Stellan Skarsgård (27.30), Alan Tudyk (27.15), Danny Mann (26.21), Harrison Ford (26.00)

\paragraph{Deconfounder, One at a Time}
Tom Cruise (42.54), Will Smith (31.01), John Ratzenberger (30.58), Morgan Freeman (29.26), Harrison Ford (27.97), Stan Lee (27.76), Tom Hanks (25.50), Arnold Schwarzenegger (24.14), Frank Welker (23.91), Hugo Weaving (23.59), Desmond Llewelyn (23.18), Liam Neeson (22.72), Jim Carrey (22.59), Judi Dench (22.51), Ving Rhames (21.90), Ian McKellen (21.52), Bruce Willis (21.47), Robin Williams (19.48), Jeffrey Tambor (18.27), Patrick Stewart (18.06), Mark Wahlberg (17.77), Alan Tudyk (17.41), Lasco Atkins (17.31), Hugh Jackman (16.05), Lois Maxwell (15.98)

\end{document}